\documentclass[letterpaper, 10pt, conference]{ieeeconf}      

\IEEEoverridecommandlockouts                              
\title{\LARGE \bf Hybrid path-lifting algorithm and Equivalence of Stability results for MRP-based control strategies}

\author{Luís Martins and Carlos Cardeira and Paulo Oliveira
}

\author{Luís~Martins,
        Carlos~Cardeira,
        and~Paulo~Oliveira
\thanks{L. Martins is with the Institute of Mechanical Engineering, Instituto Superior Técnico, Universidade de Lisboa, Lisboa, Portugal {\tt\small luis.cunha.martins@tecnico.ulisboa.pt}}
\thanks{C. Cardeira is with the Institute of Mechanical Engineering, Instituto Superior Técnico, Universidade de Lisboa, Lisboa, Portugal {\tt\small carlos.cardeira@tecnico.ulisboa.pt}}
\thanks{P. Oliveira is with the Institute of Mechanical Engineering and the Institute for Systems and Robotics, Instituto Superior Técnico, Universidade de Lisboa, Lisboa, Portugal {\tt\small paulo.j.oliveira@tecnico.ulisboa.pt}}}

\usepackage{graphicx}
\usepackage{amsmath}
\usepackage{subfigure}
\usepackage{float}
\usepackage{esvect}
\usepackage{amssymb}
\usepackage{tikz}

\usepackage[colorlinks=true,allcolors=blue]{hyperref}

\makeatletter
\renewcommand*{\eqref}[1]{%
  \hyperref[{#1}]{\textup{\tagform@{\ref*{#1}}}}%
}
\makeatother

\newtheorem{lem}{Lemma}
\newtheorem{thm}{Theorem}

\newtheorem{defn}{Definition}

\DeclareMathOperator{\Tr}{\textrm{tr}}
\DeclareMathOperator{\diag}{diag}

\usetikzlibrary{shapes,arrows}
\tikzstyle{block} = [draw, rectangle, 
    minimum height=0.75cm, minimum width=1cm]

\usetikzlibrary{positioning, arrows.meta}

\begin{document}

\maketitle
\thispagestyle{empty}
\pagestyle{empty}

\begin{abstract}
The modified Rodrigues parameters (MRP) consist of two numerically different triplets that, by switching between them, yield a minimal globally non-singular attitude description with advantageous properties. The MRP space results from the Alexandroff compactification of the three-dimensional Euclidean space and is a double cover of $\mathrm{SO(3)}$. By capitalizing on instrumental properties of the covering map, this paper proposes a novel hybrid dynamic path-lifting mechanism to unambiguously and robustly extract the MRP from the attitude space. This hybrid solution allows applying an MRP-based feedback controller to the attitude dynamics in the base space while preserving its asymptotic and exponential stability properties. Furthermore, by profiting from the distinct characteristics of the MRP, the resulting interconnection is impervious to the unwinding phenomenon. The design and validation of an MRP-based controller exemplify the application of the proposed algorithm alongside the novel results for equivalence of stability between spaces. The solution renders the attitude space tracking dynamics robustly globally exponentially stable, demonstrating the potential of this novel methodology.


\end{abstract}

\section{Introduction}
\label{section1}

\subsection{Background and Motivation}

\indent Attitude control is instrumental in aerospace and robotic engineering, as it focuses on designing reliable, effective, and advanced strategies to ensure the stability, maneuverability, and precise operation of spacecraft, satellites, aircraft, and autonomous vehicles. Continuous development of these solutions further enhances the performance, accuracy, and reliability of the resulting attitude control and tracking systems, extending their capabilities and contributing to the advancement of numerous technological applications.

\par The three-dimensional special orthogonal group, $\mathrm{SO}(3)$, is the configuration manifold for the attitude dynamics of a rigid body. This compact smooth topological space is not diffeomorphic to any vector space and, as discussed in \cite[Theorem 1]{BhatBernstein2000}, entails a topological obstruction to the global asymptotic stabilization of a given equilibrium point using continuous feedback. Furthermore, according to \cite{mayhew2011topological}, this topological obstruction also prevents the robust global asymptotic stabilization of the attitude dynamics through discontinuous feedback. State-of-the-art attitude control strategies rely on the hybrid systems framework to bypass the topological particularities of the three-dimensional rotation group representation, achieving, in this way, robust and global results (see, for instance, \cite{lee2015global, berkane2017hybrid}).

\par Alternatively, one can resort to a cover manifold of the rotation group $\mathrm{SO}(3)$ to model the attitude dynamics of a rigid body. The relation between the two topological spaces is given by a covering map that is everywhere a local diffeomorphism. By relying on the unique path-lifting property of covering spaces, it is, thus, possible to obtain a less intricate representation of a given element of the rotation group space that requires fewer parameters and satisfies fewer constraints. Nevertheless, when this map is a multiple covering of the base space, the resulting representations are an overparameterization of $\mathrm{SO}(3)$. The unit quaternion representation, whose domain is a double covering manifold of $\mathrm{SO}(3)$, is an example of this case. This particularity raises some critical repercussions, as it requires stabilizing two antipodal points to obtain a global attitude stabilization/tracking result. When this need is overlooked in the design of a controller relying on the unit quaternion representation, the resulting solution is susceptible to the phenomenon of unwinding, in which the rigid body unnecessarily performs the longest available rotation even when the attitude error is minimal. For a more detailed characterization of this phenomenon, please consult \cite{BhatBernstein2000, mayhew2013}. To circumvent the described drawback and obtain a global stabilizing/tracking result, prevalent methodologies from the literature resort to the hybrid systems theory. For example, the papers \cite{mayhew2011quaternion, invernizzi2022global, Tong2023} detail quaternion-based controllers for robust global asymptotic attitude tracking.

\par The modified Rodrigues parameters, first proposed by Wiener in \cite{wiener1962}, are a distinctive formalism for attitude description that stems from the stereographic projection of the unit quaternion representation and comprises two numerically different triplets. By judiciously switching between these sets, the MRP yield a minimal globally non-singular attitude representation with unique features whose exploitation leads to less complex control structures and increased efficiency \cite{terzakis2018}. Despite the distinct characteristics, the potential of this sophisticated representation remains under-explored in the literature. Since the MRP derive from the unit quaternion representation, the respective state space is, inevitably, a multiple covering manifold of $\mathrm{SO}(3)$. Hence, designing MRP-based strategies without carefully handling the ramifications of using local coordinates produces, at best, an almost global stability result, inasmuch as the region of attraction excludes a set of measure zero, potentially leading, in this way, to the unwinding phenomenon. Analogously to quaternion-based controllers, works from the literature resort to the hybrid systems theory to address this critical issue and simultaneously endow the strategy with some robustness properties. Furthermore, the formulation provides a suitable framework to capitalize on the unique characteristics of the MRP and to capture the discrete evolution arising from the switching between the two triplets. Martins et al., in \cite{martins2021CDC}, proposed an MRP-based global asymptotic trajectory tracking hybrid controller for an unmanned aerial vehicle.

\subsection{Related Work}

\par Global asymptotic stability of a dynamic system is highly contingent upon the topology of its state space, as the region of attraction of an asymptotically stable result is homeomorphic to the real coordinate space of equivalent dimension \cite{BhatBernstein2000}. Consequently, under the penalty of arriving at dubious claims, evaluating global stability results in $\mathrm{SO}(3)$ entails a sophisticated and thorough approach. In the vast majority of attitude control solutions from the literature that rely on local coordinates, the stability analysis is carried out exclusively in the covering manifold, thereby focusing on the lifted dynamics and lacking a more in-depth examination to ascertain whether an equivalent global stability result also holds in the base space and, thus, for the actual system. Indeed, a closed-loop system relying on local coordinates may not yield a well-defined system in the base space due to the inconsistency of having multiple values assigned to a given rigid-body attitude. This vital detail justifies the relevance of studying this stability equivalence and devising mechanisms to ensure it. On these grounds, Mayhew et al., in \cite{mayhew2013}, highlighted the criticality of pairing inconsistent unit quaternion-based feedback laws with a dynamic extraction scheme, responsible for uniquely and smoothly lifting an attitude path to the unit quaternion space, by showing that coupling them with a static quaternion selection mechanism gives rise to a non-robust dynamical system liable to the unsought chattering effect. To address the inconsistency, the authors proposed a novel hybrid dynamic path-lifting algorithm that, in addition to the judicious extraction, enables using quaternion-based feedback laws in the rotation group space without compromising the asymptotic stability properties obtained in the covering space. The solutions reported in \cite{casau2015, Naldi2017} rely on this hybrid mechanism.

\par Given the inextricable link of MRP to unit quaternion representation, similar problems may arise when resorting to them. Thus, to address these crucial issues and motivated by the under-explored potential of the MRP representation and the lack of results in this regard in the literature, inspired by the work presented in \cite{mayhew2013}, this paper formulates an innovative hybrid dynamic path-lifting algorithm for MRP extraction that allows translating MRP based controllers devised in the covering space to the rigid body attitude manifold while conserving asymptotic and exponential stability properties. To this end, first, considering the Alexandroff compactification \cite[p. 246]{dugundji1966topology} of $\mathbb{R}^3$, $\mathbb{\bar{R}}^3$, as the MRP space, which is a double cover of $\mathrm{SO}(3)$, the authors demonstrate that the corresponding covering map is everywhere a local diffeomorphism. Then, profiting from the path-lifting property of covering maps, the hybrid dynamic extraction algorithm is constructed to uniquely lift the MRP representation from the attitude rigid body space. The algorithm presented in this study expands upon the one proposed in \cite{mayhew2013} by incorporating the mathematical operation of stereographic projection. This extension also includes a memory state, indicating the MRP set in use and instrumental for selecting the projection pole, and a constant hysteresis parameter, introduced to endow the algorithm with robustness against measurement disturbances. The output of this approach evolves within a three-dimensional unit sphere enclosed by a hysteresis region adjustable through the mentioned parameter. This region is crucial for ensuring the unambiguous selection of the MRP and avoiding noise-induced chattering when transitioning between the original and shadow MRP sets. The switching logic of the novel hybrid methodology is predicated on a unique and valuable characteristic inherent to the MRP representation. Specifically, the triplet with the smallest norm invariably describes the shortest rotational direction \cite[p.~120]{junkins_2009}. This pivotal characteristic serves as the linchpin for the dynamic lifting mechanism, enabling the selection of the shortest principal rotation based on the norm of the MRP triplets. Furthermore, by devising a control strategy that capitalizes on this distinctive characteristic, the approach elegantly excludes any possibility of unwinding as it makes expendable additional control variables to ensure this sought behavior, being, thus, a clear advantage in comparison with quaternion-based solutions (cf. \cite{mayhew2011quaternion}). With this hybrid dynamic mechanism in place, the authors develop a theoretical basis for equivalence of asymptotic and exponential stability between results in the base space $\mathrm{SO}(3)$ and the covering space $\mathbb{\bar{R}}^3$. To this end, two closed-loop hybrid systems are formulated: one encompassing the closed-loop lifted attitude dynamics and the other encapsulating the interconnection between the attitude dynamics described in the base space, the hybrid path-lifting algorithm, and the MRP-based controller. Additionally, the evaluation of the equivalence of stability properties considers the memory states of the hybrid path-lifting mechanism. Finally, the work exemplifies the practical application of the resulting stability theorems through the design, analysis, and validation of a solution for the challenging rigid body attitude problem of robust global trajectory tracking.

\subsection{Contributions}

\par The main contributions of this paper revolve around the novel hybrid dynamic path-lifting mechanism and the equivalence of stability properties sustained by its application. Specifically, these contributions can be summarized as follows.
\begin{enumerate}
    \item To the best knowledge of the authors, the proposed mechanism is the first dynamic methodology for uniquely lifting a path from the rotation group to the MRP space. The hybrid mechanism capitalizes on the distinct properties of the MRP representation, is robust to measurement disturbances, allows appending dynamical equations for the output, and can be paired with an MRP-based feedback law without yielding a control structure susceptible to the unwinding phenomenon. 
    \item The latter feature is a noteworthy contribution and advantage compared to quaternion-based solutions as the dynamic lifting mechanism ensures, on its own, that the interconnection with an MRP-based feedback law has the anti-unwinding property, thereby bypassing the design of ancillary control methodologies to guarantee that the MRP-based feedback stabilizes all local representations of the desired attitude. Compared with other anti-unwinding MRP-based solutions from the literature, the work presented in \cite{dong2021} is focused on rest-to-test maneuvers, only effectively dealing with the unwinding phenomenon for the initial condition and not accounting for the ambiguity that arises from measurement disturbances, whereas the mechanism proposed in this paper robustly rules out any possibility of unwinding at any given instant. 
    \item Apart from solving the ambiguity in the selection of the MRP used for feedback, the hybrid dynamic algorithm ensures that the controller designed in the covering space $\mathbb{\bar{R}}^3$ is well-defined in the base space $\mathrm{SO}(3)$.
    \item Furthermore, as the primary contribution, this paper demonstrates that an asymptotic/exponential stability result obtained with a feedback controller designed and analyzed in the covering space for the closed-lo lifted attitude dynamics translates into an equivalent asymptotic/exponential stability result in the base space for a hybrid closed-loop system encapsulating the interconnection between the rigid body attitude dynamics, the hybrid path-lifting algorithm, and the same MRP-based feedback controller. It is worth emphasizing that the existing literature lacks theoretical results regarding the relation between stability analysis conducted in the MRP space and the stability results for the rigid body attitude system when resorting to a controller designed in the covering space. Even within the broader spectrum of solutions relying on parameterizations of $\mathrm{SO}(3)$, compared to \cite{mayhew2013}, this paper distinguishes itself by delving deeper. While the work from the literature exclusively focuses on establishing the equivalence of asymptotic stability, this paper advances beyond by introducing a new theorem for translating exponential stability results between the covering and base spaces.
    \item These novel theoretical results lay the foundation for devising control strategies in a covering manifold that, in comparison with the rotation group $\mathrm{SO}(3)$, or even the unit quaternion space $\mathbb{S}^3$, has a simpler structure, requiring fewer parameters and satisfying fewer constraints to describe its elements. Moreover, harnessing these new insights along with the distinctive properties of the MRP representation enables achieving a robust global stability result with a less intricate control structure that does not require formulating multiple potential or error functions, granting a significant advantage over solutions directly developed on $\mathrm{SO}(3)$ (cf. \cite{lee2015global, berkane2017hybrid}).
    \item In this direction, to showcase the potential of the proposed theorems, the paper presents a novel control solution for the attitude tracking problem in the rotation group. This innovative control strategy combines the hybrid dynamic path-lifting algorithm with an MRP-based feedback law designed in the covering space that, excluding a trivial feedforward canceling term, is strictly composed of linear terms. The resulting control structure elegantly solves the problem by rendering the rigid body attitude tracking dynamics robustly globally exponentially stable. Furthermore, it is straightforward to extend this solution to tackle similar challenges that involve 6-DOF rigid bodies.

\end{enumerate}   

\subsection{Organization}

\par The paper is organized as follows: \hyperref[section2]{section~\ref*{section2}} details the notation used and introduces some key concepts regarding attitude description, hybrid systems, and attitude dynamics and kinematics; \hyperref[section3]{section~\ref*{section3}} demonstrates instrumental properties of the function mapping from the MRP space to the rotation group ; \hyperref[section4]{section~\ref*{section4}} formulates and dissects the hybrid dynamic path-lifting algorithm for uniquely and consistently extract the MRP representation from a given rotation matrix; \hyperref[section5]{section~\ref*{section5}} proposes the novel theoretical framework that enables bridging the gap in terms of equivalence of stability between results obtained in the covering and base spaces;  \hyperref[section6]{section~\ref*{section6}} exemplifies the application of these theoretical results through the design in the covering space and validation of a hybrid controller that yields robust global exponential attitude tracking result in the base space; finally, \hyperref[section7]{section~\ref*{section7}} concludes the paper with some remarks.

\section{Notation and Preliminaries}
\label{section2}


Let $\mathbb{R}^n$ represent the $n$-dimensional Euclidean space, $\mathbb{R}_{\geq 0}$ express the set of non-negative real numbers, $\mathbb{N}$ symbolize the set of natural numbers, $K\mathbb{B}^n$ denote the closed ball of radius $K$ centered at the origin of $\mathbb{R}^n$, $\mathbb{R}^{n\times m}$ correspond to the set of $n \times m$ matrices, $\mathbb{S}^n = \left\{\mathbf{x} \in \mathbb{R}^{n+1}: \mathbf{x}^{\!\top}\mathbf{x} = 1\right\}$ symbolize the $n$-dimensional unit sphere, and $\mathbb{\bar{R}}^n = \mathbb{R}^n \cup \{\boldsymbol{\infty}\}$ denote the Alexandroff compactification of $\mathbb{R}^n$. Given two functions $\mathbf{f}:\mathcal{X} \mapsto \mathcal{Y}$ and $\mathbf{g}:\mathcal{Y} \mapsto \mathcal{Z}$, $\mathbf{g}\circ\mathbf{f} = \mathbf{g}\left(\mathbf{f}\right): \mathcal{X} \mapsto \mathcal{Z}$ denotes the composite function $\mathbf{g}$ of $\mathbf{f}$; $\mathbf{F}:\mathcal{X} \rightrightarrows \mathcal{Y}$ represents the set-valued map $\mathbf{F}$ from $\mathcal{X}$ to $\mathcal{Y}$ and, for $\mathbf{x} \in \mathcal{X}$, the relation $\mathbf{\dot{x}} \in \mathbf{F}(\mathbf{x})$ expresses a differential inclusion; $\textrm{dom }V$ symbolizes the domain of the function $V$; $\mathbf{I_n} \in \mathbb{R}^{n \times n}$ represents the $n$-dimensional identity matrix; $\mathbf{e_i} \in \mathbb{R}^3$ denotes a vector of zeros except for the i$^{\textrm{th}}$ entry which is 1; $\|\cdot\|$ represents the Euclidean norm; \color{black} the operator $\left[\cdot\right]_\times: \mathbb{R}^3 \mapsto \{\mathbf{S} \in \mathbb{R}^{3 \times 3}: \mathbf{S}^\top = -\mathbf{S}\}$ denotes the mapping of a vector in $\mathbb{R}^3$ to a $3 \times 3$ skew-symmetric matrix such that $\left[\boldsymbol{\omega}\right]_\times\boldsymbol{s} = \boldsymbol{\omega} \times \mathbf{s}$ for any $\mathbf{s},\boldsymbol{\omega} \in \mathbb{R}^3$, where $\times$ denotes the cross product \cite{mayhew2013}; \color{black} for a given square matrix $\mathbf{A} \in \mathbb{R}^{n\times n}$, $\lambda_{\min}(\mathbf{A})$ and $\lambda_{\max}(\mathbf{A})$ denote the minimum and maximum eigenvalues, respectively, 
$\|\mathbf{A}\| = (\lambda_{max} (\mathbf{A}^{\!\top} \mathbf{A}))^{1/2}$ corresponds to the spectral norm, $\Tr\left(\mathbf{A}\right)$ symbolizes the trace and $\|A\|_{F} \triangleq (\Tr(A^{\!\top}\!A))
^{1/2}$ defines the Frobenius norm; $\diag\left(\mathbf{s}\right)$ is such that $\diag\left(\mathbf{s}\right) \triangleq \sum_{i=1}^n (  \mathbf{e_i}\mathbf{e_i^{\!\top}})(\mathbf{e_i^{\!\top}}\mathbf{s})$ for $\mathbf{s} \in \mathbb{R}^n$; $\|\mathbf{s}\|_{\mathcal{A}}$ denotes the distance of $\mathbf{s}$ to $\mathcal{A}$ and is given by $\|\mathbf{s}\|_{\mathcal{A}} := \inf_{\mathbf{x} \in \mathcal{A}} \|\mathbf{s} - \mathbf{x}\|$. The cardinality of a given set $\mathcal{A} \subset \mathbb{R}^n$ is denoted $|\mathcal{A}|$. The saturation function considered in this work is aligned with the following definition:
\begin{defn}
\label{def:SaturationFunction}
\color{black} The mapping $\boldsymbol{\sigma}:\mathbb{R}^m \mapsto \mathbb{R}^m$ is an independent function, i.e.,  $\boldsymbol{\sigma}\left(\mathbf{s}\right) \triangleq \left[\sigma_1\left(s_1\right) \cdots \sigma_m\left(s_m\right) \right]$, \color{black} where each $\sigma_i$ is a smooth strictly increasing odd function satisfying: (1) $\sigma_i\left(0\right) = 0$; (2) $s_i\sigma_i\left(s_i\right) > 0\;\; \forall \;\;s_i \neq 0$; (3) $\lim_{s_i\to \pm\infty} \sigma_i\left(s_i\right) = \pm M$, with $M \in \mathbb{R}_{>0}$; (4) $\dot{\sigma}(s_i) \leq 1$; (5) $\ddot{\sigma}(s_i) \leq 0, \; \textrm{for}\;\; s_i \geq 0$. $\hfill \square$
\end{defn}

\par Concerning rigid-body attitude description, $\mathbf{R}$ represents an element of the three-dimensional special orthogonal group $\textrm{SO(3)}$ and corresponds to the rotation matrix from the body-fixed frame to the inertial frame. The underlying kinematic and dynamic equations for the rotation of a rigid body are, respectively, 
\begin{subequations}
    \label{eq:AttitudeDynamics}
    \begin{equation}
        \mathbf{\dot{R}} = \mathbf{R}\left[\boldsymbol{\omega}\right]_{\times}
    \end{equation}
    \begin{equation}
    \label{eq:AttitudeDynamicEquation}
        \mathbf{J}\boldsymbol{\dot{\omega}} = \left[\mathbf{J}\boldsymbol{\omega}\right]_{\times}\boldsymbol{\omega} + \boldsymbol{\tau}
    \end{equation}
\end{subequations}
where $\boldsymbol{\omega} \in \mathbb{R}^3$ symbolizes the angular velocity expressed in the body-fixed frame, $\boldsymbol{\tau} \in \mathbb{R}^3$ denotes the torque input, $\mathbf{J} \in \mathbb{R}^{3\times3}$ models the symmetric tensor of inertia of the rigid body.
\par The vector $\mathbf{q} \in \mathbb{S}^3$ denotes the unit quaternion and is defined by the pair $\left(q_0, \mathbf{q_1}\right)$, where $q_0 \in \mathbb{R}$ and $\mathbf{q_1} \in \mathbb{R}^3$ correspond, respectively, to the scalar and vector components. \color{black} A given unit quaternion represents an element of $\mathrm{SO}(3)$ through the map $\mathcal{R}: \mathbb{S}^3 \mapsto \mathrm{SO}(3)$ defined as \cite[Eq. 5]{mayhew2013}
\begin{equation}
    \mathcal{R}(\mathbf{q}) = \mathbf{I_3} + 2 q_0\left[\mathbf{q_1}\right]_\times + 2\left[\mathbf{q_1}\right]_\times^2,
\end{equation}
which satisfies $\mathcal{R}(\mathbf{q}) = \mathcal{R}(\mathbf{-q})$ \color{black}. The double-valued inverse map $\mathcal{Q}: \mathrm{SO}(3) \rightrightarrows \mathbb{S}^3$ is characterized as 
\begin{equation}
\mathcal{Q}(\mathbf{R}) = \left\{\mathbf{q} \in \mathbb{S}^3: \mathcal{R}(\mathbf{q}) = \mathbf{R}\right\}.    
\end{equation} 
Alternatively to the previous representations, the MRP vector, $\boldsymbol{\vartheta} \in \mathbb{\bar{R}}^3$, can also be used to parameterize the attitude. Each $\boldsymbol{\vartheta}$ has a shadow MRP associated, $\boldsymbol{\vartheta}^s \in \mathbb{\bar{R}}^3$. Both $\boldsymbol{\vartheta}$ and $\boldsymbol{\vartheta}^s$ are related to a given unit quaternion through
\begin{subequations}
\label{eq:StereographicProjection}
\begin{equation}
\label{eq:quat2OriginalMRP}
\boldsymbol{\vartheta} = \left\{\begin{array}{cl}
     \frac{\mathbf{q_1}}{1 + q_0} & \!\!, \; \textrm{for} \; \mathbf{q} \in \mathbb{S}^3 \setminus \{\mathbf{s}\}\\
     \boldsymbol{\infty} & \!\!, \; \textrm{for} \; \color{black} \mathbf{q} = \mathbf{s} \color{black}
\end{array}\right.
\end{equation}
\begin{equation}
\label{eq:quat2ShadowMRP}
\boldsymbol{\vartheta}^s = \left\{\begin{array}{cl}
     \frac{\mathbf{-q_1}}{1 - q_0} & \!\!, \; \textrm{for} \; \mathbf{q} \in \mathbb{S}^3 \setminus \{\mathbf{n}\}\\
     \boldsymbol{\infty} & \!\!, \; \textrm{for} \; \color{black} \mathbf{q} = \mathbf{n} \color{black} 
\end{array}\right.,
\end{equation}
\end{subequations}
where $\mathbf{s} = (-1,0,0,0)$ and  $\mathbf{n} = (1,0,0,0)$ are, respectively, the south and north poles of the three-dimensional sphere. In virtue of the original and shadow sets being singular for different rotations, judiciously switching between the original and shadow sets yields a minimal non-singular attitude representation \cite{junkins_2009}.  The shadow set can be obtained from the original set by resorting to the map $\boldsymbol{\Upsilon}: \mathbb{\bar{R}}^3 \mapsto \mathbb{\bar{R}}^3$:
\begin{equation}
    \label{eq:ShadowMap}
    \boldsymbol{\vartheta^s} =  \boldsymbol{\Upsilon}\!\left(\boldsymbol{\vartheta}\right) =  \left\{ \begin{array}{cl}
         -\boldsymbol{\vartheta}\|\boldsymbol{\vartheta}\|^{-2}
         & \!\!, \; \textrm{for} \; \boldsymbol{\vartheta} \in \mathbb{R}^3 \setminus \{\boldsymbol{0}\}  \\
         \boldsymbol{\infty} & \!\!, \; \textrm{for} \; \boldsymbol{\vartheta} \in \{\boldsymbol{0}\}  \\
    \boldsymbol{0} & \!\!, \; \textrm{for} \; \boldsymbol{\vartheta} \in \{\boldsymbol{\infty}\}
    \end{array} 
    \right..
\end{equation}
Both the original and the shadow MRP respect the following kinematic equation \cite{junkins_2009}
\begin{equation}
    \label{eq:AttitudeMRPKinematics}
    \boldsymbol{\dot{\vartheta}} \!=\! \mathbf{T}\!(\!\boldsymbol{\vartheta}\!)\boldsymbol{\omega} \!=\! \left\{\!\!\!\!\begin{array}{cl}
        \frac{(1-\|\boldsymbol{\vartheta}\|^2)\mathbf{I_3} + 2\left[\boldsymbol{\vartheta}\right]_{\!\times} \!+ 2 \boldsymbol{\vartheta} \boldsymbol{\vartheta}^{\!\top}\!}{4}\boldsymbol{\omega} &  \!\!\!, \; \textrm{for} \; \boldsymbol{\vartheta} \!\in\! \mathbb{R}^3 \\
         \boldsymbol{\infty} & \!\!\!, \; \textrm{for} \; \boldsymbol{\vartheta} \!\in\! \{\!\boldsymbol{\infty}\!\} 
    \end{array}\right. .
\end{equation}
The mapping $\mathcal{R}_{\boldsymbol{\vartheta}}\!\left(\boldsymbol{\vartheta}\right):\mathbb{\bar{R}}^3 \mapsto \textrm{SO(3)}$
\begin{equation}
    \label{eq:rotationmatrixMRP}
    \mathcal{R}_{\boldsymbol{\vartheta}}\!\left(\!\boldsymbol{\vartheta}\!\right) := \!\! \left\{ \!\!\!\begin{array}{cl}
         \mathbf{I_3} + \frac{ 8\left[\boldsymbol{\vartheta}\right]_{\!\times}^2 + 4(1 - \|\boldsymbol{\vartheta}\|^2)\!\left[\boldsymbol{\vartheta}\right]_{\!\times}}{(1 + \|\boldsymbol{\vartheta}\|^2)^{2}} &  \!\!, \; \textrm{for} \; \boldsymbol{\vartheta} \in \mathbb{R}^3\\
         \mathbf{I_3} & \!\!, \; \textrm{for} \; \boldsymbol{\vartheta} \in \{\boldsymbol{\infty}\}
    \end{array}\right.
\end{equation}
maps a given $\boldsymbol{\vartheta}$ to the corresponding rotation matrix. This map has the property $\mathcal{R}_{\boldsymbol{\vartheta}}(\boldsymbol{\vartheta}) = \mathcal{R}_{\boldsymbol{\vartheta}}(\boldsymbol{\vartheta^s})$. For further details, the reader is referred to \cite{junkins_2009}.
\par A hybrid system $\mathcal{H}$ is characterized by the quadruplet $\left(\mathbf{C},\;\mathbf{F},\;\mathbf{D},\;\mathbf{G}\right)$. Its model can be represented by
\begin{equation}
    \mathcal{H} \left \{\begin{array}{ll}
        \mathbf{\dot{x}} \in \mathbf{F}\left(\mathbf{x}\right) & ,\quad \mathbf{x} \in \mathbf{C} \\ 
        \mathbf{x^+} \in \mathbf{G}\left(\mathbf{x}\right) & ,\quad \mathbf{x^+} \in \mathbf{D}
    \end{array}
    \right. .
\end{equation}
\noindent The hybrid system evolves according to the set-valued map $\mathbf{F}:\mathbb{R}^n \rightrightarrows \mathbb{R}^n$ while in the flow set $\mathbf{C} \subset \mathbb{R}^n$ and instantaneously changes under the set-valued map $\mathbf{G}:\mathbb{R}^n \rightrightarrows \mathbb{R}^n$ while in the jump set $\mathbf{D} \subset \mathbb{R}^n$. A solution $\mathbf{x}(t,j)$ to $\mathcal{\mathcal{H}}$, with $t$ and $j$ denoting, respectively, ordinary time and jump time, is a function $\mathbf{x}:\textrm{dom}\; \mathbf{x}\mapsto\mathbb{R}^n$, where $\textrm{dom}\;\mathbf{x} \subset \mathbb{R}_{\geq0}\times \mathbb{N}$ is a hybrid time domain. For each fixed $j \in \mathbb{N}$ the function $t \mapsto \mathbf{x}(t,j)$ is locally absolutely continuous on the interval $\mathcal{T}_j:= \left\{t: (t,j) \in \textrm{dom } \mathbf{x} \right\}$. For a given hybrid time domain, $\bar{T}(\mathbf{x}) = \sup \left\{t \in \mathbb{R}_{\geq 0}: \exists\; j \in \mathbb{N} \textrm{ such that } (t,j) \in \textrm{dom}\;\mathbf{x}\right\}$. The function $\mathbf{x_{\downarrow t}}:\left[0, \bar{T}(\mathbf{x})\right) \mapsto \mathbb{R}^n$ symbolizes the time projection of $\mathbf{x}(t,j)$ given by $\mathbf{x_{\downarrow t}} = \mathbf{x}(t, J(t))$, with $J(t) = \max\left\{j \in \mathbb{N}: (t,j) \in \textrm{dom }\mathbf{x}\right\}$. For further details, see \cite{goebel_2012}.




\section{Properties of the map $\mathcal{R}_{\!\boldsymbol{\vartheta}}:\mathbf{\bar{R}}^3 \mapsto \mathrm{SO}(3)$}
\label{section3}

\par To devise a hybrid methodology for dynamic path-lifting from $\mathrm{SO(3)}$ to $\mathbb{\bar{R}}^3$ and establishing parallelisms in terms of stability results in these spaces, one has to ascertain whether the map $\mathcal{R}_{\!\boldsymbol{\vartheta}}:\mathbf{\bar{R}}^3 \mapsto \mathrm{SO}(3)$ possesses the necessary characteristics for those effects. Bearing in mind that $\mathcal{R}_{\!\boldsymbol{\vartheta}}$ consists of the composition of $\mathcal{R}$ and the inverse of the stereographic projections described in \eqref{eq:StereographicProjection} and that the properties of $\mathcal{R}$ are well-known from the literature, the focus is on showing that the stereographic projection from $\mathbb{S}^3$ to $\mathbb{\bar{R}}^3$ is a diffeomorphism. On the basis of this diffeomorphic relation, known arguments from the literature for composite maps can be applied to draw conclusions regarding the required characteristics of $\mathcal{R}_{\!\boldsymbol{\vartheta}}$.

\par Let the manifold $\mathbb{S}^3$ be described by the atlas $\mathcal{A}_1 = \left\{(\mathbf{U_n}, \boldsymbol{\varphi}_{\mathbf{n}}), (\mathbf{U_s}, \boldsymbol{\varphi}_{\mathbf{s}})\right\}$ where 
\begin{subequations}
 \label{eq:chartsS3}
    \begin{equation*}
    \mathbf{U_n} := \mathbb{S}^3 \setminus \{\mathbf{n}\}, \quad \boldsymbol{\varphi}_{\mathbf{n}}: \mathbf{U_n} \mapsto \mathbb{R}^3, \quad \boldsymbol{\varphi}_{\mathbf{n}} (\mathbf{q}) := \frac{-\mathbf{q_1}}{1 - q_0}
    \end{equation*}
    
    \begin{equation*}
    \mathbf{U_s} := \mathbb{S}^3 \setminus \{\mathbf{s}\}, \quad \boldsymbol{\varphi}_{\mathbf{s}}: \mathbf{U_s} \mapsto \mathbb{R}^3, \quad \boldsymbol{\varphi}_{\mathbf{s}} (\mathbf{q}) := \frac{\mathbf{q_1}}{1 + q_0}
    \end{equation*}
\end{subequations}
The functions $\boldsymbol{\varphi}_{\mathbf{n}}(\mathbf{q})$ and $\boldsymbol{\varphi}_{\mathbf{s}}(\mathbf{q})$ correspond, respectively, to the stereographic projection of $\mathbf{q}$ from the north pole of the sphere multiplied by $-1$ and the stereographic projection from the south pole. The referred functions verify the equality $\boldsymbol{\varphi}_{\mathbf{n}}(\mathbf{q}) = \boldsymbol{\varphi}_{\mathbf{s}}(-\mathbf{q})$. The inverse functions $\boldsymbol{\varphi}_{\mathbf{n}}^{-1}: \mathbb{R}^3 \mapsto \mathbf{U_n}$ and $\boldsymbol{\varphi}_{\mathbf{s}}^{-1}: \mathbb{R}^3 \mapsto \mathbf{U_s}$ are given by 
\begin{equation}
    \boldsymbol{\varphi}_{\mathbf{n}}^{-1} (\boldsymbol{\vartheta}) = \left(\frac{ \|\boldsymbol{\vartheta}\|^2 - 1}{1 + \|\boldsymbol{\vartheta}\|^2}, \frac{-2\boldsymbol{\vartheta}}{1 + \|\boldsymbol{\vartheta}\|^2}\right)
\end{equation}
\begin{equation}
\label{eq:InverseSouthPole}
    \boldsymbol{\varphi}_{\mathbf{s}}^{-1} (\boldsymbol{\vartheta}) = \left(\frac{1 - \|\boldsymbol{\vartheta}\|^2}{1 + \|\boldsymbol{\vartheta}\|^2}, \frac{2\boldsymbol{\vartheta}}{1 + \|\boldsymbol{\vartheta}\|^2}\right)
\end{equation}
Further, $\boldsymbol{\varphi}_{\mathbf{n}}$ and $\boldsymbol{\varphi}_{\mathbf{s}}$ are continuous bijective maps with continuous inverse functions, being, thereby, homeomorphisms. Furthermore, since $\boldsymbol{\varphi}_{\mathbf{n}}$ and $\boldsymbol{\varphi}_{\mathbf{s}}$ map from one-punctured spheres to $\mathbb{R}^3$, it follows that the pairs $(\mathbf{U_n}, \boldsymbol{\varphi}_{\mathbf{n}})$ and $(\mathbf{U_s}, \boldsymbol{\varphi}_{\mathbf{s}})$ are charts according to \cite[p.4]{Lee2013}. Regarding the sets $\mathbf{U_n}$ and $\mathbf{U_s}$, $\mathbf{U_n} \cap \mathbf{U_s} \neq \emptyset$ and $\mathbf{U_n} \cup \mathbf{U_s}$ covers $\mathbb{S}^3$. Hence, the indexed family of charts $\{(\mathbf{U_n}, \boldsymbol{\varphi}_{\mathbf{n}}), (\mathbf{U_s}, \boldsymbol{\varphi}_{\mathbf{s}})\}$ forms an atlas on $\mathbb{S}^3$. The transition map $\boldsymbol{\Psi}_{\mathbf{1}}(\boldsymbol{\vartheta}): \mathbb{R}^3 \setminus \{\boldsymbol{0}\} \mapsto  \mathbb{R}^3 \setminus \{\boldsymbol{0}\}$ yields
\begin{equation*}
    \boldsymbol{\Psi}_{\mathbf{1}}(\boldsymbol{\vartheta}) = \boldsymbol{\varphi}_{\mathbf{s}} \circ \boldsymbol{\varphi}_{\mathbf{n}}^{-1}(\boldsymbol{\vartheta}) = -\boldsymbol{\vartheta}\|\boldsymbol{\vartheta}\|^{-2},
\end{equation*}
which is identical to the mapping between the original and shadow sets, given in \eqref{eq:ShadowMap}, for $\boldsymbol{\vartheta} \in \mathbb{R}^3 \setminus \{\boldsymbol{0}\}$. Note that the transition map verifies $\boldsymbol{\Psi}_{\mathbf{1}}(\boldsymbol{\vartheta}) = \boldsymbol{\Psi}_{\mathbf{1}}^{-1}(\boldsymbol{\vartheta})$. Thus, the transition map and its inverse are smooth and, consequently, the two charts are smoothly compatible and $\mathcal{A}_1$ is a smooth atlas for $\mathbb{S}^3$ \cite[p.12]{Lee2013}.

\par Let the atlas $\mathcal{A}_2 = \left\{(\mathbf{U_a}, \boldsymbol{\varphi}_{\mathbf{a}}), (\mathbf{U_b}, \boldsymbol{\varphi}_{\mathbf{b}})\right\}$ with 
\begin{subequations}
 \label{eq:chartsRbar3}
    \begin{equation*}
    \mathbf{U_a} \!:=\! \mathbb{R}^3, \; \boldsymbol{\varphi}_{\mathbf{a}}\!: \mathbf{U_a} \mapsto \mathbb{R}^3, \; \boldsymbol{\varphi}_{\mathbf{a}} (\boldsymbol{\vartheta}) \!:=\! \boldsymbol{\vartheta}
    \end{equation*}
    \begin{equation*}
    \mathbf{U_b} \!:=\! \mathbb{\bar{R}}^3 \setminus \{\boldsymbol{0}\}, \; \boldsymbol{\varphi}_{\mathbf{b}}\!: \mathbf{U_b} \mapsto \mathbb{R}^3, \; \boldsymbol{\varphi}_{\mathbf{b}} (\!\boldsymbol{\vartheta}\!) \!:=\! \boldsymbol{\Upsilon}\!\left(\boldsymbol{\vartheta}\right), \; \textrm{for} \; \boldsymbol{\vartheta} \!\in\! 
    \mathbf{U_b}
    \end{equation*}
\end{subequations}
describe the manifold $\mathbb{\bar{R}}^3$. To shed some light on the reasoning behind the choice of these functions, one might think of $\boldsymbol{\varphi}_{\mathbf{a}}$ as the identity map for one of the MRP sets and $\boldsymbol{\varphi}_{\mathbf{b}}$ as the map that enables computing the counterpart representation. The functions $\boldsymbol{\varphi}_{\mathbf{a}}$ and $\boldsymbol{\varphi}_{\mathbf{b}}$ are continuous bijective open mappings. The inverse of these functions, given by

    \begin{equation*}
   \boldsymbol{\varphi}^{\!-\!1\!}_{\mathbf{a}}: \mathbb{R}^3 \mapsto \mathbf{U_a}, \quad \boldsymbol{\varphi}^{\!-\!1\!}_{\mathbf{a}} (\boldsymbol{\vartheta}) = \boldsymbol{\vartheta}
    \end{equation*}
    \begin{equation*}
    \boldsymbol{\varphi}^{\!-\!1\!}_{\mathbf{b}}: \mathbb{R}^3 \mapsto  \mathbf{U_b} , \quad \boldsymbol{\varphi}^{\!-\!1\!}_{\mathbf{b}} (\boldsymbol{\vartheta}) = \boldsymbol{\Upsilon}\!\left(\boldsymbol{\vartheta}\right), \; \textrm{for} \; \boldsymbol{\vartheta} \in 
    \mathbb{R}^3
    \end{equation*}
are continuous. Thus, $\boldsymbol{\varphi}_{\mathbf{a}}$ and $\boldsymbol{\varphi}_{\mathbf{b}}$ are homeomorphisms. The union $\mathbf{U_a} \cup \mathbf{U_b}$ covers $\mathbb{\bar{R}}^3$. Hence, in light of these properties, the charts $(\mathbf{U_a}, \boldsymbol{\varphi}_{\mathbf{a}})$ and $(\mathbf{U_b}$, $ \boldsymbol{\varphi}_{\mathbf{b}})$ form an atlas on 
$\mathbb{\bar{R}}^3$. The transition map $\boldsymbol{\Psi}_{\mathbf{2}}(\boldsymbol{\vartheta}): \mathbb{R}^3 \setminus \{\boldsymbol{0}\} \mapsto  \mathbb{R}^3 \setminus \{\boldsymbol{0}\}$ satisfies
\begin{equation*}
    \boldsymbol{\Psi}_{\mathbf{2}}(\boldsymbol{\vartheta}) = \boldsymbol{\varphi}_{\mathbf{b}} \circ \boldsymbol{\varphi}_{\mathbf{a}}^{-1}(\boldsymbol{\vartheta}) = -\boldsymbol{\vartheta}\|\boldsymbol{\vartheta}\|^{-2},
\end{equation*}
\begin{equation*}
        \boldsymbol{\Psi}_{\mathbf{2}}^{-1}(\boldsymbol{\vartheta}) = \boldsymbol{\varphi}_{\mathbf{a}} \circ \boldsymbol{\varphi}_{\mathbf{b}}^{-1}(\boldsymbol{\vartheta}) =-\boldsymbol{\vartheta}\|\boldsymbol{\vartheta}\|^{-2}.
\end{equation*}
By virtue of $\boldsymbol{\Psi}_{\mathbf{2}}$ and $\boldsymbol{\Psi}_{\mathbf{2}}^{-1}$ being smooth functions, the transition map is a diffeomorphism. Therefore, the two charts are smoothly compatible and $\mathcal{A}_2$ is a smooth atlas \cite[p.12]{Lee2013}.

\par Consider the map $\boldsymbol{\varphi}(\boldsymbol{\vartheta}): \mathbf{\Bar{R}}^3 \mapsto \mathbb{S}^3$ defined as
\begin{equation}
    \boldsymbol{\varphi}(\boldsymbol{\vartheta}) := \left\{\begin{array}{cl}
       \boldsymbol{\varphi}_{\mathbf{s}}^{-1}(\boldsymbol{\vartheta})  &, \; \textrm{for} \; \boldsymbol{\vartheta} \in \mathbb{R}^3  \\
        \mathbf{s} &, \; \textrm{for} \; \boldsymbol{\vartheta} \in \{\boldsymbol{\infty}\}  
    \end{array}\right. .
\end{equation}
For the purpose of characterizing the smoothness of $\boldsymbol{\varphi}$, the smooth atlas $\mathcal{A}_1$ and $\mathcal{A}_2$ are resorted to. In this direction, the following composite maps are constructed:
\begin{equation*}
    \boldsymbol{\Theta}_{\mathbf{1}}\!: \boldsymbol{\varphi}_{\mathbf{b}}\!\left(\mathbf{U_b} \cap \boldsymbol{\varphi}^{\!-\!1\!}(\mathbf{U_n})\right) \mapsto \boldsymbol{\varphi}_{\mathbf{n}}(\mathbf{U_n}), \; \boldsymbol{\Theta}_{\mathbf{1}} := \boldsymbol{\varphi}_{\mathbf{n}} \circ \boldsymbol{\varphi}_{\mathbf{b}}^{-1}(\boldsymbol{\vartheta})
\end{equation*}
\begin{equation*}
    \boldsymbol{\Theta}_{\mathbf{2}}\!: \boldsymbol{\varphi}_{\mathbf{a}}\!\left(\mathbf{U_a} \cap \boldsymbol{\varphi}^{\!-\!1\!}(\mathbf{U_s})\right) \mapsto \boldsymbol{\varphi}_{\mathbf{s}}(\mathbf{U_s}), \; \boldsymbol{\Theta}_{\mathbf{2}} := \boldsymbol{\varphi}_{\mathbf{s}} \circ \boldsymbol{\varphi} \circ \boldsymbol{\varphi}_{\mathbf{a}}^{-1}(\boldsymbol{\vartheta})
\end{equation*}
which leads to $\boldsymbol{\Theta}_{\mathbf{1}}(\boldsymbol{\vartheta}) = \boldsymbol{\vartheta}$ and $\boldsymbol{\Theta}_{\mathbf{2}}(\boldsymbol{\vartheta}) = \boldsymbol{\vartheta}$. Hence, $\boldsymbol{\Theta}_{\mathbf{1}}$ and $\boldsymbol{\Theta}_{\mathbf{2}}$ are smooth maps from $\mathbb{R}^3$ to $\mathbb{R}^3$. It follows from \cite[Proposition 2.5]{Lee2013} that the map $\boldsymbol{\varphi}$ is smooth. Concerning the inverse function $\boldsymbol{\varphi}^{\!-\!1\!}:\mathbb{S}^3 \mapsto \mathbb{\bar{R}}^3$, expressed as 
\begin{equation}
    \label{eq:StereographicProjection_varphi}
    \boldsymbol{\varphi}^{\!-\!1\!}(\mathbf{q}) = \left\{\begin{array}{cl}
       \boldsymbol{\varphi}_{\mathbf{s}} (\mathbf{q})  &, \; \textrm{for} \; \mathbf{q} \in \mathbb{S}^3 \setminus \{\mathbf{s}\} \\
        \boldsymbol{\infty} &, \; \textrm{for} \; \color{black} \mathbf{q} = \mathbf{s} \color{black}
\end{array}\right., 
\end{equation}
in light of
\begin{equation*}
   \boldsymbol{\varphi}_{\mathbf{b}} \circ \boldsymbol{\varphi}^{\!-\!1\!} \circ \boldsymbol{\varphi}_{\mathbf{n}}^{-1}(\boldsymbol{\vartheta}) = 
\left(\boldsymbol{\varphi}_{\mathbf{n}} \circ \boldsymbol{\varphi}_{\mathbf{b}}^{-1}\right)^{\!-1}\!\!(\boldsymbol{\vartheta})
 = \boldsymbol{\Phi}_{\mathbf{1}}^{-1}(\boldsymbol{\vartheta}) = \boldsymbol{\vartheta}
\end{equation*}
and 
\begin{equation*}
   \boldsymbol{\varphi}_{\mathbf{a}} \circ \boldsymbol{\varphi}^{\!-\!1\!} \circ \boldsymbol{\varphi}_{\mathbf{s}}^{-1}(\boldsymbol{\vartheta}) = \left(\boldsymbol{\varphi}_{\mathbf{s}} \circ \boldsymbol{\varphi} \circ \boldsymbol{\varphi}_{\mathbf{a}}^{-1}\right)^{\!-1}\!\!\!\!(\boldsymbol{\vartheta
   }) = \boldsymbol{\Phi}_{\mathbf{2}}^{-1}(\boldsymbol{\vartheta}) = \boldsymbol{\vartheta},
\end{equation*}
the map $\boldsymbol{\varphi}^{\!-\!1\!}$ is also smooth \cite[Proposition 2013]{Lee2013}. Therefore, one concludes that $\boldsymbol{\varphi}$ is a diffeomorphism from $\mathbb{\bar{R}}^3$ to $\mathbb{S}^3$. As a result, the smooth manifolds $\mathbb{\bar{R}}^3$ and $\mathbb{S}^3$ are diffeomorphic.

\par The demonstrated equivalence relation between the smooth manifolds $\mathbb{\bar{R}}^3$ and $\mathbb{S}^3$ rooted in the diffeomorphism $\boldsymbol{\varphi}(\boldsymbol{\vartheta}):\mathbb{\bar{R}}^3 \mapsto \mathbb{S}^3$ is instrumental to derive some crucial properties of the map $\mathcal{R}_{\!\boldsymbol{\vartheta }}: \mathbb{\bar{R}}^3 \mapsto \mathrm{SO}(3)$. In this direction, the derivation starts from a fairly known result: the manifold $\mathbb{S}^3$ is a covering space for $\mathrm{SO}(3)$ with $\mathcal{R}$ as covering map. In light of the map $\boldsymbol{\varphi}(\boldsymbol{\vartheta}): \mathbb{\bar{R}}^3 \mapsto \mathbb{S}^3$ being a diffeomorphism, it follows from \cite[Proposition 2.15]{Lee2013} that this map is also a homeomorphism. Then, since any homeomorphism is trivially a covering map, the Alexandroff compactification $\mathbb{\bar{R}}^3$ is a covering space for $\mathbb{S}^3$ with $\boldsymbol{\varphi}(\boldsymbol{\vartheta})$ as covering map. With these properties in mind, in conjunction with the notion that the map $\mathcal{R}_{\!\boldsymbol{\vartheta}}$ results from the composition $\mathcal{R}\circ\boldsymbol{\varphi}(\boldsymbol{\vartheta}): \mathbb{\bar{R}}^3 \mapsto \mathrm{SO}(3)$ \cite[p.120]{junkins_2009} and the fact that the inverse map of $\mathcal{R}(\mathbf{q})$, $\mathcal{Q}: \mathrm{SO}(3) \rightrightarrows \mathbb{S}^3$, is finite for every $\mathbf{R} \in \mathrm{SO}(3)$, it follows from \cite[p. 341]{munkres2000topology} that $\mathcal{R}_{\!\boldsymbol{\vartheta }}: \mathbb{\bar{R}}^3 \mapsto \mathrm{SO}(3)$ is a covering map with $\mathbb{\bar{R}}^3$ and $\mathrm{SO}(3)$ as covering and base spaces, respectively. Furthermore, according to \cite[Proposition 4.6]{Lee2013}, the map $\boldsymbol{\varphi}(\boldsymbol{\vartheta})$ is a local diffeomorphism and, by virtue of $\mathcal{R}$ being also a local diffeomorphism, the composition map $\mathcal{R}_{\!\boldsymbol{\vartheta}}(\boldsymbol{\vartheta})$ is itself a local diffeomorphism.

\par
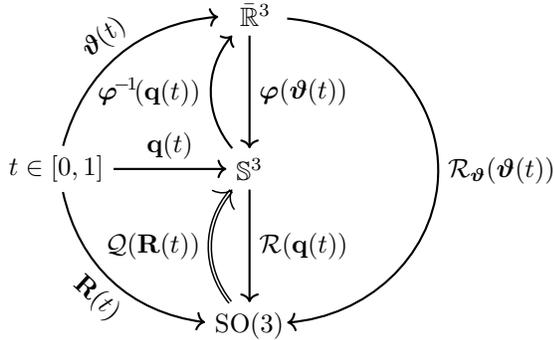
\begin{figure}[!htb]
\centering
    \begin{tikzpicture}[node distance = 1.5cm, thick]%
        \node (1) {$t \in [0,1]$};
        \node (2) [right=of 1] {$\mathbb{S}^3$};
        \node (3) [above = of 2] {$\:\:\mathbb{\bar{R}}^3$};
        \node (4) [below = of 2] {$\mathrm{SO}(3)$};
        \draw[->] (1) -- node [midway,above] {$\mathbf{q}(t)$} (2);
        \draw[->] (1) to [bend left = 35] node [sloped, midway,above]{$\boldsymbol{\vartheta}(t)$} (3);
        \draw[->] (1) to [bend right = 35] node [sloped, midway,below]{$\mathbf{R}(t)$} (4);
        \draw[->] (3) -- node [midway,right] {$\boldsymbol{\varphi}(\boldsymbol{\vartheta}(t))$} (2);
        \draw[->] (2) -- node [midway,right] {$\mathcal{R}(\mathbf{q}(t))$} (4);
        \draw[double, ->, line width=0.2mm] (4) to [bend left = 40] node [midway,left]{$\mathcal{Q}(\mathbf{R}(t))$} (2);
        \draw[->] (2) to [bend left = 40] node [midway,left]{$\boldsymbol{\varphi}^{\!-\!1}\!(\mathbf{q}(t))$} (3);
    \draw[->] (3)+(.5,-0.05) arc
    [start angle = 85, end angle= -85, x radius= 2.2cm, y radius = 2.03cm] node [ midway,right]{$\mathcal{R}_{\boldsymbol{\vartheta}}(\boldsymbol{\vartheta}(t))$} (4) ;
    \end{tikzpicture}%
\caption{Scheme of the maps between the topological spaces $\mathbb{\bar{R}}^3$, $\mathbb{S}^3$ and $\mathrm{SO}(3)$ and of the respective elements describing the associated attitude representation for a given time instant.}
\label{fig:PathLifting}
\end{figure}

 The proved properties of the map $\mathcal{R}_{\!\boldsymbol{\vartheta}}(\boldsymbol{\vartheta})$, namely, being a covering map and a local diffeomorphism everywhere, are quite relevant: the former opens the door for the application of unique path-lifting methodologies \cite[Lemma 54.1.]{munkres2000topology}, whereas the latter allows preserving everywhere the local differentiable structure when mapping between $\mathbb{\bar{R}}^3$ and $\mathrm{SO}(3)$. In \hyperref[fig:PathLifting]{Fig.~\ref*{fig:PathLifting}}, the continuous mapping between the topological spaces  $\mathbb{\bar{R}}^3$, $\mathbb{S}^3$ and $\mathrm{SO}(3)$, as well as the corresponding elements describing the related attitude representation for a certain time instant, are schematically presented. \color{black} Due to the rotation matrix path $\mathbf{R}(t)$ being a continuous mapping of $[0,1]$ into $\mathrm{SO}(3)$ \color{black} and given that $\mathcal{R}_{\boldsymbol{\vartheta}}(\boldsymbol{\vartheta})$ and $\mathcal{R}(\mathbf{q})$ are covering maps, $\boldsymbol{\vartheta}(t)$ and $\mathbf{q}(t)$ are lifts of $\mathbf{R}(t)$. It is worth mentioning that the closed interval $[0,1]$ was considered due to its extensive use in topology to represent a bounded time interval. Any other closed and bounded time interval could be considered for this purpose.

\section{Hybrid Dynamic Path-lifting Algorithm for MRP extraction}
\label{section4}

For the purpose of extracting the MRP representation of the attitude, a hybrid algorithm is constructed to ensure the uniqueness and consistency of this operation. The algorithm extends the dynamic path-lifting solution proposed by \cite{mayhew2013} with the inclusion of an additional state and application of stereographic projection to the smoothly lifted quaternion. \color{black} Let the stereographic projection be defined by the smooth inverse function $\boldsymbol{\varphi}^{\!-\!1\!}(\mathbf{q}): \mathbb{S}^3 \mapsto \mathbb{\bar{R}}^3$ given in \eqref{eq:StereographicProjection_varphi}. \color{black} Consider a discrete state $m \in \{-1,1\}$ and a hysteretic parameter $\delta \in \mathbb{R}_{>0}$. Then, the following flow and jump sets are defined  
\begin{equation}
\begin{aligned}
    \mathbf{C_m} \!:=\! \{\left( \mathbf{\hat{q}}, \mathbf{R}, m\right) \in \; & \mathbb{S}^3 \!\times \mathrm{SO}(3) \times \{-1,1\} : \\ & \|\boldsymbol{\varphi}^{\!-\!1\!}(m\boldsymbol{\Phi}(\mathbf{\hat{q}}, \mathbf{R}))\| \leq 1 + \delta\}
\end{aligned}
\end{equation}
\begin{equation}
\begin{aligned}
    \mathbf{D_m} := \{\left( \mathbf{\hat{q}}, \mathbf{R}, m\right) \in \; & \mathbb{S}^3 \!\times \mathrm{SO}(3) \times \{-1,1\} : \\ & \|\boldsymbol{\varphi}^{\!-\!1\!}(m\boldsymbol{\Phi}(\mathbf{\hat{q}}, \mathbf{R}))\| \geq 1 + \delta \}
\end{aligned}
\end{equation}
where, as proposed in \cite{mayhew2013}, $\boldsymbol{\Phi}:\mathbb{S}^3 \times \mathrm{SO}(3) \rightrightarrows \mathbb{S}^3$ denotes the map

\begin{equation}
    \boldsymbol{\Phi}(\mathbf{\hat{q}}, \mathbf{R}) := \!\!\begin{array}{cc}
        & \\ \textrm{argmax} &  \!\mathbf{\hat{q}}^{\!\top}\mathbf{p} \\ 
        \mathbf{p} \in \mathcal{Q}(\mathbf{R}) & 
    \end{array},
\end{equation}
and $\mathbf{\hat{q}} := (\hat{q}_0, \mathbf{\hat{q}_1}) \in \mathbb{S}^3 $ symbolizes the memory state, which is updated in accordance with the following 
flow and jump sets
\begin{equation*}
    \mathbf{C_l} \!:=\! \{\!\left( \mathbf{\hat{q}}, \mathbf{R}, m\right) \! \in \mathbb{S}^3 \!\times\! \mathrm{SO}(3) \!\times\! \{-1,\!1\!\} \!:  \mathrm{dist}(\mathbf{\hat{q}}, \mathcal{Q}(\mathbf{R}) \!) \!\leq \!\alpha\}, 
\end{equation*}
\begin{equation*}
    \mathbf{D_l} \!:=\! \{\!\left( \mathbf{\hat{q}}, \mathbf{R}, m\right) \!\in \mathbb{S}^3 \!\times\! \mathrm{SO}(3) \!\times\! \{-1,\!1\!\} \!:  \mathrm{dist}(\mathbf{\hat{q}}, \mathcal{Q}(\mathbf{R}) \!) \!\geq \!\alpha\}, 
\end{equation*}
with $\mathrm{dist}(\mathbf{\hat{q}}, \mathcal{Q}(\mathbf{R})) = \mathrm{inf}\left\{1-\mathbf{\hat{q}}^{\!\top}\mathbf{p}: \mathbf{p} \in \mathcal{Q}(\mathbf{R})\right\}$ and $\alpha \in (0,1)$. With these definitions in mind, the hybrid algorithm for the extraction of the MRP representation is described by the system $\mathcal{H}_{\boldsymbol{\vartheta}}:= \left(\mathbf{C}_{\boldsymbol{\vartheta}},\;\mathbf{F}_{\boldsymbol{\vartheta}},\;\mathbf{D}_{\boldsymbol{\vartheta}},\;\mathbf{G}_{\boldsymbol{\vartheta}}\right)$:
\begin{subequations}
    \begin{equation}
        \mathbf{C}_{\boldsymbol{\vartheta}} := \mathbf{C_l} \cap \mathbf{C_m}
    \end{equation}
    \begin{equation}
        \mathbf{F}_{\boldsymbol{\vartheta}} \left\{ \begin{array}{c}
             \mathbf{\dot{\hat{q}}} = \boldsymbol{0} \\ \dot{m} = 0 
        \end{array}\right.
    \end{equation}
        \begin{equation}
        \mathbf{D}_{\boldsymbol{\vartheta}} := \mathbf{D_l} \cup \mathbf{D_m}
    \end{equation}
    \begin{equation}
        \mathbf{G}_{\boldsymbol{\vartheta}} \left\{ \begin{array}{cl}
            \mathbf{\hat{q}^+} \in \boldsymbol{\Phi}(\mathbf{\hat{q}}, \mathbf{R}), \; m^+ = m \!\!\!\! &, \; \left( \mathbf{\hat{q}}, \mathbf{R}, m\right) \!\in\! \mathbf{D_l} \\
        \mathbf{\hat{q}^+} = \mathbf{\hat{q}}, \; m^+=-m \!\!\!\! &, \; \left( \mathbf{\hat{q}}, \mathbf{R}, m\right) \!\in\! \mathbf{D_m}
        \end{array}\right.
    \end{equation}
\end{subequations}

with continuous input $\mathbf{R}:\mathbb{R}_{\geq} 0 \mapsto \mathrm{SO}(3)$ and output 
\begin{equation}
    \label{eq:OutputLiftingMRP}
    \boldsymbol{\vartheta} := \left\{ \begin{array}{cl}
        \boldsymbol{\varphi}^{\!-\!1\!}(m\boldsymbol{\Phi}(\mathbf{\hat{q}}, \mathbf{R})) &, \quad \left( \mathbf{\hat{q}}, \mathbf{R}, m\right) \!\in\! \mathbf{C_m} \cap \mathbf{C_l} \\
         \emptyset &, \quad \left( \mathbf{\hat{q}}, \mathbf{R}, m\right) \notin \mathbf{C_m} \cap \mathbf{C_l}
    \end{array}\right.
\end{equation}
In the event of the conditions for $\mathbf{D_m}$ and $\mathbf{D_l}$ being concurrently met, $ \left(\mathbf{\hat{q}}, \mathbf{R}, m\right) \in \mathbf{D_m} \cap \mathbf{D_l}$, the formulation allows for either jump. Anticipating this two-fold possibility is required to ensure the fulfillment of the hybrid basic conditions. To study the properties of this hybrid algorithm, consider the autonomous system $ \mathcal{H}^*_{\boldsymbol{\vartheta}} := \left(\mathbf{C}^*_{\boldsymbol{\vartheta}},\;\mathbf{F}^*_{\boldsymbol{\vartheta}},\;\mathbf{D}^*_{\boldsymbol{\vartheta}},\;\mathbf{G}^*_{\boldsymbol{\vartheta}}\right)$ given by 
\begin{subequations}
    \label{eq:HybridSystemLiftingMRP}
    \begin{equation}
        \mathbf{C}^*_{\boldsymbol{\vartheta}} := \mathbf{C}_{\boldsymbol{\vartheta}}
    \end{equation}
    \begin{equation}
        \mathbf{F}_{\boldsymbol{\vartheta}}^* \left\{ \begin{array}{c}
             \mathbf{F}_{\boldsymbol{\vartheta}} \\ \mathbf{\dot{R}} \in \mathbf{R}\left[K\mathbb{B}^3\right]_\times
        \end{array}\right.
    \end{equation}
        \begin{equation}
        \mathbf{D}^*_{\boldsymbol{\vartheta}} := \mathbf{D}_{\boldsymbol{\vartheta}}
    \end{equation}
    \begin{equation}
        \mathbf{G}^*_{\boldsymbol{\vartheta}} \left\{ \begin{array}{cl}
            \mathbf{\hat{q}^+} \in \boldsymbol{\Phi}(\mathbf{\hat{q}}, \; \mathbf{R}), \mathbf{R^+} \!=\! \mathbf{R}, \; m^+ \!=\! m \!\!\!\! &, \; \left( \mathbf{\hat{q}}, \mathbf{R}, m\right) \!\in\! \mathbf{D_l} \\
        \mathbf{\hat{q}^+} \!=\! \mathbf{\hat{q}}, \; \mathbf{R^+} \!=\! \mathbf{R}, \; m^+ \!=\!- m \!\!\!\! &, \; \left( \mathbf{\hat{q}}, \mathbf{R}, m\right) \!\in\! \mathbf{D_m}
        \end{array}\right.
    \end{equation}
\end{subequations}
with the output $\boldsymbol{\vartheta}$ specified in \eqref{eq:OutputLiftingMRP}. In comparison with $\mathcal{H}_{\boldsymbol{\vartheta}}$, the latter system also encapsulates the dynamics of trajectories in $\mathrm{SO}(3)$. \hyperref[lemH*]{Lemma~\ref*{lemH*}} details and demonstrates a few key attributes of the hybrid system $\mathcal{H}_{\boldsymbol{\vartheta}}^*$. 

\begin{lem}
 
\label{lemH*}
Let $\mathrm{dist}(\mathbf{\hat{q}}, \mathcal{Q}(\mathbf{R}))|_{(0, 0)} < 1$.The hybrid system $\mathcal{H}_{\boldsymbol{\vartheta}}^*$, described in \eqref{eq:HybridSystemLiftingMRP}, has the following properties:
\begin{enumerate}
\item $\mathcal{H}_{\boldsymbol{\vartheta}}^*$ is well-posed \cite[Definition 6.29.]{goebel_2012};
\item Every maximal solution \cite[Definition 2.7.]{goebel_2012} to $\mathcal{H}_{\boldsymbol{\vartheta}}^*$ is complete;
\item \color{black} For any solution $\left(\mathbf{\hat{q}}, \mathbf{R}, m\right)(t,j)$, the condition $\left.(\mathbf{\hat{q}}, \mathbf{R}, m)\right|_{t,j} \notin \mathbf{C_m} \cap \mathbf{C_l}$ can only occur for $(t,j) = (0,0)$, i.e., \color{black}
\begin{equation*}
    \left\{(t,j): \left.( \mathbf{\hat{q}}, \mathbf{R}, m)\right|_{t,j} \notin \mathbf{C_m} \cap \mathbf{C_l} \right\} \subset \{(0, 0)\}
\end{equation*}
\item The output $\boldsymbol{\vartheta}(t,j)$ verifies $\mathbf{R_{\downarrow t}}(t) = \mathcal{R}_{\boldsymbol{\vartheta}}\left(\boldsymbol{\vartheta}_{\mathbf{\downarrow t}}(t)\right)$
\item The output verifies the bound $\|\boldsymbol{\vartheta}(t,j)\| \leq 1 + \delta$.
\end{enumerate}
\end{lem}
\begin{proof} See \hyperref[appendix:ProofLemma1]{Appendix~\ref{appendix:ProofLemma1}}.  \end{proof}

\par As demonstrated in \hyperref[lemH*]{Lemma~\ref*{lemH*}}, the output $\boldsymbol{\vartheta}$ is defined on a three-dimensional unit sphere enclosed by a hysteresis region defined with the parameter $\delta$, i.e., $\boldsymbol{\vartheta} \in (1 + \delta)\mathbb{B}^3$. This region plays a pivotal role in averting noise-induced chattering when switching between the original and shadow MRP representations.

\section{Equivalence of stability results in the covering and base spaces}
\label{section5}
The previous section discussed the design and properties of a hybrid approach for lifting a path from the base space $\mathrm{SO}(3)$ to the covering space $\mathbb{\bar{R}}^3$. Now, considering this methodology, the current section aims to assess the equivalence of stability between a closed-loop hybrid system that encompasses the differential equations of the rigid body, detailed in \eqref{eq:AttitudeDynamics}, and the hybrid path-lifting algorithm, and a hybrid closed-loop system that encapsulates the dynamics of the rigid body attitude expressed in terms of MRP, which comprise \eqref{eq:AttitudeDynamicEquation} and \eqref{eq:AttitudeMRPKinematics}. \color{black} Within this framework, consider an arbitrary MRP-based controller that has $\boldsymbol{\rho} \in \boldsymbol{\Lambda} \subset \mathbb{R}^n$ as state, $\boldsymbol{\vartheta}$ and $\boldsymbol{\omega}$ as inputs, and the torque $\boldsymbol{\tau}(\boldsymbol{\vartheta}, \boldsymbol{\omega}, \boldsymbol{\rho}):\mathbb{\bar{R}}^3 \times \mathbb{R}^3 \times \boldsymbol{\Lambda} \mapsto \mathbb{R}^3$ as output. The dynamics of the controller state $\boldsymbol{\rho}$ are governed by the differential equation 
\begin{equation}
    \label{eq:DynamicsMRPController}
    \boldsymbol{\dot{\rho}} = \mathbf{f}(\boldsymbol{\vartheta}, \boldsymbol{\omega}, \boldsymbol{\rho}).
\end{equation}
To shed some light, the controller state $\boldsymbol{\rho}$ can be, for instance, an integral state similar to the one proposed in \cite[Equation 49]{martins2024}. \color{black} Due to the equivalence between the shortest rotational direction available and the MRP set with the lowest norm, MRP-based controllers do not require additional control mechanisms to determine the mentioned direction as it exploits this unique feature to prevent the unwinding phenomenon (see, for instance, \cite{martins2021CDC}). Hence, instead of the controller state having a discrete evolution, the attitude representation jumps by selectively transitioning between the original and shadow sets. Conversely, hybrid quaternion-based solutions rely on a discrete logic state, whose update depends on the scalar component of the quaternion, to impose the shortest rotation sought (\cite{mayhew2011quaternion, casau2015}). 

\par The first closed-loop hybrid system, which consists of an interconnection between the rigid-body dynamic model \eqref{eq:AttitudeDynamics}, the output $\boldsymbol{\tau}(\boldsymbol{\vartheta}, \boldsymbol{\omega}, \boldsymbol{\rho}):\mathbb{\bar{R}}^3 \times \mathbb{R}^3 \times \boldsymbol{\Lambda} \mapsto \mathbb{R}^3$ and dynamics \eqref{eq:DynamicsMRPController} of the MRP-based controller, and the hybrid algorithm $\mathcal{H}_{\boldsymbol{\vartheta}}$, is denoted by $\mathcal{H}_{\mathbf{1}}:= \left(\mathbf{C_1}, \mathbf{F_1}, \mathbf{D_1}, \mathbf{G_1}\right)$, has the vector $ \mathbf{x_1} := (\mathbf{R_1}, \mathbf{\hat{q}_1}, m_1, \boldsymbol{\omega}_{\mathbf{1}}, \boldsymbol{\rho}_{\mathbf{1}}) \in \boldsymbol{\chi}_{\mathbf{1}} := \mathrm{SO}(3) \times \mathbb{S}^3 \times \{-1,1\} \times \mathbb{R}^3 \times \boldsymbol{\Lambda}$ as state, and is defined as
 
\begin{subequations}
\label{eq:HybridSystemSO(3)}
\begin{equation}
    \mathbf{C_1}(\mathbf{x_1}) := \left\{\mathbf{x_1} \in \boldsymbol{\chi}_\mathbf{1}: \left( \mathbf{R}_{\mathbf{1}}, \mathbf{\hat{q}}_{\mathbf{1}}, m_1\right) \in \mathbf{C_m} \cap \mathbf{C_l} \right\}
\end{equation}
\begin{equation}
    \mathbf{F_1}(\mathbf{x_{1}}) \!:=\!\! \left(\!\!\!\!\! \begin{array}{c}
         \mathbf{R}_{\mathbf{1}}\!\left[\boldsymbol{\omega}_{\mathbf{1}}\right]_{\times} \\ 
        \boldsymbol{0} \\
        0 \\ \mathbf{J}^{-1}\!\!\left(\left[\mathbf{J}\boldsymbol{\omega}_{\mathbf{1}}\!\right]_{\!\times}\!\boldsymbol{\omega}_{\mathbf{1}} \!+\! \boldsymbol{\tau}(\boldsymbol{\varphi}^{\!-\!1\!}(m_1\boldsymbol{\Phi}(\mathbf{\hat{q}}_{\mathbf{1}}, \! \mathbf{R}_{\mathbf{1}}\!)), \!\boldsymbol{\omega}_{\mathbf{1}}, \!\boldsymbol{\rho}_{\mathbf{1}})\!\right) \\
        \mathbf{f}(\boldsymbol{\vartheta}_{\mathbf{1}}, \boldsymbol{\omega}_{\mathbf{1}}, \boldsymbol{\rho}_{\mathbf{1}})
        \end{array}\!\!\!\!\!\right)
\end{equation}
\begin{equation}
     \mathbf{D_1}(\mathbf{x_1}) := \left\{\mathbf{x_1} \in \boldsymbol{\chi}_\mathbf{1}: \left( \mathbf{R}_{\mathbf{1}}, \mathbf{\hat{q}}_{\mathbf{1}}, m_1\right) \in \mathbf{D_m} \cup \mathbf{D_l} \right\}
\end{equation}
\begin{equation}
    \mathbf{G_1}(\mathbf{x_1}) \!:=\!\! \left\{\!\!\!\!\begin{array}{ll}
            (\mathbf{R}_{\mathbf{1}}, \boldsymbol{\Phi}(\mathbf{\hat{q}}_{\mathbf{1}}, \!\mathbf{R}_{\mathbf{1}}\!), m_1, \boldsymbol{\omega}_{\mathbf{1}}, \boldsymbol{\rho}_{\mathbf{1}}\!) &  \!\!\!\!\! , \left( \!\mathbf{R_1}, \!\mathbf{\hat{q}_1}, \!m_1\!\right) \!\in\! \mathbf{D_l} \\
            (\mathbf{R}_{\mathbf{1}}, \mathbf{\hat{q}_1}, -m_1, \boldsymbol{\omega}_{\mathbf{1}}, \boldsymbol{\rho}_{\mathbf{1}}\!) &  \!\!\!\!\! , \left( \!\mathbf{R_1}, \!\mathbf{\hat{q}_1}, \!m_1\!\right) \!\in\! \mathbf{D_m}
    \end{array}\right.
\end{equation}
\end{subequations}

\par
\begin{figure}[!htb]
	\centering
        \label{fig:mrpHysteresisDiagram}
	\includegraphics[height=5cm]{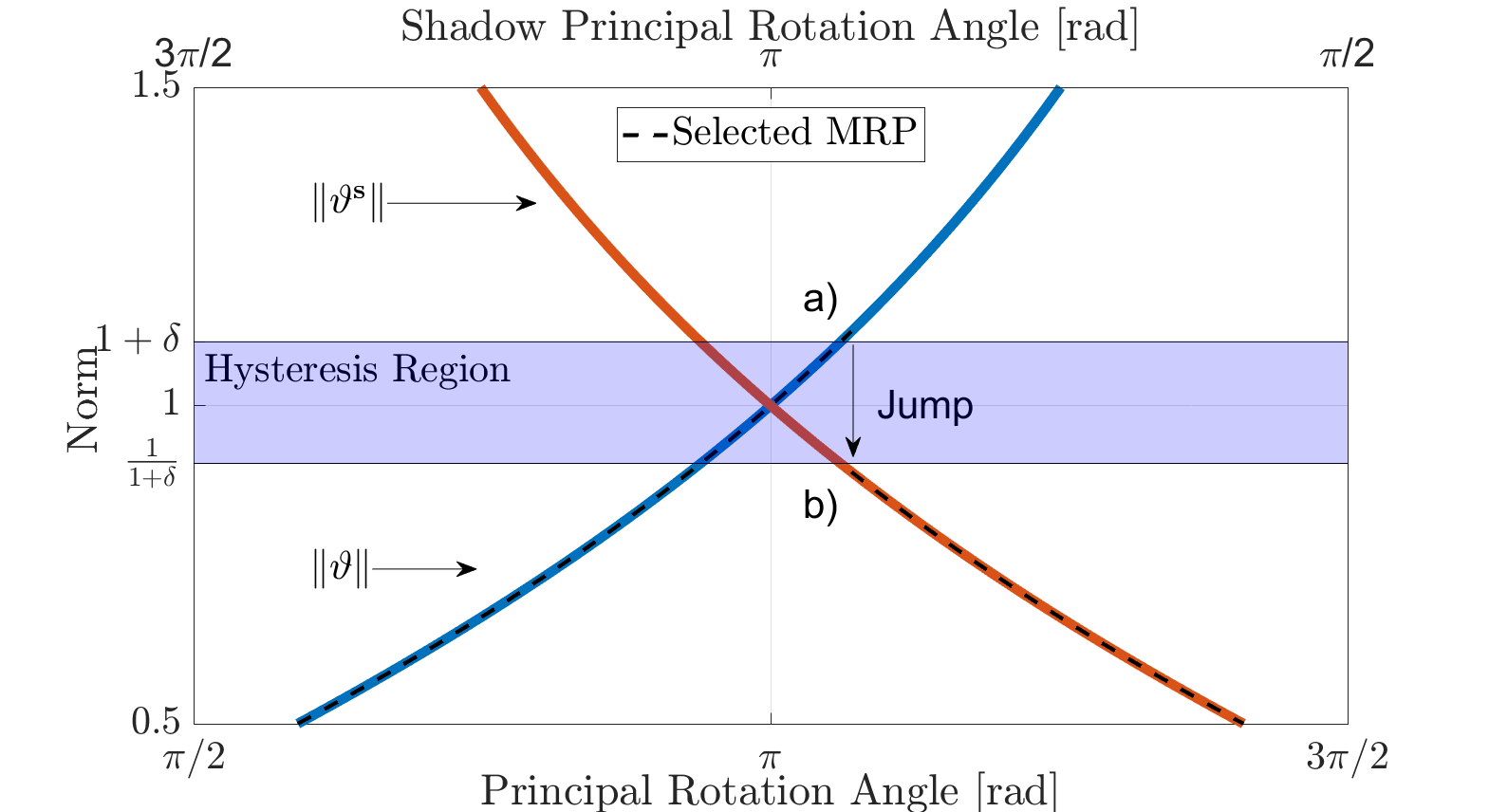}
	 \caption{\color{black} Depiction of the MRP switching logic: when $\|\boldsymbol{\vartheta}_{\mathbf{2}}\|$ is equal or greater than $1 + \delta$, \textbf{a)}, the state jumps to the shadow MRP representation through the map $\boldsymbol{\Upsilon}(\boldsymbol{\vartheta}_{\mathbf{2}})$. The resulting MRP description, \textbf{b)}, has a norm equal or lower than $(1 + \delta)^{-1}$, and corresponds to the shortest principal rotation available \cite[p.~120]{junkins_2009}. The switch is hysteretic due to the inclusion of the parameter $\delta$. \color{black}}  
\end{figure}

 Notice that the hybrid system $\mathcal{H}_{\mathbf{1}}$ behaves similarly to the lifting system $\mathcal{H}_{\boldsymbol{\vartheta}}$ in terms of flows and jumps. As previously discussed, the primary purpose lies in evaluating the equivalence of stability between the solutions of the hybrid system $\mathcal{H}_{\mathbf{1}}$ and another that resorts to the MRP description to model the attitude dynamics. \color{black} The MRP-based solutions found in the literature are devised directly on $\mathbb{\bar{R}}^3$ without explicitly including a lifting system in the control architecture (see, for instance, \cite{martins2021CDC, martins2023, martins2024}). Thus, the second closed-loop hybrid system, characterized by the data $\mathcal{H}_{\mathbf{2}}:= \left(\mathbf{C_2}, \mathbf{F_2}, \mathbf{D_2}, \mathbf{G_2}\right)$, in terms of flow dynamics, is strictly composed of the MRP kinematic equation \eqref{eq:AttitudeMRPKinematics}, the attitude dynamic equation \eqref{eq:AttitudeDynamicEquation}, and the controller state dynamic equation \eqref{eq:DynamicsMRPController}, and its jump dynamics include the mapping between the MRP shadow and original set representations \eqref{eq:ShadowMap}. \color{black} With this in mind, let $\mathbf{x_2} := \left(\boldsymbol{\vartheta}_{\mathbf{2}}, \boldsymbol{\omega}_{\mathbf{2}},  \boldsymbol{\rho}_{\mathbf{2}}\right) \in \boldsymbol{\chi}_{\mathbf{2}}:= \mathbb{\bar{R}}^3 \times \mathbb{R}^3 \times \boldsymbol{\Lambda}$ define the state vector in order to formalize the hybrid system $\mathcal{H}_{\mathbf{2}}$ as follows \color{black}

\begin{subequations}
\label{eq:HybridSystemLiftedDynamics}
\begin{equation}
    \mathbf{C}_{\mathbf{2}}(\mathbf{x_2}) := \left\{\mathbf{x_2} \in \boldsymbol{\chi}_\mathbf{2}: \|\boldsymbol{\vartheta}_{\mathbf{2}}\| \leq 1 + \delta \right\}
\end{equation}
\begin{equation}
    \mathbf{F_2}(\mathbf{x_{2}}) := \left(\begin{array}{c} 
     \mathbf{T}(\boldsymbol{\vartheta}_{\mathbf{2}}) \boldsymbol{\omega}_{\mathbf{2}}\\
        \mathbf{J^{-1}}\left(\left[\mathbf{J}\boldsymbol{\omega}_{\mathbf{2}}\right]_{\times}\!\boldsymbol{\omega}_{\mathbf{2}} + \boldsymbol{\tau}(\boldsymbol{\vartheta}_{\mathbf{2}},\boldsymbol{\omega}_{\mathbf{2}}, \boldsymbol{\rho}_{\mathbf{2}})\right) \\
        \boldsymbol{\dot{\rho}}_{\mathbf{2}} = \mathbf{f}(\boldsymbol{\vartheta}_{\mathbf{2}}, \boldsymbol{\omega}_{\mathbf{2}}, \boldsymbol{\rho}_{\mathbf{2}})
        \end{array}\right)
\end{equation}
\begin{equation}
     \mathbf{D_2}(\mathbf{x_2}) := \left\{\mathbf{x_2} \in \boldsymbol{\chi}_\mathbf{2}: \|\boldsymbol{\vartheta}_{\mathbf{2}}\| \geq 1 + \delta \right\}
\end{equation}
\begin{equation}
    \mathbf{G_2}(\mathbf{x_2}) := \left( \boldsymbol{\Upsilon}(\boldsymbol{\vartheta}_{\mathbf{2}}), \boldsymbol{\omega_{\mathbf{2}}}, \boldsymbol{\rho}_{\mathbf{2}}\right)
\end{equation}
\end{subequations}
In this way, $\mathcal{H}_{\mathbf{2}}$ can be perceived as a closed-loop hybrid system that results from the application of the MRP-based controller, that has $\boldsymbol{\rho}_{\mathbf{2}}$ as state and the torque $\boldsymbol{\tau}(\boldsymbol{\vartheta}_{\mathbf{2}},\boldsymbol{\omega}_{\mathbf{2}}, \boldsymbol{\rho}_{\mathbf{2}})$ as output, to the rigid-body attitude dynamics described in terms of MRP. It is worth emphasizing that $\mathbf{C_2} \cap \mathbf{D_2} \neq \emptyset$ leads to non-uniqueness of solutions to $\mathcal{H}_{\mathbf{2}}$: for $\mathbf{x_2} \in \mathbf{C_2} \cap \mathbf{D_2}$, the state can either flow or jump. However, according to \cite[p. 87]{goebel_2012}, such data formulation does not affect asymptotic stability and robustness. Furthermore, this is required to guarantee the fulfillment of the hybrid basic conditions \cite[Assumption 6.5]{goebel_2012}. Regarding the jump logic, in \hyperref[fig:mrpHysteresisDiagram]{Figure~\ref {fig:mrpHysteresisDiagram}}, the MRP switching logic and the hysteresis region are illustrated.

\color{black}
\par To lay the groundwork for comparing the stability of the hybrid systems $\mathcal{H}_{\mathbf{1}}$ and $\mathcal{H}_{\mathbf{2}}$, 
\hyperref[lemSolutionsEquivalence]{Lemma~\ref*{lemSolutionsEquivalence}} establishes parallelisms between the solutions of these two systems. Inspired by the approach proposed in \cite[Lemma 8]{mayhew2013}, the proof of this Lemma follows a recursive procedure to construct the solution of one of the hybrid systems and its hybrid time domain in terms of the solution of the other. Bear in mind that $\mathcal{H}_{\mathbf{1}}$ encapsulates the jumps caused by the update of the memory state $\mathbf{\hat{q}}$ and by the MRP norm, whereas $\mathcal{H}_{\mathbf{2}}$ only encompasses the latter. In each step, the approach verifies whether the constructed solution complies with the flow and jump dynamics of $\mathcal{H}_{\mathbf{2}}$.

\begin{lem}
\label{lemSolutionsEquivalence}
Let $\mathrm{dist}(\mathbf{\hat{q}_1}, \!\mathcal{Q}(\mathbf{R_1})\!)|_{(0, 0)} < 1$. For every solution $\mathbf{x_1}$ to $\mathcal{H}_\mathbf{1}$, there exists a solution $\mathbf{x_2}$ to $\mathcal{H}_\mathbf{2}$ that, for every $(t,j) \in \mathrm{dom}\; \mathbf{x_1}$, yields
\begin{equation}
    \label{eq:SolutionsH1H2}
    \begin{aligned}
    \left(\mathbf{R_1}, \boldsymbol{\varphi}^{\!-\!1\!}\! \left(m_1 \!\!\right.\right.& \left.\left. \boldsymbol{\Phi}\!\left(\mathbf{\hat{q}_1}, \!\mathbf{R_1}\!\right)\right), \boldsymbol{\omega}_{\mathbf{1}}, \boldsymbol{\rho}_{\mathbf{1}}\right)|_{(t,j)} =  \\ & = \left(\mathcal{R}_{\boldsymbol{\vartheta}}(\boldsymbol{\vartheta}_2), \boldsymbol{\vartheta}_{\mathbf{2}}, \boldsymbol{\omega}_{\mathbf{2}}, \boldsymbol{\rho}_{\mathbf{2}}\right)|_{(t,j')},
\end{aligned}
\end{equation}
with $(t,j') \in \mathrm{dom}\;\mathbf{x_2}$ and $j' \in \mathbb
{N}$ satisfying $j' \leq j$. Inversely, for every solution $\mathbf{x_2}$ to $\mathcal{H}_{\mathbf{2}}$, there exists a solution $\mathbf{x_1}$ to $\mathcal{H}_{\mathbf{1}}$ such that, for every $(t,j') \in \mathrm{dom}\;\mathbf{x_2}$, \eqref{eq:SolutionsH1H2} holds with $(t,j) \in \mathrm{dom}\;\mathbf{x_1}$ and $j \geq j'$.
\end{lem}
\begin{proof} See \hyperref[appendix:ProofLemma2]{Appendix~\ref{appendix:ProofLemma2}}.  \end{proof}

\par The relation provided in \hyperref[lemSolutionsEquivalence]{Lemma~\ref*{lemSolutionsEquivalence}} allows one to develop a theoretical basis for bridging 
the stability results of the hybrid systems $\mathcal{H}_{\mathbf{1}}$ and $\mathcal{H}_\mathbf{2}$. In this direction, by building on the previous results, \hyperref[thm:EquivalenceAsymptoticStability]{Theorem~\ref*{thm:EquivalenceAsymptoticStability}} focuses on demonstrating that a feedback controller designed in the covering space $\mathbb{\bar{R}}^3$ yields identical stability results for a system defined in the MRP space and another defined in the base space $\mathrm{SO}(3)$ when the latter resorts to the hybrid dynamic path-lifting algorithm $\mathcal{H}_{\boldsymbol{\vartheta}} $ to lift the attitude representation from $\mathrm{SO}(3)$ to $\mathbb{\bar{R}}^3$. 

\begin{thm}
\label{thm:EquivalenceAsymptoticStability}
Let $\alpha \in \; (0,1)$ and $\delta \in \mathbb{R}_{> 0}$. Suppose that a compact set $\mathcal{A}_{\boldsymbol{\vartheta}} \in \boldsymbol{\chi}_{\mathbf{2}}$ is asymptotically stable for the hybrid system $\mathcal{H}_{\mathbf{2}}$ with $\mathcal{B}_{\boldsymbol{\vartheta}}$ as basin of attraction. Then, the compact set
\begin{equation*}
    \begin{aligned}
    \mathcal{A} = \{ & \mathbf{x_1} \in  \boldsymbol{\chi}_{\mathbf{1}}:  \left(\boldsymbol{\varphi}^{\!-\!1\!}\!\left(m_1\boldsymbol{\Phi}\!\left(\mathbf{\hat{q}_1}, \mathbf{R_1}\!\right)\!\right), \boldsymbol{\omega_1}, \boldsymbol{\rho_1}\right) \in \mathcal{A}_{\boldsymbol{\vartheta}},  \\ &
 \|\boldsymbol{\varphi}^{\!-\!1\!}(m_1\boldsymbol{\Phi}(\mathbf{\hat{q}_1}, \!\mathbf{R_1}\!)\!)\| \leq 1 + \delta, \; 
\mathrm{dist}(\mathbf{\hat{q}_1}, \mathcal{Q}(\mathbf{R_1}\!) \!) \leq \alpha\}
\end{aligned}
\end{equation*}
is asymptotically stable for the hybrid system $\mathcal{H}_{\mathbf{1}}$ with 
\begin{equation*}
\begin{aligned}
    \mathcal{B} = \{ & \mathbf{x_1} \in \boldsymbol{\chi}_{\mathbf{1}}: \left(\boldsymbol{\varphi}^{\!-\!1\!}\!\left(m_1\boldsymbol{\Phi}\!\left(\mathbf{\hat{q}_1}, \mathbf{R_1}\!\right)\!\right), \boldsymbol{\omega_1}, \boldsymbol{\rho_1}\right) \in \mathcal{B}_{\boldsymbol{\vartheta}}, \\ & \mathrm{dist}(\mathbf{\hat{q}_1}, \!\mathcal{Q}(\mathbf{R_1})\!) < 1\}
\end{aligned}
\end{equation*}
as basin of attraction. Conversely, if the compact set $\mathcal{A}$ is asymptotically stable for $\mathcal{H}_{\mathbf{1}}$ with $\mathcal{B}$ as basin of attraction, then the compact set $\mathcal{A}_{\boldsymbol{\vartheta}}$ is asymptotically stable for the hybrid system $\mathcal{H}_{\mathbf{2}}$ with $\mathcal{B}_{\boldsymbol{\vartheta}}$ as basin of attraction.
\end{thm}
\begin{proof} See \hyperref[appendix:ProofTheorem1]{Appendix~\ref{appendix:ProofTheorem1}}.  \end{proof}

\par In addition to the previous parallel asymptotic stability result for the hybrid systems $\mathcal{H}_\mathbf{1}$ and $\mathcal{H}_\mathbf{2}$, \hyperref[thm:EquivalenceExponentialStability]{Theorem~\ref*{thm:EquivalenceExponentialStability}} demonstrates that an analogous equivalence can also be established for the case of exponential stability.

\begin{thm}
\label{thm:EquivalenceExponentialStability}
Let $\alpha \in \; (0,1)$ and $\delta \in \mathbb{R}_{> 0}$. Suppose that the compact set 
\begin{equation}
    \mathcal{A}_{\boldsymbol{\vartheta}} = \left\{ \mathbf{x_2} \in \boldsymbol{\chi}_{\mathbf{2}}: \boldsymbol{\vartheta}_{\mathbf{2}} = \boldsymbol{0}, \boldsymbol{\omega}_{\mathbf{2}} = \boldsymbol{0}, \boldsymbol{\rho}_{\mathbf{2}} \in \mathcal{A}_{\rho} \right\} 
\end{equation} is exponentially stable for the hybrid system $\mathcal{H}_{\mathbf{2}}$ for all $\mathbf{x_2}(0, 0) \in  \mathcal{B}_{\boldsymbol{\vartheta}} = \{\mathbf{x_2} \in \boldsymbol{\chi}_{\mathbf{2}}: \|\mathbf{x_2}\|_{\mathcal{A}_{\vartheta}} < \mu \}$. Then, the compact set
\begin{equation*}
\begin{aligned}
    \mathcal{A} = \{\mathbf{x_1} \in \mathcal{A}_R:  
 \|\boldsymbol{\varphi}^{\!-\!1\!}(m_1\boldsymbol{\Phi}(\mathbf{\hat{q}_1}, \!\mathbf{R_1}\!)\!)\| \leq 1 + \delta, & \\ 
\mathrm{dist}(\mathbf{\hat{q}_1}, \mathcal{Q}(\mathbf{R_1}\!) \!) \leq & \alpha\}  , 
\end{aligned}
\end{equation*}
where $\mathcal{A}_R = \{\mathbf{x_1} \in \boldsymbol{\chi}_{\mathbf{1}}: \mathbf{R_1} = \mathbf{I_3}, \boldsymbol{\omega}_{\mathbf{1}} = \boldsymbol{0}, \boldsymbol{\rho}_{\mathbf{1}} \in \mathcal{A}_\rho\}$, is exponentially stable for the hybrid system $\mathcal{H}_{\mathbf{1}}$ for all $\mathbf{x_1}(0, 0) \in \mathcal{B}$ with
\begin{equation*}
\begin{aligned}
    \mathcal{B} = \left\{\mathbf{x_1} \in \boldsymbol{\chi}_{\mathbf{1}}: \left(\boldsymbol{\varphi}^{\!-\!1\!}\!\left(m_1\boldsymbol{\Phi}\!\left(\mathbf{\hat{q}_1}, \mathbf{R_1}\!\right)\!\right), \boldsymbol{\omega_1}, \boldsymbol{\rho_1}\right) \in \mathcal{B}_{\boldsymbol{\vartheta}}, \right. & \\ \left.\mathrm{dist}(\mathbf{\hat{q}_1}, \!\mathcal{Q}(\mathbf{R_1})\!) < \right. & \left. 1 \right\}.
\end{aligned}
\end{equation*}
Furthermore, the converse also holds true.
\end{thm}
\begin{proof} See \hyperref[appendix:ProofTheorem2]{Appendix~\ref{appendix:ProofTheorem2}}.  \end{proof}

\section{Application to Robust Global Exponential Attitude Tracking}
\label{section6}

\par This section focuses on the application of the theoretical basis developed in the previous sections for equivalence of stability between results in the base space $\mathrm{SO}(3)$ and the covering space $\mathbb{\bar{R}}^3$. To this end, the following provides an approach to tackle the robust global exponential tracking problem for the attitude of a rigid body. The design of the proposed control strategy relies on the MRP attitude description to obtain a robust global exponential result for the attitude tracking dynamics in the covering space $\mathbb{\bar{R}}_3$. Then, the previously presented theoretical framework to attain a parallel robust global exponential result for the attitude tracking dynamics parameterized in the base space $\mathrm{SO}(3)$. 

\subsection{Problem Statement}

\par Let the map $\mathbf{r}(t):\mathbb{R}_{\geq 0} \mapsto \boldsymbol{\Omega}$, given by $\mathbf{r}(t) := (\mathbf{R_d}, \boldsymbol{\omega}_{\mathbf{d}})(t)$, define the reference trajectory encompassing the desired rotation matrix, $\mathbf{R_d} \in \mathrm{SO}(3)$, and the desired angular velocity $\boldsymbol{\omega}_{\mathbf{d}} \in \mathbb{R}^3$. The trajectory $\mathbf{r}(t)$ belongs to the compact set $\boldsymbol{\Omega} \subset \mathrm{SO}(3) \times \mathbb{R}^{3}$ and is governed by the differential inclusion
\begin{equation*}
\mathbf{\dot{r}} \in  \mathbf{F_r}(\mathbf{r}) := (\mathbf{R_{d}}[\boldsymbol{\omega}_{\mathbf{d}}]_\times,K_{\omega}\mathbb{B}^3),
\end{equation*}
\noindent with $K_\omega \in \mathbb{R}_{>0}$. With this definition in place, the control objective consists of devising a controller to render the compact set 
\begin{equation}
    \mathcal{A} = \{(\mathbf{r}, \mathbf{x}) \in \boldsymbol{\Omega} \times \boldsymbol{\chi}: \mathbf{R} = \mathbf{R_d}, \boldsymbol{\omega} = \boldsymbol{\omega}_{\mathbf{d}}\},
\end{equation}
with $\mathbf{x}:= (\mathbf{R}, \boldsymbol{\omega}) \in \boldsymbol{\chi} := \mathrm{SO}(3) \times \mathbb{R}^{3}$, robustly globally exponentially stable for the attitude dynamics presented in \eqref{eq:AttitudeDynamics}.

\subsection{MRP-based control design}

\par Let $\mathbf{\Tilde{R}} \in \textrm{SO}(3)$ represent the rotation matrix error given by $\mathbf{\Tilde{R}} = \mathbf{R_d^{\!\top}}\mathbf{R}$ and satisfying $\mathbf{\Tilde{R}} = \mathcal{R}_{\boldsymbol{\vartheta}}(\boldsymbol{\Tilde{\vartheta}})$, where $\boldsymbol{\Tilde{\vartheta}} \in (1+\delta)\mathbb{B}$ denotes the unique and consistent MRP error representation that results from lifting the attitude error $\mathbf{\Tilde{R}}$ to $\mathbb{\bar{R}}^3$ using the hybrid dynamic path-lifting algorithm $\mathcal{H}_{\boldsymbol{\vartheta}}$. The continuous evolution of $\boldsymbol{\Tilde{\vartheta}}$ is governed by
\begin{equation}
\label{eq:AttitudeErrorDynamics}
\boldsymbol{\dot{\Tilde{\vartheta}}} = \mathbf{T}(\boldsymbol{\Tilde{\vartheta}})\boldsymbol{\Tilde{\omega}} = \mathbf{T}(\boldsymbol{\Tilde{\vartheta}})(\boldsymbol{\omega} - \mathbf{\Tilde{R}}^{\!\top}\boldsymbol{\omega}_\mathbf{d})
\end{equation}
where $\boldsymbol{\Tilde{\omega}} \in \mathbb{R}^3$ symbolizes the angular velocity error. Recalling \eqref{eq:AttitudeDynamics}, the underlying differential equation of $\boldsymbol{\Tilde{\omega}}$ has the form 
\begin{equation*}
    \mathbf{J}\boldsymbol{\dot{\Tilde{\omega}}} = \left[\mathbf{J}\boldsymbol{\omega}\right]_{\!\times}\!\boldsymbol{\omega} + \boldsymbol{\tau} - \mathbf{J}(\mathbf{\Tilde{R}}^{\!\top}\boldsymbol{\dot{\omega}}_{\mathbf{d}} - \left[ \boldsymbol{\Tilde{\omega}}\right]_{\!\times}\mathbf{\Tilde{R}}^{\!\top}\!\boldsymbol{\omega}_{\mathbf{d}})
\end{equation*}
\par The design of the control solution relies on the hybrid systems framework to capture the transitions between the original and shadow sets and the resulting repercussions in terms of discrete evolution. It is worth underlining that the MRP description possesses the distinctive characteristic of the shortest principal rotation available corresponding to the set with the smallest norm \cite[p.~120]{junkins_2009}. Devising a control structure that hinges on this inherent characteristic enables handling the unwinding phenomenon automatically, thereby suppressing the requirement of implementing additional mechanisms to solve this issue, which is an advantage compared to quaternion-based solutions (cf.\cite{mayhew2011quaternion}). Furthermore, compared to solutions design on $\mathrm{SO}(3)$, a strategy that leverages this differentiated feature has the potential of achieving a global stability result without multiple potential or error functions at its core (cf. \cite{lee2015global}). Let $\mathbf{x^*_2} := (\mathbf{r}, \mathbf{\Tilde{x}}_{\boldsymbol{\vartheta}}) \in \boldsymbol{\Omega} \times \boldsymbol{\chi_\vartheta}$, with $\mathbf{\Tilde{x}}_{\boldsymbol{\vartheta}} = (\boldsymbol{\Tilde{\vartheta}}, \boldsymbol{\Tilde{\omega}})$ and $\boldsymbol{\chi_\vartheta} = \mathbb{\bar{R}}^3 \times \mathbb{R}^3$, and the control input $\boldsymbol{\tau}: \boldsymbol{\Omega} \times \boldsymbol{\chi_\vartheta} \mapsto \mathbb{R}^3$ be given by:
\begin{equation}
    \label{eq:TauInput}
    \boldsymbol{\tau}(\mathbf{r},\boldsymbol{\Tilde{\vartheta}}, \boldsymbol{\Tilde{\omega}}) = -k_\vartheta \boldsymbol{\Tilde{\vartheta}} -k_\omega \boldsymbol{\Tilde{\omega}} - \left[\mathbf{J}\boldsymbol{\omega}\right]_{\!\times}\!\boldsymbol{\omega} + \mathbf{J}(\mathbf{\Tilde{R}}^{\!\top}\boldsymbol{\dot{\omega}}_{\mathbf{d}} - \left[ \boldsymbol{\Tilde{\omega}}\right]_{\!\times}\mathbf{\Tilde{R}}^{\!\top}\!\boldsymbol{\omega}_{\mathbf{d}})  
\end{equation}
with $k_\vartheta, k_\omega \in \mathbb{R}_{>0}$. With these definitions in place, to define the hybrid attitude tracking system $\mathcal{H}^*_{\mathbf{2}}:= (\mathbf{C^*_2}, \mathbf{F^*_2}, \mathbf{D^*_2}, \mathbf{G^*_2})$, consider the following data:
\begin{subequations}
\label{eq:HybridSystemErrorDynamicsMRP}
\begin{equation}
   \mathbf{F^*_2} (\mathbf{x^*_2}) := \left(\begin{array}{c}
        \mathbf{F_r}(\mathbf{r}) \\ 
        \mathbf{T}(\boldsymbol{\Tilde{\vartheta}})\boldsymbol{\Tilde{\omega}} \\
        \mathbf{J^{-1}} (  -k_\vartheta \boldsymbol{\Tilde{\vartheta}} -k_\omega \boldsymbol{\Tilde{\omega}})
        \end{array}\!\!\right)
\end{equation}
\begin{equation}
    \mathbf{C^*_2} (\mathbf{x^*_2}) := \left\{ \mathbf{x^*_2} \in \boldsymbol{\Omega} \times \boldsymbol{\chi_{\vartheta}}: \|\boldsymbol{\Tilde{\vartheta}}\| \leq 1 + \delta \right\}
\end{equation}
\begin{equation}
    \mathbf{G^*_2} (\mathbf{x^*_2}) := (\mathbf{r}, \boldsymbol{\Upsilon}(\boldsymbol{\Tilde{\vartheta}}),\boldsymbol{\Tilde{\omega}})
\end{equation}
\begin{equation}
     \mathbf{D^*_2} (\mathbf{x^*_2}) := \left\{ \mathbf{x^*_2} \in \boldsymbol{\Omega} \times \boldsymbol{\chi_{\vartheta}}: \|\boldsymbol{\Tilde{\vartheta}}\| \geq 1 + \delta \right\}
\end{equation}
\end{subequations}
\noindent for some $\delta \in \mathbb{R}_{>0}$. \color{black} Note that $\mathcal{H}^*_{\mathbf{2}}$ consists of a particular case of the hybrid system $\mathcal{H}_\mathbf{2}$ presented in \hyperref[section5]{section~\ref*{section5}} insofar as it incorporates the lifted attitude tracking dynamics, which are related to \eqref{eq:AttitudeMRPKinematics}, \eqref{eq:AttitudeDynamicEquation}, and the mapping between the MRP original and shadow sets given in \eqref{eq:ShadowMap}. On the other hand, $\mathcal{H}^*_{\mathbf{2}}$ does not have any controller states analogous to $\boldsymbol{\rho}_{\mathbf{2}}$. Furthermore, given the intention to solve the attitude tracking problem, $\mathcal{H}^*_\mathbf{2}$ also encompasses the differential inclusion governing the attitude trajectory, $\mathbf{F_r}(\mathbf{r})$, constituting another difference from $\mathcal{H}_{\mathbf{2}}$. \color{black} \color{black} Thereby, $\mathcal{H}^*_{\mathbf{2}}$ can be interpreted as a hybrid attitude tracking system that encapsulates the desired attitude trajectory dynamics and the closed-loop that stems from applying the MRP-based feedback law \eqref{eq:TauInput} to the attitude tracking dynamics described using the MRP representation. The jumps of $\mathcal{H}^*_{\mathbf{2}}$ share the same logic of $\mathcal{H}_{\mathbf{2}}$. \color{black} As previously discussed, due to the inclusion of the parameter $\delta$, the jumps between the two MRP sets are robust to measurement noise \cite{martins2021} and, thereby, the solution circumvents the potentially disruptive behavior that arises from noise-induced chattering. \hyperref[thm:StabilityAttitudeTrackingSystemMRP]{Theorem~\ref*{thm:StabilityAttitudeTrackingSystemMRP}} demonstrates the global exponential stability result of the set $\mathcal{A}_\vartheta := \{\mathbf{x^*_2} \in \boldsymbol{\Omega} \times \boldsymbol{\chi_\vartheta}: \boldsymbol{\Tilde{\vartheta}} = \boldsymbol{0},\,\boldsymbol{\Tilde{\omega}} = \boldsymbol{0}\}$ for the hybrid attitude tracking system $\mathcal{H}^\mathbf{*}_\mathbf{2}$.

\begin{thm}
\label{thm:StabilityAttitudeTrackingSystemMRP}
The solutions of the hybrid system $\mathcal{H}^\mathbf{*}_\mathbf{2}$ are complete and bounded, and the compact set $\mathcal{A}_\vartheta$ is globally exponentially stable.
\end{thm}
\begin{proof} See \hyperref[appendix:ProofTheorem3]{Appendix~\ref{appendix:ProofTheorem3}}.  \end{proof}

\subsection{Robust Global Exponential Tracking on $\mathrm{SO}(3)$}

\par With the demonstration of the global exponential stability result of the set $\mathcal{A}_{\vartheta}$ for the lifted tracking system $\mathcal{H}^*_{\mathbf{2}}$, the focus is now on resorting to the theoretical basis previously developed to obtain an equivalent exponential stability result in the base space $\mathrm{SO}(3)$. In this direction, let $\mathcal{H}^*_{\mathbf{1}}:= (\mathbf{C^*_1}, \mathbf{F^*_1}, \mathbf{D^*_1}, \mathbf{G^*_1})$ represent the hybrid system encompassing the reference trajectory and the interconnection between the attitude tracking dynamics described in $\mathrm{SO}(3)$, the dynamic path lifting algorithm $\mathcal{H}_{\vartheta}$, and the MRP-based feedback controller $\boldsymbol{\tau}(\mathbf{r}, \boldsymbol{\Tilde{\vartheta}}, \boldsymbol{\Tilde{\omega}})$. The hybrid system $\mathcal{H}^*_{\mathbf{1}}$ has $\mathbf{x^*_1} := (\mathbf{r}, \mathbf{\Tilde{x}_R}) \in \boldsymbol{\Omega} \times \boldsymbol{\chi}_{\mathbf{R}}$, with $\mathbf{\Tilde{x}_R} = (\mathbf{\Tilde{R}}, \mathbf{\hat{q}}, m, \boldsymbol{\Tilde{\omega}})$ and $\boldsymbol{\chi}_{\mathbf{R}} = \mathrm{SO}(3) \times \mathbb{S}^3 \times \{-1,1\} \times \mathbb{R}^3$, as state vector and its data is given by:

\begin{subequations}
\label{eq:HybridSystemErrorDynamicsR}
\begin{equation}
   \mathbf{F^*_1} (\mathbf{x_1^*}) := \left(\begin{array}{c}
        \mathbf{F_r}(\mathbf{r}) \\
        \mathbf{\Tilde{R}}\left[\boldsymbol{\Tilde{\omega}}\right]_\times \\ \boldsymbol{0} \\ 
        0 \\
        \mathbf{J^{-1}}\left(  -k_\vartheta \boldsymbol{\varphi}^{\!-\!1\!}(m\boldsymbol{\Phi}(\mathbf{\hat{q}}, \mathbf{\Tilde{R}})) - k_\omega \boldsymbol{\Tilde{\omega}}\right)
        \end{array}\!\!\right)
\end{equation}
\begin{equation}
    \mathbf{C^*_1} (\mathbf{x_1^*}) \!:=\! \left\{\mathbf{x^*_1} \!\in\! \boldsymbol{\Omega} \times \boldsymbol{\chi}_{\mathbf{R}}: ( \mathbf{\Tilde{R}}, \mathbf{\hat{q}}, m) \in \mathbf{C_m} \cap \mathbf{C_l} \right\}
\end{equation}
\begin{equation}
    \mathbf{G^*_1} (\mathbf{x_1^*}) \!:=\!
    \left\{\!\!\begin{array}{ll}
      \!\!\!(\mathbf{r}, \mathbf{\Tilde{R}}, \boldsymbol{\Phi}(\mathbf{\hat{q}}, \mathbf{\Tilde{R}}), m, \boldsymbol{\Tilde{\omega}})
\!\!\!&\!\!,  ( \mathbf{\Tilde{R}}, \mathbf{\hat{q}}, m) \!\in\! \mathbf{D_l} \\
        (\mathbf{r}, \mathbf{\Tilde{R}}, \mathbf{\hat{q}}, -m, \boldsymbol{\Tilde{\omega}}) \!\!\!&\!\!, ( \mathbf{\Tilde{R}}, \mathbf{\hat{q}}, m) \!\in\! \mathbf{D_m}
    \end{array}\right.
\end{equation}
\begin{equation}
     \mathbf{D^*_1} (\mathbf{x_1^*}) \!:=\! \left\{\mathbf{x^*_1} \in \boldsymbol{\Omega} \times \boldsymbol{\chi}_{\mathbf{R}}: ( \mathbf{\Tilde{R}}, \mathbf{\hat{q}}, m ) \in \mathbf{D_m} \cup \mathbf{D_l} \right\}
\end{equation}
\end{subequations}

\par \color{black} Analogous to the pair $\mathcal{H}_{\mathbf{2}}$ and $\mathcal{H}_{\mathbf{2}}^{\mathbf{*}}$, the hybrid system $\mathcal{H}_\mathbf{1}^{\mathbf{*}}$ also constitutes a special case of the closed-loop system $\mathcal{H}_{\mathbf{1}}$ previously presented. Similar to $\mathcal{H}_{\mathbf{2}}^{\mathbf{*}}$, $\mathcal{H}_{\mathbf{1}}^{\mathbf{*}}$ encompasses the differential inclusion $\mathbf{F_r}(r)$ and does not have any controller states analogous to $\boldsymbol{\rho}_{\mathbf{1}}$. Since the reference trajectory $\mathbf{r}$ is constant during jumps, the exponential stability equivalence between $\mathcal{H}^*_{\mathbf{1}}$ and $\mathcal{H}^*_{\mathbf{2}}$ is altogether analogous to the one between $\mathcal{H}_{\mathbf{1}}$ and $\mathcal{H}_{\mathbf{2}}$. \color{black} Therefore, one can apply one of the main theoretical contributions of this paper, \hyperref[thm:EquivalenceExponentialStability]{Theorem~\ref*{thm:EquivalenceExponentialStability}}, to demonstrate the equivalence between $\mathcal{H}_\mathbf{1}^{\mathbf{*}}$ and $\mathcal{H}_\mathbf{2}^{\mathbf{*}}$ in terms of global exponential stability results. \hyperref[thm:StabilityAttitudeTrackingSystemR]{Theorem~\ref*{thm:StabilityAttitudeTrackingSystemR}} explores this line of reasoning. In addition, it also shows that $\mathcal{H}_\mathbf{1}^{\mathbf{*}}$ meets the hybrid basic conditions, ensuring, thereby, its well-posedness and, consequently, the robustness of the global exponential result for this hybrid system.

\begin{thm}
\label{thm:StabilityAttitudeTrackingSystemR}
The hybrid system $\mathcal{H}_\mathbf{1}^{\mathbf{*}}$ is well-posed \cite[Definition 6.29.]{goebel_2012}. Furthermore, the compact set $\mathcal{A}_R := \{\mathbf{x^*_1} \in \boldsymbol{\Omega} \times \boldsymbol{\chi}_{\mathbf{R}}: \mathbf{\Tilde{R}} = \mathbf{I_3},\,\boldsymbol{\Tilde{\omega}} = \boldsymbol{0}\}$ is robustly globally exponentially stable.
\end{thm}
\begin{proof} See \hyperref[appendix:ProofTheorem4]{Appendix~\ref{appendix:ProofTheorem4}}.  \end{proof}

\par Bearing in mind the control objective defined at the outset of this section, it is worth noting that $\mathbf{\Tilde{R}} = \mathbf{I_3}$ and $\boldsymbol{\Tilde{\omega}} = \boldsymbol{0}$ imply, respectively, $\mathbf{R} = \mathbf{R_d}$ and $\boldsymbol{\omega} = \boldsymbol{\omega}_{\mathbf{d}}$. Consequently, based on the stability result derived in \hyperref[thm:StabilityAttitudeTrackingSystemR]{Theorem~\ref*{thm:StabilityAttitudeTrackingSystemR}}, it can be inferred that the devised control solution, comprising the MRP-based feedback law $\boldsymbol{\tau}(\mathbf{r}, \boldsymbol{\Tilde{\vartheta}}, \boldsymbol{\Tilde{\omega}})$ and the hybrid dynamic path-lifting algorithm $\mathcal{H}^*_{\boldsymbol{\vartheta}}$, ensures the robust global exponential stability of the compact set $\mathcal{A}$ for the attitude dynamics presented in \eqref{eq:AttitudeDynamics}. Hence, the proposed solution effectively accomplishes the control objective.

\subsection{Simulation results}

\par A comprehensive attitude trajectory tracking experiment was conducted in simulation to demonstrate the global nature of the control solution and the underlying switching logic. The model used for this purpose encompasses the differential equations of the attitude dynamics and kinematics \eqref{eq:AttitudeDynamics}, accounts for measurement noise, and constrains torque actuation within maximum values. This simulation model has a sampling time of $0.01$ seconds and $\mathbf{J} = \textrm{diag}(2.24\times10^{-3},2.90\times10^{-3}, 5.30\times10^{-3}) [\textrm{kg}\textrm{m}^2]$ defined as inertia matrix. Regarding the bounds on the torque input, the model considers the values $|\mathbf{e_1^{\!\top}}\boldsymbol{\tau}|, |\mathbf{e_2^{\!\top}}\boldsymbol{\tau}| \leq 0.45$[Nm] and $|\mathbf{e_3^{\!\top}}\boldsymbol{\tau}| \leq 0.15$ [Nm]. In more detail, the test aimed to assess the capability of the strategy to accurately track an aggressive trajectory, which included challenging flip maneuvers, subjected to an initial condition involving a significant attitude error. The trajectory was defined in terms of Euler angles (roll $\varphi$, pitch $\theta$, and yaw $\psi$) due to its intuitiveness:
\begin{equation*}
\begin{aligned}
    \varphi(t) \!=\! - & \pi(\tanh(1.5\pi (t-2)\!) - \tanh(1.5\pi(t-6)\!) +  \\  & + \tanh(9\pi(t-10)\!) + 1\!), \quad \theta(t) = 0 , 
\end{aligned}
\end{equation*}
\begin{equation*}
    \psi(t) = -\pi(\tanh(\pi(t-4)) - \tanh(\pi(t-10)))
\end{equation*}
The initial attitude was set considering $(\varphi, \theta, \psi)(0) = (-179, 0, 260) [{}^\circ]$ and the control parameters were selected with the values $k_\vartheta = 5, k_\omega = 0.1,$ and $\delta = 0.02$.
 
\par
\begin{figure}[!htb]%

\centering
\subfigure{
\label{fig:SimPosAtt_a}%
\includegraphics[height=3.25cm]{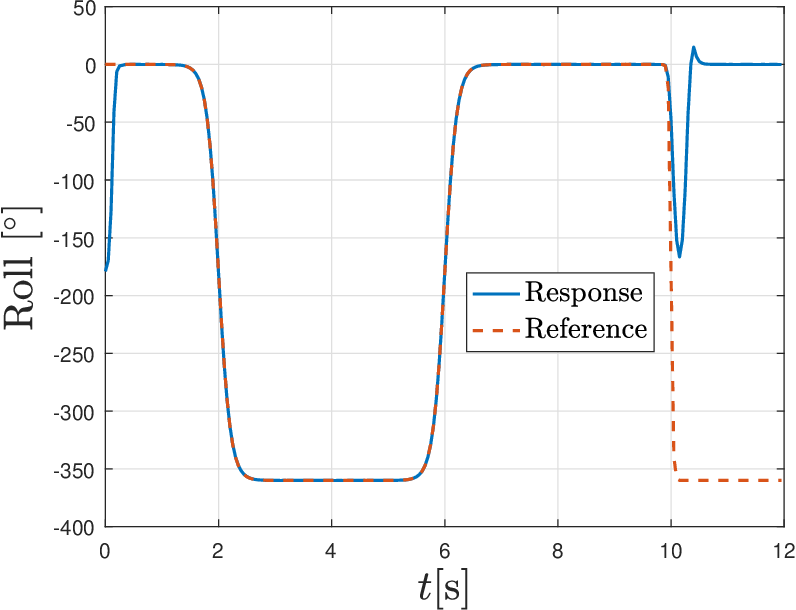}}%
\hspace{-0.0cm}%
\subfigure{%
\label{fig:SimPosAtt_b}%
\includegraphics[height=3.25cm]{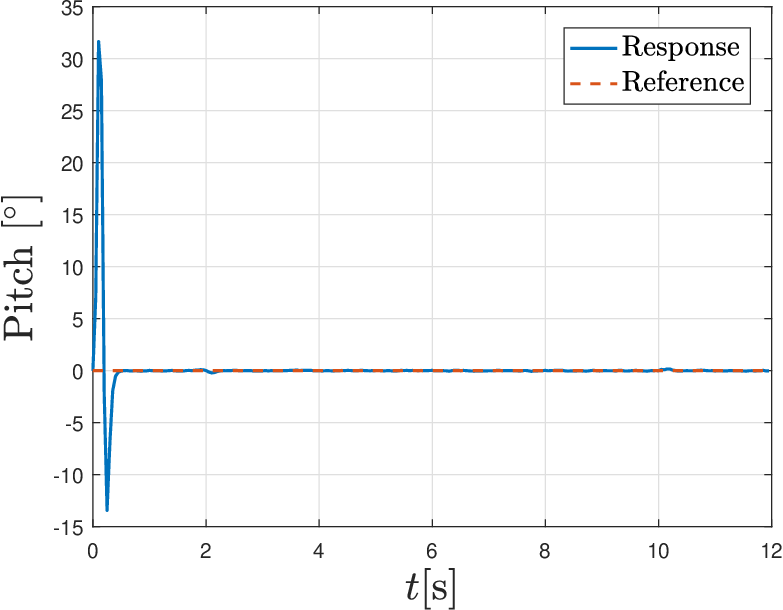}}
\subfigure{
\label{fig:SimPosAtt_c}%
\includegraphics[height=3.25cm]{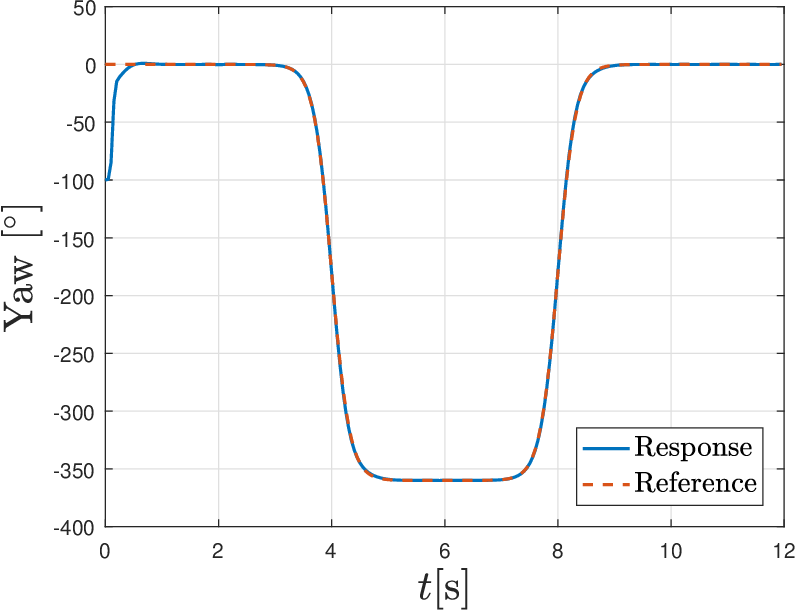}}%
\hspace{-0.05cm}%
\subfigure{%
\label{fig:SimPosAtt_d}%
\includegraphics[height=3.25cm]{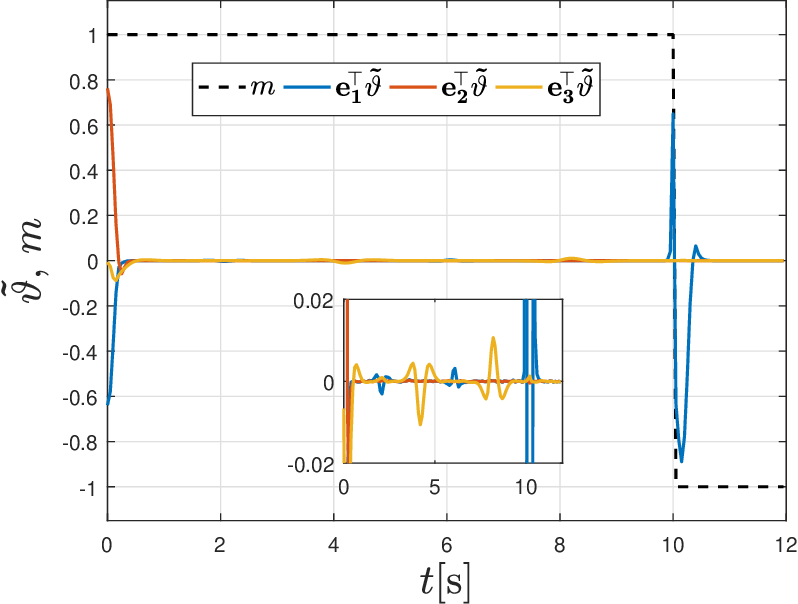}}
\caption{Attitude responses obtained in simulation. From top to bottom, left to right:
\subref{fig:SimPosAtt_a} roll, \subref{fig:SimPosAtt_b} pitch, \subref{fig:SimPosAtt_c} yaw, and \subref{fig:SimPosAtt_d} attitude error in MRP in conjunction with the discrete state $m$.}
\label{fig:SimPosAtt}%
\end{figure}

 The results of the simulation test are illustrated in \hyperref[fig:SimPosAtt]{Fig.~\ref*{fig:SimPosAtt}}. Even though the solution considered the MRP to describe the attitude, the attitude response is also presented in Euler angles to ease its interpretation. These results demonstrate that the rigid body successfully overcame the initial downward-facing orientation and accurately tracked the reference trajectory with the proposed strategy, thereby illustrating the global nature of the solution. Focusing on the roll response, for the third side flip command, provided slightly before the instant $t = 10$s, the rigid body initiated the desired maneuver, but did not complete it. Due to the more aggressive nature of this third command, evident from the steeper slope of the reference, and the limits on the actuation, the MRP error began to increase until the point in which the jump set condition was eventually triggered, causing the rigid body to rotate in the opposite direction. Stemming from the trigger of the jump set condition, in \hyperref[fig:SimPosAtt]{Fig.~\ref*{fig:SimPosAtt_d}}, one can easily observe that the discrete state $m$, which specifies the MRP set in use, changed its value from $1$ to $-1$ and the MRP error also experienced a discrete change. In other words, the rigid body followed the reference rotation as long as it represented the shortest rotation available. Once an equivalent orientation emerged as the new closest target, the direction of rotation changed accordingly. By showcasing this behavior, rooted in the exploitation of the unique properties of the MRP, the responses demonstrate that the devised methodology deals effectively with tumbling situations and the \textit{unwinding phenomenon}. Furthermore, shifting the focus to the yaw response, one notes that the proposed methodology handled the initial tumbling situation by executing the shortest rotation available.

\par
\begin{figure}[!htb]%

\centering
\subfigure{
\label{fig:SimErrorNormInput_a}%
\includegraphics[height=3.25cm]{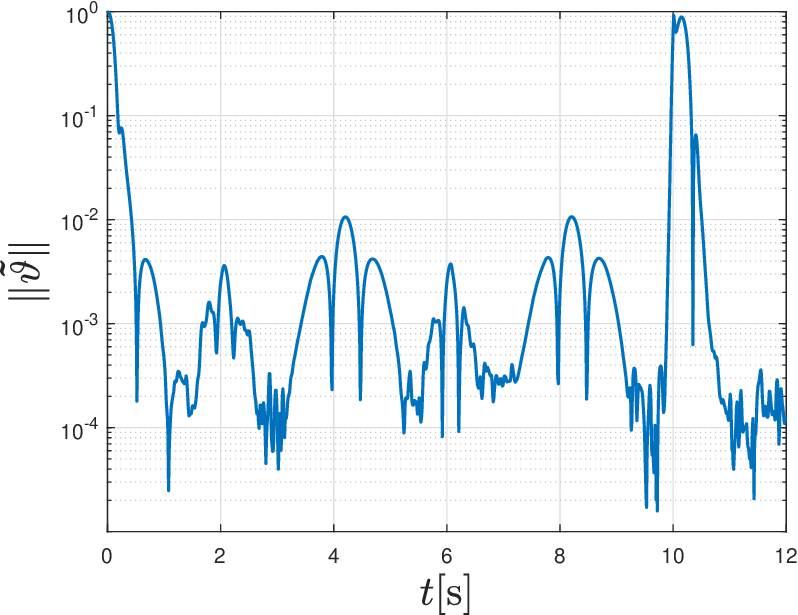}}%
\hspace{-0.0cm}%
\subfigure{%
\label{fig:SimErrorNormInput_b}%
\includegraphics[height=3.25cm]{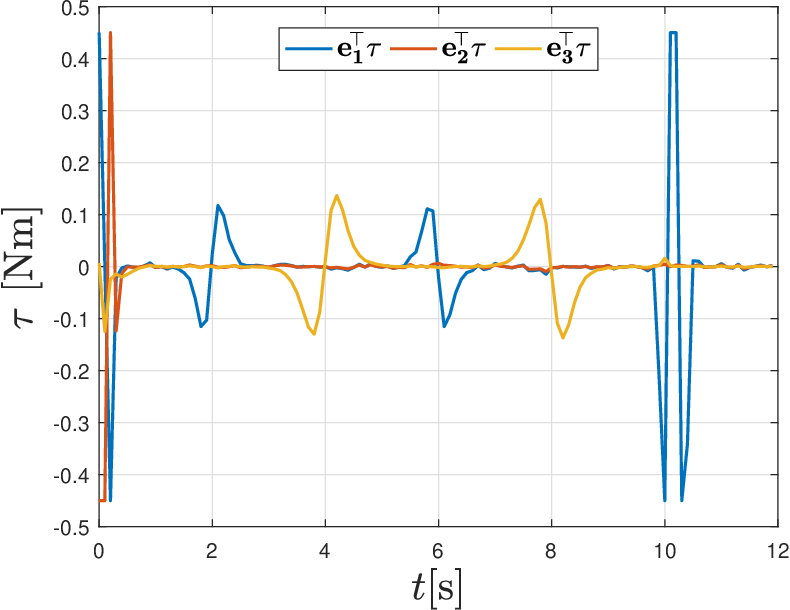}}
\caption{MRP error norm and actuation obtained during the attitude tracking test in simulation. From left to right:
\subref{fig:SimErrorNormInput_a} $\|\boldsymbol{\Tilde{\vartheta}}\|$,
\subref{fig:SimErrorNormInput_b} $\boldsymbol{\tau}$.}
\label{fig:SimErrorNormInput}%
\end{figure}

 \hyperref[fig:SimErrorNormInput]{Figure \ref*{fig:SimErrorNormInput}} exhibits the evolution of the MRP error norm and actuation during the attitude tracking test. The MRP error norm started just below 1, indicating that the rigid body executed the shortest rotation possible to correct the significant initial attitude error. After each side flip, the error norm converged to values smaller than $0.001$. Moreover, it is noteworthy that the control strategy tracked the first two flip maneuvers with a maximum MRP error norm of roughly $0.01$, highlighting the success and global capacity of the approach. Facing the initial attitude error and tracking the reference for the third side flip maneuver led to saturated values of the actuation. In the latter scenario, saturation favored the increase of the MRP error norm, ultimately triggering the jump set condition, as previously discussed.

\vspace{-0.50cm}

\section{Conclusion}
\label{section7}

\par This study proposed a novel hybrid dynamic algorithm to uniquely lift a path from the three-dimensional special orthogonal group $\mathrm{SO}(3)$ to the MRP space $\mathbb{\bar{R}}^3$. 

\par To construct the hybrid algorithm, first, the authors demonstrated that the mapping $\mathcal{R}_{\!\boldsymbol{\vartheta }}: \mathbb{\bar{R}}^3 \mapsto \mathrm{SO}(3)$ is a covering map and a local diffeomorphism everywhere. The path-lifting mechanism robustly solves the ambiguous triplet selection, yielding an output confined within a three-dimensional unit sphere enclosed by a hysteresis region and for which one can append dynamical equations. Moreover, by leveraging the inherent characteristics of the MRP representation, the algorithm can be integrated with an MRP-based controller to yield a closed-loop invulnerable to the unwinding phenomenon, ensuring, thereby, the well-definedness in the base space of a controller designed in the covering manifold. Furthermore, when applying this controller to the rigid-body attitude dynamics, the hybrid algorithm preserves the asymptotic and exponential stability properties.

\par The illustrated application of the path-lifting methodology to tackle the robust global exponential tracking problem for the rigid body attitude dynamics, through the design and validation of an MRP-based controller, evidenced the potential of this novel framework for the equivalence of stability.

\section*{Acknowledgment}

This work was supported by FCT, through IDMEC, under LAETA, project UIDB/50022/2020 and from FCT project CAPTURE (PTDC/EEI-AUT/1732/2020) funded by the Lisboa 2020 and PIDDAC programs, and Project POCI-01-0247- FEDER-046102, co-financed by Programa Operacional Competitividade e Internacionalização and Programa Operacional Regional de Lisboa, through Fundo Europeu de Desenvolvimento Regional (FEDER) and by National Funds through FCT — Fundação para a Ciência e Tecnologia. Luís Martins holds a PhD scholarship 2022.14126.BD from FCT. 

\appendices

\section{Proof of Lemma 1}
\label{appendix:ProofLemma1}

 The sets $\mathbf{C_l}$, $\mathbf{C_m}$, $\mathbf{D_l}$, and $\mathbf{D_m}$ are closed. Since the intersection of closed sets is closed and the union of two closed sets is closed as well, the flow and jump maps, $\mathbf{C}^*_{\boldsymbol{\vartheta}}$ and $\mathbf{D}^*_{\boldsymbol{\vartheta}}$, respectively, are closed. The differential inclusion $\mathbf{\dot{R}} \in \mathbf{R}\left[K\mathbb{B}^3\right]_\times$ comprises a nonempty, outer-semicontinuous, locally bounded, and convex-valued map. From \cite[Theorem 7]{mayhew2013}, the map $\boldsymbol{\Phi}(\mathbf{\hat{q}}, \mathbf{R})$ is nonempty, outer semicontinuous, and locally bounded. Further, the difference equations are given by continuous functions. Thus, the hybrid system $\mathcal{H}_{\boldsymbol{\vartheta}}^*$ verifies the hybrid basic conditions stated in \cite[Assumption 6.5]{goebel_2012}. Therefore, it follows from \cite[Theorem 6.30]{goebel_2012} that $\mathcal{H}_{\boldsymbol{\vartheta}}^*$ is well-posed.
\par Focusing on the completeness of the maximal solutions, by virtue of the set $\mathbf{C}^*_{\boldsymbol{\vartheta}} \cup \mathbf{D}^*_{\boldsymbol{\vartheta}} =  \mathbb{S}^3 \times \mathrm{SO}(3) \times \{-1,1\}$ being compact, every solution is bounded and, consequently, does not escape to infinity. Furthermore, any jump from $\mathbf{D}^*_{\boldsymbol{\vartheta}}$ lands at a point in $\mathbf{C}^*_{\boldsymbol{\vartheta}} \cup \mathbf{D}^*_{\boldsymbol{\vartheta}}$. Therefore, every maximal solution is complete.
\par The set $\mathbf{C_m}$ is invariant since any jump from its boundary, which corresponds to the intersection between the sets $\mathbf{C_m}$ and $\mathbf{D_m}$, maps the state to $\mathbf{C_m} \setminus \mathbf{D_m}$. Furthermore, according to \cite[Theorem 7]{mayhew2013}, the set $\mathbf{C_l}$ is invariant as well. Therefore, in light of the intersection between invariant sets being also invariant, $\mathbf{C}^*_{\boldsymbol{\vartheta}}$ is an invariant set. Hence, in virtue of any given state in $\mathbf{D}^*_{\boldsymbol{\vartheta}}$ being mapped to $\mathbf{C}^*_{\boldsymbol{\vartheta}}$, it follows that any solution $\left(\mathbf{\hat{q}}, \mathbf{R}, m\right)(t,j)$ to $\mathcal{H}_{\boldsymbol{\vartheta}}$ verifies $\{(t,j): \left.( \mathbf{\hat{q}}, \mathbf{R}, m)\right|_{t,j} \notin \mathbf{C_m} \cap \mathbf{C_l} \} \subset \{(0, 0)\}$.

\par In virtue of $\mathcal{R}_{\!\boldsymbol{\vartheta}}: \mathbb{\bar{R}}^3 \mapsto \mathrm{SO}(3)$ being a covering map, it follows from \cite[Lemma 54.1]{munkres2000topology} that, for each fixed $j \in \mathbb{N}$, a given continuous path $\mathbf{R_{\downarrow t}}: \mathcal{T}_j \mapsto \mathrm{SO}(3)$ satisfying $\mathbf{R_{\downarrow t}}(min\{\mathcal{T}_j\}) = \mathcal{R}_{\!\boldsymbol{\vartheta}}(\boldsymbol{\vartheta}_{\mathbf{\downarrow t}}(min\{\mathcal{T}_j\}))$ has a unique lifting to a continuous path $\boldsymbol{\vartheta}_{\mathbf{\downarrow t}}(t)$ in the covering space $\mathbb{\bar{R}}^3$ verifying 
\begin{equation*}
    \boldsymbol{\vartheta}_{\mathbf{\downarrow t}}(\min\{\mathcal{T}_j\}) = \boldsymbol{\varphi}^{\!-\!1\!}(m\boldsymbol{\Phi}(\mathbf{\hat{q}}, \mathbf{R}))|_{(\min\{\mathcal{T}_j\},j)}.
\end{equation*}
With this in mind, invariably, the ensuing step is to demonstrate that for $(t,j) \in \mathcal{T}_j \times j$, the lifted path indeed verifies 
\begin{equation*}
    \boldsymbol{\vartheta}_{\mathbf{\downarrow t}}(t) = \boldsymbol{\varphi}^{\!-\!1\!}(m\boldsymbol{\Phi}(\mathbf{\hat{q}}, \mathbf{R}))|_{(t,j)}.
\end{equation*}
In this direction, suppose that, for $\left( \mathbf{\hat{q}}, \mathbf{R}, m\right) \!\in\! \mathbf{C_m} \cap \mathbf{C_l}$, a continuous $\mathbf{R_{\downarrow t}}(t): \mathcal{T}_j \mapsto \mathrm{SO}(3)$, and a continuous $\boldsymbol{\vartheta}_{\mathbf{\downarrow t}}(t): \mathcal{T}_j \mapsto \mathbb{\bar{R}}^3$ satisfying $\mathcal{R}_{\!\boldsymbol{\vartheta}}(\boldsymbol{\vartheta}_{\mathbf{\downarrow t}}(t)) = \mathbf{R}_{\mathbf{\downarrow t}}(t)$, there exists an instant $t' \in \mathcal{T}_j$ such that 
\begin{equation*}
\boldsymbol{\varphi}^{\!-\!1\!}(m_{\downarrow t}(t')\boldsymbol{\Phi}(\mathbf{\hat{q}}_{\mathbf{\downarrow t}}(t'), \mathbf{R}_{\mathbf{\downarrow t}}(t'))) = \boldsymbol{\Upsilon}(\boldsymbol{\vartheta}_{\mathbf{\downarrow t}}(t')).    
\end{equation*}
Note that the latter equality is equivalent to 
\begin{equation*}
    \boldsymbol{\varphi}(\boldsymbol{\vartheta}_{\mathbf{\downarrow t}}(t')) = - \boldsymbol{\Phi}(\mathbf{\hat{q}}_{\mathbf{\downarrow t}}(t'), \mathbf{R}_{\mathbf{\downarrow t}}(t')),
\end{equation*}
which would imply that either 
\begin{equation}
    \label{eq:scenario1}
    1+\mathbf{\hat{q}}_{\mathbf{\downarrow t}}^{\!\top}(t')\boldsymbol{\varphi}(\boldsymbol{\vartheta}_{\mathbf{\downarrow t}}(t')) > 1 
\end{equation}
or 
\begin{equation}
    \label{eq:scenario2}
    m_{\downarrow t}(t') = -m_{\downarrow t}\left(\min\{\mathcal{T}_j\}\right)
\end{equation}
Since the function (1 + $\mathbf{\hat{q}^{\!\top}}\mathbf{q}):\mathbb{S}^3 \times \mathbb{S}^3 \mapsto \mathbb{R}$ is continuous, the inequality \eqref{eq:scenario1} would translate to the existence of an instant $t^* \in \left[\min\{\mathcal{T}_j\}\; t'\right]$ such that 
\begin{equation*}
    \mathrm{dist}(\mathbf{\hat{q}}_{\mathbf{\downarrow t}}(t^*), \mathcal{Q} (\mathbf{R}_{\mathbf{\downarrow t}}(t^*))) = 1.
\end{equation*}
Alternatively, \eqref{eq:scenario2} presupposes the existence of an instant $t^* \in \left[\min\{\mathcal{T}_j\}\; t'\right]$ such that \begin{equation*}
    \|\boldsymbol{\varphi}^{\!-\!1\!}\left(m_{\downarrow t} (t^*)\boldsymbol{\Phi}\left(\mathbf{\hat{q}}_{\mathbf{\downarrow t}}(t^*), \mathbf{R}_{\mathbf{\downarrow t}}(t^*)\right)\right)\| > 1 + \delta.
\end{equation*}
Thus, given that the function $\boldsymbol{\Phi}(\mathbf{\hat{q}}, R)$ is double-valued for $\mathrm{dist}(\mathbf{\hat{q}}, \mathcal{Q}(\mathbf{R})) = 1$, both scenarios would violate the flow assumption $\left( \mathbf{\hat{q}}, \mathbf{R}, m\right) \!\in\! \mathbf{C_m} \cap \mathbf{C_l}$. Hence, for $(t,j) \in \mathcal{T}_j \times j$, the map composition $\boldsymbol{\varphi}^{\!-\!1\!}(m\boldsymbol{\Phi}(\mathbf{\hat{q}}, \mathbf{R}))$ is continuous and the lifted path satisfies 
\begin{equation*}
    \boldsymbol{\vartheta}(t,j) = \boldsymbol{\varphi}^{\!-\!1\!}(m\boldsymbol{\Phi}(\mathbf{\hat{q}}, \mathbf{R}))|_{(t,j)}.
\end{equation*}
As a result, for a fixed $j \in \mathbb{N}$, the output $\boldsymbol{\vartheta}(t,j)$ verifies $\mathbf{R_{\downarrow t}}(t) = \mathcal{R}_{\boldsymbol{\vartheta}}\left(\boldsymbol{\vartheta}_{\mathbf{\downarrow t}}(t)\right)$. To examine the behavior of the output during jumps, the following property of the map $\mathcal{R}_{\!\boldsymbol{\vartheta}}$ is exploited:
\begin{equation}
    \label{eq:PropertyRtheta}
    \mathcal{R}_{\!\boldsymbol{\vartheta}}(\boldsymbol{\vartheta'}) = \mathcal{R}_{\!\boldsymbol{\vartheta}}(\boldsymbol{\vartheta}) \quad \mathrm{iff} \quad \boldsymbol{\vartheta'} = \boldsymbol{\Upsilon}(\boldsymbol{\vartheta})
\end{equation}
Let $\{(t,j), (t, j+1)\} \subset \mathrm{dom}\;\left( \mathbf{\hat{q}}, \mathbf{R}, m\right)$. There are two scenarios for $\boldsymbol{\vartheta}(t, j+1)$: the jump either stems from the fulfilment of the condition defining $\mathbf{D_m}$, yielding $\boldsymbol{\vartheta}(t, j+1) = \boldsymbol{\varphi}^{\!-\!1\!}(-m\boldsymbol{\Phi}(\mathbf{\hat{q}}, \mathbf{R}))|_{(t,j)}$, or from the verification of the inequality governing $\mathbf{D_l}$, resulting in $\boldsymbol{\vartheta}(t, j+1) = \boldsymbol{\varphi}^{\!-\!1\!}(m\boldsymbol{\Phi}(\boldsymbol{\Phi}(\mathbf{\hat{q}}, \mathbf{R}), \mathbf{R}))|_{(t,j)}$. Regarding the first scenario, observe that the map $\boldsymbol{\varphi}$ satisfies the equality
\begin{equation}
 \label{eq:PropertyJumpStereographicProjection}
    \boldsymbol{\varphi}^{\!-\!1\!}(-m\boldsymbol{\Phi}(\mathbf{\hat{q}}, \mathbf{R}))|_{(t,j)} = \boldsymbol{\Upsilon}\left(\boldsymbol{\varphi}^{\!-\!1\!}(m\boldsymbol{\Phi}(\mathbf{\hat{q}}, \mathbf{R}))|_{(t,j)}\right)
\end{equation}
Thus, $\boldsymbol{\vartheta}(t,j+1) = \boldsymbol{\Upsilon}(\boldsymbol{\vartheta}(t,j))$, and, recalling \eqref{eq:PropertyRtheta}, $\mathcal{R}_{\!\boldsymbol{\vartheta}}(\boldsymbol{\vartheta}(t,j+1)) = \mathcal{R}_{\!\boldsymbol{\vartheta}}(\boldsymbol{\vartheta}(t,j))$. Concerning the second scenario, in virtue of the assumption $\mathrm{dist}(\mathbf{\hat{q}}, \mathcal{Q}(\mathbf{R}))|_{(0, 0)} < 1$ and, according to \cite[Theorem 7]{mayhew2013}, $\mathbf{C_l}$ being invariant, it follows from \cite[Lemma 3]{mayhew2013} that a jump out of $\mathbf{D_l}$ yields $\boldsymbol{\vartheta}(t,j) = \boldsymbol{\vartheta}(t,j+1)$. Hence, one has $\mathcal{R}_{\!\boldsymbol{\vartheta}}(\boldsymbol{\vartheta}(t,j+1)) = \mathcal{R}_{\!\boldsymbol{\vartheta}}(\boldsymbol{\vartheta}(t,j))$ as well for this jump scenario. Therefore, the map $\mathcal{R}_{\!\boldsymbol{\vartheta}}(\boldsymbol{\vartheta}(t,j))$ is constant during jumps. This conclusion, in conjunction with the fact that, for every flow, exists a unique path-lifting $\boldsymbol{\vartheta}_{\mathbf{\downarrow t}}(t)$ of $\mathbf{R_{\downarrow t}}(t)$ satisfying $\mathcal{R}_{\!\boldsymbol{\vartheta}}(\boldsymbol{\vartheta}_{\mathbf{\downarrow t}}(\min\{\mathcal{T}_j\})) = \mathbf{R}_{\mathbf{\downarrow t}}(\min\{\mathcal{T}_j\}))$ with $\boldsymbol{\vartheta}_{\mathbf{\downarrow t}}(\min\{\mathcal{T}_j\}) = \boldsymbol{\varphi}^{\!-\!1\!}(m\boldsymbol{\Phi}(\mathbf{\hat{q}}, \mathbf{R}))|_{(\min\{\mathcal{T}_j\},j)}$, allows stating that the output $\boldsymbol{\vartheta}(t,j)$ verifies $\mathbf{R_{\downarrow t}}(t) = \mathcal{R}_{\boldsymbol{\vartheta}}\left(\boldsymbol{\vartheta}_{\mathbf{\downarrow t}}(t)\right)$ for all $(t,j) \in \mathrm{dom}\;\left( \mathbf{\hat{q}}, \mathbf{R}, m\right)$.
\par In view of the above, the output $\boldsymbol{\vartheta}$ is constant during jumps from $\mathbf{D_l}$ and its norm satisfies $\|\boldsymbol{\vartheta}\| \leq 1 + \delta$ after a jump from $\mathbf{D_m}$. Furthermore, notice that $\left\{(t,j): \|\boldsymbol{\varphi}^{\!-\!1\!}(m\boldsymbol{\Phi}(\mathbf{\hat{q}}, \mathbf{R}))\| > 1 + \delta\right\} \subset \left\{(0, 0)\right\}$ and that this initial condition would mean $\left(\mathbf{\hat{q}}, \mathbf{R}, m\right)|_{(0, 0)} \in \mathbf{D_m}$. Thus, given the definition of the output $\boldsymbol{\vartheta}$, provided in \eqref{eq:OutputLiftingMRP}, the bound $\|\boldsymbol{\vartheta}(t,j)\| \leq 1 + \delta$ is attained.

\section{Proof of Lemma 2}
\label{appendix:ProofLemma2}

 Given the assumption $\mathrm{dist}(\mathbf{\hat{q}_1}, \!\mathcal{Q}(\mathbf{R_1})\!)|_{(0, 0)} < 1$ and  \cite[Theorem 7]{mayhew2013}, the function $\left.\boldsymbol{\Phi}\left(\mathbf{\hat{q}_1}, \mathbf{R_1}\right)\right|_{(t,j)}$, and, consequently, $\left.\boldsymbol{\varphi}^{\!-\!1\!}\!\left(m_1\boldsymbol{\Phi}\!\left(\mathbf{\hat{q}_1}, \mathbf{R_1}\!\right)\!\right)\right|_{(t,j)}$, are single-valued for all $(t,j) \in \mathrm{dom}\; \mathbf{x_1}$. For the purpose of recursively expressing the solution $\mathbf{x_2}$ in terms of $\mathbf{x_1}$, let $(t_0^\vartheta, j_0^\vartheta) = (0, 0)$. Then, for each $i \in \mathbb{N}$, let $\left(t^{\vartheta}_{i+1}, j^{\vartheta}_{i+1} \right) \in \mathrm{dom}\; \mathbf{x_1}$, given by
\begin{equation}
\begin{aligned}
    \left(t^{\vartheta}_{i+1} \right. & \left.\!\!\!,  \!j^{\vartheta}_{i+1} \!\right) \!=\!  \min\{\!(t,j) \!\in\! \mathrm{dom}\;\mathbf{x_1} \!\!:\! (t, j\!-\!1\!) \!\in\! \mathrm{dom}\;\mathbf{x_1}, \\ & \!j \! > \! j^{\vartheta}_{i} \! , \mathbf{\hat{q}}(t, j\!-\!1) \!=\! \mathbf{\hat{q}}(t, j), m(t, j\!-\!1\!) \!=\! -m(t,j)\!\},
\end{aligned}
\end{equation}
denote the hybrid time instant immediately after the $(i+1)^\mathrm{{th}}$ jump triggered by the MRP norm. The minimum of this set is computed based on a natural way of
ordering satisfying: for $(t,j), (t',j') \in \mathrm{dom}\; \mathbf{x_1}, (t,j) \preceq (t',j')$ if $t \leq t'$ or $t = t'$ and $j \leq j'$. There are two possible outcomes for each $\left(t^{\vartheta}_{i+1}, j^{\vartheta}_{i+1} \right)$: either $\left(t^{\vartheta}_{i+1}, j^{\vartheta}_{i+1} \right) \neq \emptyset$, corresponding to the case in which an MRP jump indeed occurs, or $\left(t^{\vartheta}_{i+1}, j^{\vartheta}_{i+1} \right) = \emptyset$, indicating that there are no further jumps caused by the MRP vector norm. Starting with the former case, for each $i \in \mathbb{N}$, the hybrid time domain of $\mathbf{x_2}$ is recursively constructed through
\begin{equation*}
\begin{aligned}
        \left\{\!(t,j) \in \mathrm{dom}\;\mathbf{x_2}\!:\! \right. & \left. (t,j) \preceq (t^\vartheta_{i+1}, i + 1 )\!\right\} \!= \\ & \!\left(\bigcup_{k = 0}^{i} \left(\left[t^\vartheta_k, \; t^\vartheta_{k+1}\right], k\right)\!\!\right) \!\cup (t^\vartheta_{i+1}, i + 1 )
\end{aligned}
\end{equation*}
and the solution $\mathbf{x_2}$, for every $(t,j) \ \in \mathrm{dom}\;\mathbf{x_1}$ verifying $ t \in \left[t_i^\vartheta,\;t^{\vartheta}_{i+1}\right]$ and $j^\vartheta_i \leq j \leq j^\vartheta_{i+1}$, is defined as
\begin{equation}
    \label{eq:SolutionX2MRPJumps}
    \mathbf{x_2}(t, i) = \left.\left(\boldsymbol{\varphi}^{\!-\!1\!}\!\left(m_1\boldsymbol{\Phi}\!\left(\mathbf{\hat{q}_1}, \mathbf{R_1}\!\right)\!\right), \boldsymbol{\omega_1}, \boldsymbol{\rho_1}\right)\right|_{(t,j)}
\end{equation}
To verify the compliance of \eqref{eq:SolutionX2MRPJumps} with the jump dynamics of $\mathcal{H}_\mathbf{2}$, observe that, for the $(i+1)^{\mathrm{th}}$ jump due to the MRP norm, one has
\begin{equation}
    \mathbf{x_2}(t^\vartheta_{i+1}, i) \!=\! \left.\left(\boldsymbol{\varphi}^{\!-\!1\!}\!\left(m_1\boldsymbol{\Phi}\!\left(\mathbf{\hat{q}_1}, \mathbf{R_1}\!\right)\!\right), \boldsymbol{\omega_1}, \boldsymbol{\rho_1} \!\right) \right|_{(t^\vartheta_{i+1},j^\vartheta_{i+1}\!-\!1)}
\end{equation}
\begin{equation}
    \mathbf{x_2}(t^\vartheta_{i+1}, i \!+\! 1\!) \!=\! \left.\left(\boldsymbol{\varphi}^{\!-\!1\!}\!\left(m_1\boldsymbol{\Phi}\!\left(\mathbf{\hat{q}_1}, \mathbf{R_1}\!\right)\!\right), \boldsymbol{\omega_1}, \boldsymbol{\rho_1} \! \right) \right|_{(t^\vartheta_{i+1},j^\vartheta_{i+1})}
\end{equation}
The jump dynamics of $\mathcal{H}_\mathbf{1}$ combined with the property \eqref{eq:PropertyJumpStereographicProjection} leads to 
\begin{equation}
    \mathbf{x_2}(t^\vartheta_{i+1}, i + 1) = \left.\left(\boldsymbol{\Upsilon}(\boldsymbol{\vartheta}_2), \boldsymbol{\omega_2}, \boldsymbol{\rho_2}\right)\right|_{(t^\vartheta_{i+1}, i)}
\end{equation}
Therefore, the solution \eqref{eq:SolutionX2MRPJumps} verifies the jump dynamics of $\mathcal{H}_{\mathbf{2}}$ for each pair $\{(t,j), (t,j+1)\} \in \mathrm{dom}\; \mathbf{x_2}$ satisfying $(t,j) \preceq (t^\vartheta_{i+1}, i)$. Regarding the flow dynamics, first, it is worth highlighting that the covering map $\mathcal{R}_{\boldsymbol{\vartheta}}: \mathbb{\bar{R}}^3 \mapsto \mathrm{SO}^3$ is everywhere a local diffeomorphism and, thus, for every element $\boldsymbol{\vartheta} \in \mathbb{\bar{R}}^3$, the open neighborhood of $\boldsymbol{\vartheta}$ is diffeomorphic to an open neighborhood of an element $\mathcal{R}_{\boldsymbol{\vartheta}}(\boldsymbol{\vartheta}) = \mathbf{R} \in \mathrm{SO}(3)$, having, thereby, equivalent local differentiable structures. In addition, notice that, if $j^\vartheta_i < j^\vartheta_{i+1} - 1$, the hybrid system $\mathcal{H}_{\mathbf{1}}$ jumps to update the memory state $\mathbf{\hat{q}_1}$ for $t \in  \left[t^\vartheta_{i}, t^\vartheta_{i+1}\right]$. Nonetheless, as demonstrated in  \hyperref[lemH*]{Lemma~\ref*{lemH*}}, the equality $\left.\boldsymbol{\varphi}^{\!-\!1\!}\!\left(m_1\boldsymbol{\Phi}\!\left(\mathbf{\hat{q}_1}, \mathbf{R_1}\!\right)\!\right)\right|_{(t,j)} = \left.\boldsymbol{\varphi}^{\!-\!1\!}\!\left(m_1\boldsymbol{\Phi}\!\left(\mathbf{\hat{q}_1}, \mathbf{R_1}\!\right)\!\right)\right|_{(t,j+1)}$ holds when updates of the memory state $\mathbf{\hat{q}_1}$ occur and, consequently, the solution proposed in \eqref{eq:SolutionX2MRPJumps} remains constant during these jumps. Furthermore, from the proof of \hyperref[lemH*]{Lemma~\ref*{lemH*}} it also follows that, for all $(t,j) \in \mathrm{dom}\; \mathbf{x_1}$ satisfying $ t \in \left[t_i^\vartheta,\;t^{\vartheta}_{i+1}\right]$ and $j^\vartheta_i \leq j \leq j^\vartheta_{i+1}$, the map $\left.\boldsymbol{\varphi}^{\!-\!1\!}\!\left(m_1\boldsymbol{\Phi}\!\left(\mathbf{\hat{q}_1}, \mathbf{R_1}\!\right)\!\right)\right|_{(t,j)}$ is continuous and verifies $ \mathcal{R}_{\boldsymbol{\vartheta}}(\left.\boldsymbol{\varphi}^{\!-\!1\!}\!\left(m_1\boldsymbol{\Phi}\!\left(\mathbf{\hat{q}_1}, \mathbf{R_1}\!\right)\!\right)\right|_{(t,j)}) = \mathbf{R_1}(t,j)$. Thus, $\boldsymbol{\vartheta_2}(t,i) = \left.\boldsymbol{\varphi}^{\!-\!1\!}\!\left(m_1\boldsymbol{\Phi}\!\left(\mathbf{\hat{q}_1}, \mathbf{R_1}\!\right)\!\right)\right|_{(t,j)}$, for all $(t,j) \in \mathrm{dom}\; \mathbf{x_1}$ satisfying $ t \in \left[t_i^\vartheta,\;t^{\vartheta}_{i+1}\right]$ and $j^\vartheta_i \leq j \leq j^\vartheta_{i+1}$, yields $\mathbf{R_1}(t,j) = \mathcal{R}_{\boldsymbol{\vartheta}}(\boldsymbol{\vartheta_2}(t,i))$ and complies with the kinematic expression \eqref{eq:AttitudeMRPKinematics}. Hence, on the basis of these results and in light of the states $\boldsymbol{\omega_2}$ and $\boldsymbol{\rho_2}$ being equal to the counterpart states of $\mathcal{H}_\mathbf{1}$, being governed by identical differential equations, and being constant during jumps, one concludes that the solution \eqref{eq:SolutionX2MRPJumps} verifies the flow dynamics of $\mathcal{H}_{\mathbf{2}}$ for each $(t,j) \in \mathrm{dom}\; \mathbf{x_2}$ satisfying $(t,j) \preceq (t^\vartheta_{i+1}, i+1)$. In the event of $\left(t^{\vartheta}_{i+1}, j^{\vartheta}_{i+1} \right)  \neq \emptyset \; \forall \; i \in \mathbb{N}$, the domain of the solution $\mathbf{x_2}(t,j)$ is given by
\begin{equation*}
    \mathrm{dom}\;\mathbf{x_2} = \bigcup^\infty_{k=0} \left([t^\vartheta_k, t^\vartheta_{k+1}], k\right),
\end{equation*}
and, in virtue of the foregoing considerations, the expression \eqref{eq:SolutionX2MRPJumps} describes a solution to $\mathcal{H}_{\mathbf{2}}$ for all $(t,j') \in \mathrm{dom}\; \mathbf{x_2}$. Moreover, for every $(t,j) \in \mathrm{dom}\;\mathbf{x_1}$, there exists $(t,j') \in \mathrm{dom}\;\mathbf{x_2}$, with $j' \leq j$, such that \eqref{eq:SolutionsH1H2} holds.
\par As previously discussed, the second case anticipates the scenario in which there exists an $i' \in \mathbb{N}$ such that $\left(t^{\vartheta}_{i'+1}, j^{\vartheta}_{i'+1} \right) = \emptyset$. In this context, the domain of $\mathbf{x_2}$ is defined as
\begin{equation}
    \label{eq:DomainX2NofurtherMRPjumps}
    \mathrm{dom}\; \mathbf{x_2} \!=\!\! \left\{ \!\!\!\!\!\begin{array}{cl}
         \left([0, \bar{T}(\mathbf{x_1})[,0\right) \!\!\!\!\!&,  \; \mathrm{for} \; i'\!=\!0\\
        \left(\bigcup \limits_{k = 0}^{i'-1} \!\!\left(\!\left[t^\vartheta_k, t^\vartheta_{k+1}\!\right]\!\!, k\right)\!\!\right) \!\!\cup\! \left([t^\vartheta_{i'}, \bar{T}(\mathbf{x_1})\!),i'\right) \!\!\!\!\!&,  \; \mathrm{otherwise}
    \end{array} \right.
\end{equation}
The previous paragraph demonstrates that \eqref{eq:SolutionX2MRPJumps} yields a solution to $\mathcal{H}_{\mathbf{2}}$ for every $\{(t,j)\} \in \mathrm{dom}\; \mathbf{x_2}$ satisfying $(t,j) \preceq (t^\vartheta_{i'}, i')$. For every $(t,i') \in \mathrm{dom}\;\mathbf{x_2}$ and every $(t,j) \in \mathrm{dom}\; \mathbf{x_2}$ satisfying $j \geq j^\vartheta_{i'}$ , one has
\begin{equation}
    \label{eq:SolutionX2NofurtherMRPjumps}
    \mathbf{x_2}(t,i') = \left.\left(\boldsymbol{\varphi}^{\!-\!1\!}\!\left(m_1\boldsymbol{\Phi}\!\left(\mathbf{\hat{q}_1}, \mathbf{R_1}\!\right)\!\right), \boldsymbol{\omega_1}, \boldsymbol{\rho_1}\right)\right|_{(t,j)}
\end{equation}
By applying a reasoning similar to that followed to demonstrate that \eqref{eq:SolutionX2MRPJumps} verifies the hybrid dynamics of $\mathcal{H}_{\mathbf{2}}$ for every $\{(t,j)\} \in \mathrm{dom}\; \mathbf{x_2}$ satisfying $(t,j) \preceq (t^\vartheta_{i'}, i')$, one concludes that $\mathbf{x_2}:\mathrm{dom}\; \mathbf{x_2} \mapsto \boldsymbol{\chi}_{\mathbf{2}}$, defined by \eqref{eq:SolutionX2MRPJumps} in combination with \eqref{eq:SolutionX2NofurtherMRPjumps}, with 
$\mathrm{dom}\; \mathbf{x_2}$ specified in \eqref{eq:DomainX2NofurtherMRPjumps}, verifies the flow and jump dynamics of $\mathcal{H}_\mathbf{2}$, constituting, thereby, a solution to this hybrid system. Further, for every $(t,j) \in \mathrm{dom}\; \mathbf{x_1}$, there exists $(t,j') \in \mathrm{dom}\; \mathbf{x_2}$, with $j'\leq j$, such that \eqref{eq:SolutionsH1H2} holds. 
\par Having demonstrated the first part of the lemma, the proof of the second part is straightforward if one uses the previous sequence of arguments as a blueprint. To elaborate, this proof follows similarly by recursively constructing the solution to $\mathcal{H}_\mathbf{1}$ in terms of the solution to $\mathcal{H}_\mathbf{2}$, with the minor nuance of adding the jumps that update the quaternion memory state $\mathbf{\hat{q}_1}$.

\section{Proof of Theorem 1}
\label{appendix:ProofTheorem1}

    The proof flows from combining the definitions of stable and attractive sets with the result presented in \hyperref[lemSolutionsEquivalence]{Lemma~\ref*{lemSolutionsEquivalence}}. In this direction, to prove the first part of the theorem, assume that the compact set $\mathcal{A}_{\boldsymbol{\vartheta}}$ is asymptotically stable for $\mathcal{H}_{\mathbf{2}}$. Then, according to \cite[Definition 7.1]{goebel_2012}, it follows that for every $\epsilon \in \mathbb{R}_{>0}$ there exists $\iota \in \mathbb{R}_{>0}$ satisfying $\iota \leq \epsilon$ such that each solution $\mathbf{x_2}(t,j)$ to $\mathcal{H}_{\mathbf{2}}$ with $\|\mathbf{x_2}(0, 0)\|_{\mathcal{A}_{\boldsymbol{\vartheta}}} < \iota$ yields $\|\mathbf{x_2}(t,j)\|_{\mathcal{A}_{\boldsymbol{\vartheta}}} < \epsilon\; \forall \; (t,j) \in \mathrm{dom}\; \mathbf{x_2}$. Furthermore, it also follows that $\mathbf{x_2}(0, 0) \in \mathcal{B}_{\boldsymbol{\vartheta}}$ yields $\lim_{(t+j) \rightarrow \infty} \|\mathbf{x_2}(t,j)\|_{\mathcal{A}_{\boldsymbol{\vartheta}}} = 0$. In light of \hyperref[lemSolutionsEquivalence]{Lemma~\ref*{lemSolutionsEquivalence}}, for every solution $\mathbf{x_2}$ to $\mathcal{H}_{\mathbf{2}}$, there exists a solution $\mathbf{x_1}$ to $\mathcal{H}_{\mathbf{1}}$ such that, for every $(t,j') \in \mathrm{dom}\;\mathbf{x_2}$, \eqref{eq:SolutionsH1H2} holds with $(t,j) \in \mathrm{dom}\;\mathbf{x_1}$ and $j \geq j'$. Note that, one can assume, without sacrificing generality, that $\mathbf{x_1}(t,j)$ satisfies $\mathrm{dist}(\mathbf{\hat{q}_1}, \!\mathcal{Q}(\mathbf{R_1})\!)|_{(0, 0)} < 1$. Therefore, $\|\mathbf{x_2}(0, 0)\|_{\mathcal{A}_{\boldsymbol{\vartheta}}} < \iota$ and $\mathbf{x_2}(0, 0) \in \mathcal{B}_{\boldsymbol{\vartheta}}$ imply, respectively,
    \begin{subequations}
    \begin{equation} 
    \label{eq:SolutionX1IC1}
    \|\left. \left( \boldsymbol{\varphi}^{\!-\!1\!} \! \left(m_1\boldsymbol{\Phi}\!\left(\mathbf{\hat{q}_1}, \!\mathbf{R_1}\!\right)\right), \boldsymbol{\omega}_{\mathbf{1}}, \boldsymbol{\rho}_{\mathbf{1}}\right)\right|_{(0, 0)}\!\|_{\mathcal{A}_{\boldsymbol{\vartheta}}} < \iota 
    \end{equation} 
    \begin{equation}
    \label{eq:SolutionX1IC2}
    \left. \left( \boldsymbol{\varphi}^{\!-\!1\!} \!\left(m_1\boldsymbol{\Phi}\!\left(\mathbf{\hat{q}_1}, \!\mathbf{R_1}\!\right)\right), \boldsymbol{\omega}_{\mathbf{1}}, \boldsymbol{\rho}_{\mathbf{1}}\right)\right|_{(0, 0)} \in \mathcal{B}_{\boldsymbol{\vartheta}}
    \end{equation}
    \end{subequations}    
Moreover, given the latter results, since $\mathcal{A}_{\boldsymbol{\vartheta}}$ is stable and attractive from $\mathcal{B}_{\boldsymbol{\vartheta}}$ for $\mathcal{H}^\mathbf{*}_\mathbf{2}$, the equivalence described by \eqref{eq:SolutionsH1H2} also presupposes that, for every $\epsilon \in \mathbb{R}_{>0}$, there exists $\iota \in \mathbb{R}_{>0}$ verifying $\iota \leq \epsilon$ such that any solution  $\mathbf{x_1}$ to $\mathcal{H}_{\mathbf{1}}$ satisfying \eqref{eq:SolutionX1IC1} yields $\|\left.\left(\boldsymbol{\varphi}^{\!-\!1\!}\!\left(m_1\boldsymbol{\Phi}\!\left(\mathbf{\hat{q}_1}, \!\mathbf{R_1}\!\right)\right), \boldsymbol{\omega}_{\mathbf{1}}, \boldsymbol{\rho}_{\mathbf{1}}\right)\right|_{(t,j)}\!\|_{\mathcal{A}_{\boldsymbol{\vartheta}}} < \epsilon \; \forall \; (t,j) \in \mathrm{dom}\;\mathbf{x_1}$, and every solution $\mathbf{x_1}$ to $\mathcal{H}_{\mathbf{1}}$ satisfying \eqref{eq:SolutionX1IC2} leads to 
\begin{equation}
        \label{eq:LimitAttractionX1}
        \lim_{(t+j) \rightarrow \infty} \|\left.\left(\boldsymbol{\varphi}^{\!-\!1\!}\!\left(m_1\boldsymbol{\Phi}\!\left(\mathbf{\hat{q}_1}, \!\mathbf{R_1}\!\right)\right), \boldsymbol{\omega}_{\mathbf{1}}, \boldsymbol{\rho}_{\mathbf{1}}\right)\right|_{(t,j)}\|_{\mathcal{A}_{\boldsymbol{\vartheta}}} = 0
\end{equation}
The properties of the map $\mathcal{R}_{\boldsymbol{\vartheta}}: \mathbb{\bar{R}}^3 \mapsto \mathrm{SO}(3)$ demonstrated in \hyperref[section3]{section~\ref*{section3}}, namely, being a covering map and everywhere a local diffeomorphism, are instrumental inasmuch as these properties enable expressing a given open neighborhood of $\mathcal{A}$ in terms of an open neighborhood of $\mathcal{A}_{\boldsymbol{\vartheta}}$. In this direction, let $\mathcal{U}^{\boldsymbol{\epsilon}} \supset \mathcal{A}$ and $\mathcal{U}^{\boldsymbol{\iota}} \supset \mathcal{A}$ denote open neighborhoods of $\mathcal{A}$ defined as
\begin{equation*}
\begin{aligned}
    \mathcal{U}^{\boldsymbol{\epsilon}} & \!=\! \{  \mathbf{x_1} \!\in\!  \boldsymbol{\chi}_\mathbf{1}: \! \| \!\left.\left(\boldsymbol{\varphi}^{\!-\!1\!}\!\left(m_1\boldsymbol{\Phi}\!\left(\mathbf{\hat{q}_1}, \!\mathbf{R_1}\!\right)\right), \boldsymbol{\omega}_{\mathbf{1}}, \boldsymbol{\rho}_{\mathbf{1}}\right)\right|_{(t,j)}\!\|_{\mathcal{A}_{\boldsymbol{\vartheta}}} \!<\! \epsilon, \\ & \|\boldsymbol{\varphi}^{\!-\!1\!}(m_1\boldsymbol{\Phi}(\mathbf{\hat{q}_1}, \!\mathbf{R_1}\!)\!)\| \!\leq\! 1 \!+\! \delta \!+\! \epsilon, \; 
\mathrm{dist}(\mathbf{\hat{q}_1}, \!\mathcal{Q}(\mathbf{R_1}\!) \!) \!\leq\! \alpha \!+\! \epsilon\},
\end{aligned}
\end{equation*}
\begin{equation*}
\begin{aligned}
    \mathcal{U}^{\boldsymbol{\iota}} & \!=\! \{ \mathbf{x_1} \!\in\!  \boldsymbol{\chi}_\mathbf{1}: \|\left.\left(\boldsymbol{\varphi}^{\!-\!1\!}\!\left(m_1\boldsymbol{\Phi}\!\left(\mathbf{\hat{q}_1}, \!\mathbf{R_1}\!\right)\right), \boldsymbol{\omega}_{\mathbf{1}}, \boldsymbol{\rho}_{\mathbf{1}}\right)\right|_{(t,j)}\!\|_{\mathcal{A}_{\boldsymbol{\vartheta}}} \!<\! \iota,  \\ &
 \|\boldsymbol{\varphi}^{\!-\!1\!}(m_1\boldsymbol{\Phi}(\mathbf{\hat{q}_1}, \!\mathbf{R_1}\!)\!)\| \! \leq \! 1 \!+\! \delta \!+\! \iota, \; 
\mathrm{dist}(\mathbf{\hat{q}_1}, \mathcal{Q}(\mathbf{R_1}\!) \!) \! \leq \! \alpha \!+\! \iota \}.
\end{aligned}
\end{equation*}
Note that, in virtue of the map $\boldsymbol{\Phi}$ being double-valued for $\mathrm{dist}(\mathbf{\hat{q}_1}, \!\mathcal{Q}(\mathbf{R_1})\!) = 1$, the open neighborhoods $\mathcal{U}^{\boldsymbol{\epsilon}}$ and $\mathcal{U}^{\boldsymbol{\iota}}$ are defined for $\iota \leq \epsilon < 1 - \alpha$. Bearing in mind \cite[Theorem 7]{mayhew2013} and \hyperref[lemH*]{Lemma~\ref*{lemH*}}, the initial condition $\mathbf{x_1}(0, 0) \in \mathcal{U}^{\boldsymbol{\iota}}$ results in $\left.\|\boldsymbol{\varphi}^{\!-\!1\!}(m_1\boldsymbol{\Phi}(\mathbf{\hat{q}_1}, \!\mathbf{R_1}\!)\!)\right|_{(t,j)}\| \leq 1 + \delta + \epsilon$ and $ \left.\mathrm{dist}(\mathbf{\hat{q}_1}, \!\mathcal{Q}(\mathbf{R_1})\!)\right|_{(t,j)} \leq \alpha + \epsilon$ for all $(t,j) \in \mathrm{dom}\;\mathbf{x_1}$. Hence, given the foregoing results, for every $\epsilon < 1 - \alpha$, there exists there exists $\iota \leq \epsilon$ such that any solution $\mathbf{x_1}$ to $\mathcal{H}_{\mathbf{1}}$ satisfying $\mathbf{x_1}(0, 0) \in \mathcal{U}^{\boldsymbol{\iota}}$ yields $\mathbf{x_1}(t,j) \in \mathcal{U}^{\boldsymbol{\epsilon}} \;\forall\; (t,j) \in \mathrm{dom}\;\mathbf{x_1}.$ Thus, the compact set $\mathcal{A}$ is stable for $\mathcal{H}_{\mathbf{1}}$. Furthermore, under the same reasoning and with particular emphasis on \eqref{eq:LimitAttractionX1}, it follows that $\mathbf{x_1}(0, 0) \in \mathcal{B}$ leads to $\lim_{(t+j) \rightarrow \infty} \|\mathbf{x_1}(t,j)\|_{\mathcal{A}} = 0 $. Therefore, the compact set $\mathcal{A}$ is asymptotically stable for $\mathcal{H}_{\mathbf{1}}$ with $\mathcal{B}$ as basin of attraction whenever the compact set $\mathcal{A}_{\boldsymbol{\vartheta}}$ is asymptotically stable for $\mathcal{H}_{\mathbf{2}}$ with $\mathcal{B}_{\boldsymbol{\vartheta}}$ as basin of attraction.
\par To demonstrate the converse, suppose now that the set $\mathcal{A}$ is asymptotically stable for $\mathcal{H}_{\mathbf{1}}$ with $\mathcal{B}$ as basin of attraction. Then,  since for $\mathbf{x_1}(0, 0) \in \mathcal{B} \cup \mathcal{U}^{\boldsymbol{\iota}}$ one has $\mathrm{dist}(\mathbf{\hat{q}_1}, \!\mathcal{Q}(\mathbf{R_1})\!)|_{(0, 0)} < 1$, in virtue of \hyperref[lemSolutionsEquivalence]{Lemma~\ref*{lemSolutionsEquivalence}}, for every solution $\mathbf{x_1}$ to $\mathcal{H}_\mathbf{1}$, there exists a solution $\mathbf{x_2}$ to $\mathcal{H}_\mathbf{2}$ that, for every $(t,j) \in \mathrm{dom}\; \mathbf{x_1}$, \eqref{eq:SolutionsH1H2} holds with $(t,j') \in \mathrm{dom}\;\mathbf{x_2}$ and $j' \in \mathbb{N}$ satisfying $j' \leq j$. In this way, $\mathbf{x_1}(0, 0) \in \mathcal{B}$ implies $\mathbf{x_2}(0, 0) \in \mathcal{B}_{\boldsymbol{\vartheta}}$ and, since $\mathcal{A}$ is attractive from $\mathcal{B}$, the equality \eqref{eq:SolutionsH1H2} allows concluding that $\lim_{(t+j') \rightarrow \infty} \|\mathbf{x_2}(t,j')\|_{\mathcal{A}_{\boldsymbol{\vartheta}}} = 0$. Similarly, stemming from \eqref{eq:SolutionsH1H2}, given that $\mathcal{A}$ is stable, for any given $\epsilon$ satisfying $\epsilon < 1-\alpha$, there exists $\iota$, verifying $\iota \leq \epsilon$, such that $\mathbf{x_1}(0, 0) \in \mathcal{U}^{\boldsymbol{\iota}}$ infers $\|\mathbf{x_2}(0, 0)\|_{\mathcal{A_{\vartheta}}} < \iota$ and $\|\mathbf{x_2}(t,j')\|_{\mathcal{A_{\vartheta}}} < \epsilon \; \forall (t,j') \in \mathrm{dom}\; \mathbf{x_2}$. Consequently, under the assumption that the compact set $\mathcal{A}$ is asymptotically stable for $\mathcal{H}_{\mathbf{1}}$ with $\mathcal{B}$ as basin of attraction, the compact set $\mathcal{A}_{\boldsymbol{\vartheta}}$ is asymptotically stable for the hybrid system $\mathcal{H}_{\mathbf{2}}$ with $\mathcal{B}_{\boldsymbol{\vartheta}}$ as basin of attraction.

\section{Proof of Theorem 2}
\label{appendix:ProofTheorem2}

\begin{figure}[!htb]
\centering
\begin{tikzpicture}
  \clip (-4,-1) rectangle ({4.5},{4.5});  
  \def\r{3} 
  \def\q{4} 
  \def\x{{\r*1/2}} 
  \def\y{{\r*sqrt(3)/2}} 
  
  \coordinate (O) at (0, 0); 
  \coordinate (A) at (0,\r); 
  \coordinate (B) at (0,-\r); 
  \coordinate (C) at (\r,0); 
  \coordinate (D) at (-\r,0); 
  \coordinate (E) at (0, \y);
  \coordinate (F) at ({1.25*\r*(sqrt(6)+sqrt(2))/4}, {-1.25*\r*(sqrt(6)-sqrt(2))/4}); 
  \coordinate (G) at ({-1.25*\r*sqrt(2)/2}, {1.25*\r*sqrt(2)/2});
  \coordinate (H) at ({1.25*\r*1/2},{1.25*\r*sqrt(3)/2});
  \coordinate (I) at (0, 0.79335*\r);
  \coordinate (Qm) at (\x,\y); 

  \coordinate (1) at (0,0.5*\r);
  \coordinate (2) at ({0.5*\r*1/2},{0.5*\r*sqrt(3)/2});
  \coordinate (3) at ({0.5*\r*(sqrt(6)+sqrt(2))/4}, {-0.5*\r*(sqrt(6)-sqrt(2))/4});
  \coordinate (4) at ({-0.6*\r*sqrt(2)/2}, {0.6*\r*sqrt(2)/2});

  \filldraw[fill=teal!20,draw=black, dashed] (0, 0) -- (3)
    arc [start angle=-15, end angle=60, radius=\r*0.5];
  \node[teal] at ({0.75*0.5*\r}, {0.3*0.5*\r}) {\small$\theta'$};
  \filldraw[fill=orange!20,draw=black, dashed] (0, 0) -- (1)
    arc [start angle=90, end angle=135, radius=\r*0.5];
  \node[orange] at ({-0.3*0.5*\r}, {0.71*0.5*\r}) {\small$\theta' \!\!\! - \! \theta^*$};
  \filldraw[fill=blue!20,draw=black, dashed] (0, 0) -- (2)
    arc [start angle=60, end angle=90, radius=\r/2];
  \node[blue] at ({0.25*0.5*\r}, {0.80*0.5*\r}) {\small$\theta^*$};

  \draw[->, thick] (0,-1.15*\r) -- (0,1.3*\r) node[right] {$q_0$};
  \draw[->, thick] (-1.15*\r,0) -- (1.3*\r,0) node[below] {$\mathbf{e_1^{\!\top}}\!\mathbf{q_1}$};
  \draw[dashed] (Qm) -- (E);
  \draw[dashed] (-0.60876*\r,0.79335*\r) -- (0.60876*\r,0.79335*\r);  
  \draw[dashed, thick] (0, 0) -- (H);
  \draw[dashed, thick] (0, 0) -- (F);
  \draw[dashed, thick] (0, 0) -- (G);
  \draw[blue!60,thick] (O) circle(\r);
  \draw [red!60,very thick,domain=90-37.5:90+37.5] plot ({\r*cos(\x)}, {\r*sin(\x)});
  \node[red!60] at ({-1}, {1.05*\r}) {\small$\mathcal{U}_{\alpha_1}$};
  \draw[red!60, very thick] ({-0.60876*(\r-0.1)},{0.79335*(\r-0.1)}) -- ({-0.60876*(\r+0.1)},{0.79335*(\r+0.1)});  
  \draw[red!60, very thick] ({0.60876*(\r-0.1)},{0.79335*(\r-0.1)}) -- ({0.60876*(\r+0.1)},{0.79335*(\r+0.1)});

  \fill[black] (A) circle(0.025) node[above left] {1};
  \fill[black] (B) circle(0.025) node[below left] {-1};
  \fill[black] (C) circle(0.025) node[below right] {1};
  \fill[black] (D) circle(0.025) node[below left] {-1};
  \fill[black] (E) circle(0.05) node[left] {\footnotesize $\cos(\!\theta^*\!)$\normalsize};
  \fill[black] (I) circle(0.05) node[below left] {\footnotesize $\alpha_1$\normalsize};  
  \fill[blue] (Qm) circle(0.05) node[right] {\small $\; \mathbf{\hat{q}_1}$\normalsize};
  \fill[blue] (A) circle(0.05) node[above right] {\small $\boldsymbol{\Phi}(\!\mathbf{\hat{q}_1}, \!\mathbf{I_3}\!)$ \normalsize};
  
\end{tikzpicture}
\caption{Scheme of the approach used to determine $T'$. For ease of comprehension, this visual representation was obtained considering $\mathbf{e_2^{\!\top}}\mathbf{q_1} = \mathbf{e_3^{\!\top}}\mathbf{q_1} = 0$. The variable $\alpha$ and the memory state $\mathbf{\hat{q}_1}$ were set with the values $1 - \cos(75^\circ)$ and $(\cos(30^\circ), \sin(30^\circ), 0, 0)$, respectively.}
\label{fig:ApproachT'}
\end{figure}
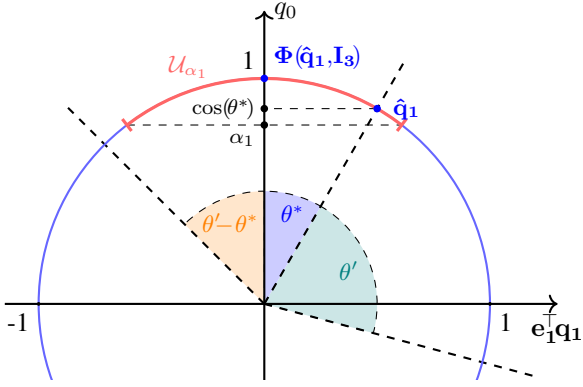

    In the same spirit of the proof of \hyperref[thm:EquivalenceAsymptoticStability]{Theorem~\ref*{thm:EquivalenceAsymptoticStability}}, the demonstration hinges on the parallelism presented in \hyperref[lemSolutionsEquivalence]{Lemma~\ref*{lemSolutionsEquivalence}} and on the notion of exponential stability for hybrid systems. To the end of showing the first part of the theorem, one starts from the premise that the $\mathcal{A}_{\boldsymbol{\vartheta}}$ is exponentially stable for $\mathcal{H}_{\mathbf{2}}$. In this way, based on the definition provided in \cite{teel2012}, there exist $\mu, \lambda, \kappa \in \mathbb{R}_{>0}$ such that each solution $\mathbf{x_2}$ satisfying $\mathbf{x_2}(0, 0) \in \mathcal{B}_\vartheta $ also satisfies
\begin{equation}
    \label{eq:ExponentialBoundH2}
    \|\mathbf{x_2}(t,j)\|_{\mathcal{A}_{\vartheta}} \leq \kappa e^{-\lambda (t + j)} \|\mathbf{x_2}(0, 0)\|_{\mathcal{A}_{\vartheta}} \;\; \forall \;\; (t,j) \in \textrm{dom } \mathbf{x_2}
\end{equation}
From \hyperref[lemSolutionsEquivalence]{Lemma~\ref*{lemSolutionsEquivalence}}, for every solution $\mathbf{x_2}$ to $\mathcal{H}_{\mathbf{2}}$, there exists a solution $\mathbf{x_1}$ to $\mathcal{H}_{\mathbf{1}}$ that, for every $(t,j') \in \mathrm{dom}\;\mathbf{x_2}$, yields the equality \eqref{eq:SolutionsH1H2} with $(t,j) \in \mathrm{dom}\;\mathbf{x_1}$ and $j \geq j'$. In this direction, by defining the vector
\begin{equation*}
    \mathbf{z_1}\!(t,j) = \left.( \boldsymbol{\varphi}^{\!-\!1\!} \! \left(m_1\boldsymbol{\Phi}\!\left(\mathbf{\hat{q}_1}, \!\mathbf{R_1}\!\right)\right), \boldsymbol{\omega}_{\mathbf{1}}, \boldsymbol{\rho}_{\mathbf{1}})\right|_{(t,j)},
\end{equation*}
the initial condition $\mathbf{x_2}(0, 0) \in \mathcal{B}_\vartheta $ leads to 
    \begin{equation} 
    \label{eq:SolutionX1ICExponential}
    \|\mathbf{z_1}\!(0, 0)\|_{\mathcal{A}_{\boldsymbol{\vartheta}}} < \mu 
    \end{equation}
In contrast to the proof of \hyperref[thm:EquivalenceAsymptoticStability]{Theorem~\ref*{thm:EquivalenceAsymptoticStability}}, in which equivalences regarding the stability bounds were directly established, for the case of exponential stability, the jumps of $\mathcal{H}_{\mathbf{1}}$ due to the update of the memory state $\mathbf{\hat{q}_1}$ hinder the immediate derivation of an exponential bound for $\|\mathbf{z_1}\!(t,j)\|$ from \eqref{eq:ExponentialBoundH2}. To isolate the hybrid time instants immediately after the jumps triggered due to the update of the quaternion memory state $\mathbf{\hat{q}_1}$, consider the set
\begin{equation}
\begin{aligned}
     & \mathcal{I}_q = \{(t,j)  \in \mathrm{dom}\;\mathbf{x_1} \!:\! (t, j-1) \in \mathrm{dom}\;\mathbf{x_1}, \\ & \mathbf{\hat{q}_1}(t, j) \in \left.\boldsymbol{\Phi}(\mathbf{\hat{q}_1}, \!\mathbf{R_1})\right|_{(t,j\!-\!1)}, \; m(t, j\!-\!1) = m(t,j)\},
\end{aligned}
\end{equation}
Without loss of generality, assume that $\mathbf{x_1}(t,j)$ verifies $\mathrm{dist}(\mathbf{\hat{q}_1}, \!\mathcal{Q}(\mathbf{R_1})\!)|_{(0, 0)} < 1$. Then, by virtue of \cite[Lemma 3]{mayhew2013}, for any $\mathbf{\hat{q}_1} \in \mathbb{S}^3$ and $\mathbf{R_1} \in \mathrm{SO}(3)$, the equality $\boldsymbol{\Phi}\!\left(\mathbf{\hat{q}_1}, \mathbf{R_1}\right) = \boldsymbol{\Phi}\!\left(\boldsymbol{\Phi}(\mathbf{\hat{q}_1}, \!\mathbf{R_1}), \mathbf{R_1}\right)$ holds. Thus, the vector $\mathbf{z_1}\!(t,j)$ remains constant for $(t,j) \in \mathcal{I}_q$ and, bearing in mind \eqref{eq:SolutionsH1H2} and \eqref{eq:SolutionX1ICExponential}, verifies an exponential bound related to \eqref{eq:ExponentialBoundH2} for every $(t,j) \in \textrm{dom } \! \mathbf{x_1} \setminus \mathcal{I}_q$. With this in mind, for all $(t,j) \in \textrm{dom } \! \mathbf{x_1}$, the following bound is derived 
\begin{equation}
    \label{eq:FirstExponentialBoundZ1}
    \|\mathbf{z_1}\!(t,j)\|_{\mathcal{A}_\vartheta} \!\leq\!  \kappa e^{-\lambda (t+j-j^q(t,j))}\|\mathbf{z_1}\!(0, 0)\|_{\mathcal{A}_\vartheta}
\end{equation}
where $j^q(t,j) = |\left\{ (t^*,j^*) \in \mathcal{I}_q: (t^*,j^*) \preceq (t,j) \right\}|$ serves the purpose of cancelling the jump times $j$ caused by the update of $\mathbf{\hat{q}_1}$. Bear in mind that if $\mathcal{I}_q = \emptyset$, which corresponds to the case in which no jumps occur due to the update of $\mathbf{\hat{q}_1}$, one can directly infer that each solution $\mathbf{z_1}$ satisfying \eqref{eq:FirstExponentialBoundZ1} also satisfies
\begin{equation}
    \label{eq:FirstExponentialBoundZ1EmptySet}
    \|\mathbf{z_1}\!(t,j)\|_{\mathcal{A}_\vartheta} \!\leq\!  \kappa e^{\!-\!\lambda (t+j)} \! \|\mathbf{z_1}\!(0, \! 0)\|_{\mathcal{A}_\vartheta} \; \forall \; (t,j) \in \textrm{dom } \! \mathbf{x_1}
\end{equation}
Otherwise, given that the term $e^{-\lambda j^q(t,j)}$ approaches $0$ as the number of these updates tends to infinity, to derive the exponential bound for $\mathbf{z_1}\!(t,j)$, one has to demonstrate that the number of such jumps is finite, i.e, $\lim_{t+j \rightarrow \infty} j^q(t,j)$ is bounded. To this end, the sublevel sets of the function $\left.\mathrm{dist}(\mathbf{\hat{q}_1}, \mathcal{Q}(\mathbf{R_1}\!) \!)\right|_{(t,j)}$ are analyzed. Since the double-valued map $\mathcal{Q}$ yields $\mathcal{Q}(\mathbf{R_1}) = \left\{-\mathbf{p_1}, \mathbf{p_1}\right\}$, one has $\mathrm{dist}(\mathbf{\hat{q}_1}, \!\mathcal{Q}(\mathbf{R_1})\!) = 1 - |\mathbf{\hat{q}_1}^{\!\top}\mathbf{p_1}| = 1 - \cos{\theta}$, with $\theta \in [0, \pi/2) $ denoting the angle between $\mathbf{\hat{q}_1}$ and $\boldsymbol{\Phi}\!\left(\mathbf{\hat{q}_1}, \mathbf{R_1}\right)$. Bearing in mind \eqref{eq:InverseSouthPole}, the scalar component of $\left.\boldsymbol{\Phi}\!\left(\mathbf{\hat{q}_1}, \mathbf{R_1}\right)\right|_{(t,j)}$ satisfies 
\begin{equation*}
    \label{eq:scalarq1}
    |[\mathbf{e_1^{\!\top}} 0] \!\left.\boldsymbol{\Phi}\!\left(\mathbf{\hat{q}_1}, \mathbf{R_1}\right)\right|_{(t,j)}\!| \!=\! \frac{1 \!-\! \| \! \left.\boldsymbol{\varphi}^{\!-\!1\!} \! \left(m_1\boldsymbol{\Phi}\!\left(\mathbf{\hat{q}_1}, \!\mathbf{R_1}\!\right)\right)\right|_{(t,j)}\!\|^2}{1 \! +\! \| \! \left.\boldsymbol{\varphi}^{\!-\!1\!} \! \left(m_1\boldsymbol{\Phi}\!\left(\mathbf{\hat{q}_1}, \!\mathbf{R_1}\!\right)\right)\right|_{(t,j)}\!\|^2}
\end{equation*}
Then, by virtue of \eqref{eq:FirstExponentialBoundZ1}, for all $(t,j) \in \textrm{dom } \! \mathbf{x_1}$, one has
\begin{equation}
\label{eq:scalarq1bound}
    |[\mathbf{e_1^{\!\top}} 0] \!\left.\boldsymbol{\Phi}\!\left(\mathbf{\hat{q}_1}, \!\mathbf{R_1}\!\right)\right|_{(t,j)}\!| \!\geq\! \frac{1 \!-\! \kappa^2 e^{-2\lambda (t+j-j^q(t,j))}\|\mathbf{z_1}\!(0, 0)\|^2_{\mathcal{A}_\vartheta}}{1 \!+\! \kappa^2 e^{-2\lambda (t+j-j^q(t,j))}\|\mathbf{z_1}\!(0, 0)\|^2_{\mathcal{A}_\vartheta}}
\end{equation}
In this way, there is a finite hybrid time instant $(\bar{t}, \bar{j})$ such that $\mathbf{x_1}(t,j)$ ultimately enters the sublevel set 
\begin{equation*}
    \mathcal{U}_{\alpha_0} = \{ \mathbf{x_1} \in \boldsymbol{\chi}_{\mathbf{1}}: |[\mathbf{e_1^{\!\top}} 0] \left.\boldsymbol{\Phi}\!\left(\mathbf{\hat{q}_1}, \mathbf{R_1}\right)\right|_{(t,j)}| \geq \alpha_0\},
\end{equation*}
with $\alpha_0 \in [0,1]$, and will never leave it again, i.e., $\mathbf{x_1}(t,j) \in \mathcal{U}_{\alpha_0} \; \forall \; (t,j) \succeq (\bar{t}, \bar{j})$. With the understanding that the equality $\left.\boldsymbol{\varphi}^{\!-\!1\!} \! \left(m_1\boldsymbol{\Phi}\!\left(\mathbf{\hat{q}_1}, \!\mathbf{R_1}\!\right)\right)\right|_{(t,j)} = \boldsymbol{0}$ implies $\left.\boldsymbol{\Phi}\!\left(\mathbf{\hat{q}_1}, \mathbf{R_1}\right)\right|_{(t,j)} = (\pm 1,\boldsymbol{0})$ and, given \eqref{eq:SolutionsH1H2} and \eqref{eq:rotationmatrixMRP}, $\mathbf{R_1} = \mathbf{I_3}$, consider the function $\mathrm{dist}(\mathbf{\hat{q}_1}, \mathcal{Q}(\mathbf{I_3})) = 1 - |[\mathbf{e_1^{\!\top}} 0]\mathbf{\hat{q}_1}|$. Then, it follows that $\cos(\theta^*) = |[\mathbf{e_1^{\!\top}} 0]\mathbf{\hat{q}_1}|$, with $\theta^* \in [0, \pi/2] $ denoting the angle between $\mathbf{\hat{q}_1}$ and $\boldsymbol{\Phi}\!\left(\mathbf{\hat{q}_1}, \mathbf{I_3}\right)$. Since the update of $\mathbf{\hat{q}_1}$ is given by $\mathbf{\hat{q}_1}(t, j) \in \left.\boldsymbol{\Phi}(\mathbf{\hat{q}_1}, \mathcal{Q}(\mathbf{R_1})))\right|_{(t, j-1)}$ for $(t,j) \in \mathcal{I}_q$, the function $|[\mathbf{e_1^{\!\top}} 0]\mathbf{\hat{q}_1}(t,j)| = \cos(\theta^*(t,j))$ is bounded by \eqref{eq:scalarq1bound}
for all $(t,j) \in \mathcal{I}_q$ and constant for all $(t,j) \in \textrm{dom } \! \mathbf{x_1} \setminus \mathcal{I}_q$. In this way, for the case in which $\mathcal{I}_q \neq \emptyset$, consider the hybrid time instant $(\hat{t},\hat{j})$ satisfying  
\begin{equation*}
\begin{aligned}
        (\hat{t},\hat{j}) \!\in\! \textrm{dom } \! \mathbf{x_1}\!: & \mathbf{\hat{q}_1}(t, j) \in \left.\boldsymbol{\Phi}(\mathbf{\hat{q}_1}, \!\mathbf{R_1})\right|_{(t,j\!-\!1)}, \\ & \theta' \!>\! 2\theta^*(t,j),  \mathbf{x_1}(t,j) \!\in \mathcal{U}_{\alpha_1} \; \forall \; (t,j) \succeq (\hat{t},\hat{j}),
\end{aligned}
\end{equation*}
with $\theta' = \cos^{-1}(1-\alpha)$ and $\alpha_1 = \cos(\theta'/2)$. The condition $\theta' > 2\theta^*(t',j')$ leads to $ |[\mathbf{e_1^{\!\top}} 0]\mathbf{\hat{q}_1}(\hat{t},\hat{j})| > \alpha_1$. Then, since $\mathbf{x_1}(t,j) \in \mathcal{U}_{\alpha_1} \; \forall \; (t,j) \succeq (\hat{t},\hat{j})$ results in $\mathbf{x_1}(t,j) \in \mathbf{C_1}(\mathbf{x_1}) \; \forall \; (t,j) \succeq (\hat{t},\hat{j})$, it follows that $(\hat{t},\hat{j})$ corresponds to the instant in which the last jump triggered to update the memory state $\mathbf{\hat{q}_1}$ occurs. The approach described and the quantities considered are schematically represented in \hyperref[fig:ApproachT']{Figure~\ref*{fig:ApproachT'}}. Bearing in mind \eqref{eq:FirstExponentialBoundZ1}, $(\hat{t},\hat{j})$ is a finite hybrid time. Therefore, one concludes that there exists $M_j \in \mathbb{N}$ such that $\lim_{t+j \rightarrow \infty} j^q(t,j) \leq M_j$. To quantify $M_j$, first, by letting $T(t,j) = t + j - j^q(t,j)$, one determines the hybrid time $T' \in \mathbb{R}_{\geq 0}$ that ensures $|[\mathbf{e_1^{\!\top}} 0] \!\left.\boldsymbol{\Phi}\!\left(\mathbf{\hat{q}_1}, \!\mathbf{R_1}\!\right)\right|_{(t,j)}\!| > \alpha_1 \;\forall\; T(t,j) > T'$. In this direction, using the bound \eqref{eq:scalarq1bound}, it follows that
\begin{equation}
    T' = \frac{1}{2\lambda} \max\left\{0, \ln\left(\frac{(1+\alpha_1)\kappa^2\|\mathbf{z_1}\!(0, 0)\|^2_{\mathcal{A}_\vartheta}}{1-\alpha_1}\right)\right\} 
\end{equation}

Note that \eqref{eq:FirstExponentialBoundZ1}, for all $(t,j) \in \textrm{dom } \! \mathbf{x_1}$, also enables writing 
\begin{equation}
    \label{eq:BoundAngularVelocity}
    \|\boldsymbol{\omega}_\mathbf{1}(t,j)\|_{\mathcal{A}_\vartheta} \leq  \kappa e^{- \lambda T(t,j)}\|\mathbf{z_1}\!(0, 0)\|_{\mathcal{A}_\vartheta}
\end{equation}
\noindent The scenario in which the number of jumps due to the update of $\mathbf{\hat{q}}$ is maximum occurs when 
\begin{equation}
    \left.\mathrm{dist}(\mathbf{\hat{q}_1}, \!\mathcal{Q}(\mathbf{R_1})\!)\right|_{(0, 0)} \geq \alpha. 
\end{equation}
This initial condition triggers the update condition, leading to
\begin{equation}
        \left.\mathrm{dist}(\mathbf{\hat{q}_1}, \!\mathcal{Q}(\mathbf{R_1})\!)\right|_{(0,1)} = 0.
\end{equation}
Let $(t^u_1, j^u_1) = (0,1)$ and, for $i = 2, ..., M_j$, consider $(t^u_i, j^u_i) = \min\{ (t,j) \in \mathcal{I}_q: (t,j) \succeq (t^u_{i-1}, j^u_{i-1}) \}$.  From the proof of \cite[Theorem 7]{mayhew2013}, assuming that $\mathcal{I}^u_i = \{(t,j) \in \mathrm{dom } \; \mathbf{x_1}: (t^u_{i-1}, j^u_{i-1}) \preceq (t,j) \preceq (t^u_i, j^u_i-1) \}$ has a nonempty interior for $i = 2,...,M_j$, the following inequality yields for all $(t,j) \in \mathcal{I}^u_i$:  
\begin{equation*}
    \frac{d}{dt} \left( \mathrm{dist}(\mathbf{\hat{q}_1}, \!\mathcal{Q}(\mathbf{R_1})\!) \right) \leq \frac{1}{2} \|\boldsymbol{\omega}_{\mathbf{1}}(t,j)\| 
\end{equation*}
Then, in light of \eqref{eq:BoundAngularVelocity}, for $i = 2,...,M_j$, it follows that 
\begin{equation}
    \label{eq:ExpBoundDerivativeDist}
    \frac{d}{dt} \!\left( \mathrm{dist}(\mathbf{\hat{q}_1}, \!\mathcal{Q}(\mathbf{R_1})\!) \right) \!\leq\! \frac{1}{2}\kappa e^{\!-\lambda T(t,j)}\|\mathbf{z_1}\!(0, 0)\|_{\mathcal{A}_\vartheta} \; \forall \; (t,j) \!\in\! \mathcal{I}^u_i
\end{equation}
Let $T_1 = t^u_{1} + j^u_{1} - j^q(t^u_{1},j^u_{1}) = 0$. With this definition and recalling that 
\begin{equation}
    \left.\mathrm{dist}(\mathbf{\hat{q}_1}, \!\mathcal{Q}(\mathbf{R_1})\!)\right|_{(t,j+1)} = 0 \;\; \forall \;\; (t,j) \in \mathcal{I}_q,
\end{equation}
for $i = 2,...,M_j$, the latter exponential bound leads to
\begin{equation*}
    \label{eq:DistFunctionBound}
    \left.\mathrm{dist}(\mathbf{\hat{q}_1}, \!\mathcal{Q}(\mathbf{R_1}\!)\!)\right|_{(t,j)} \!\!\leq\! \frac{\kappa}{2\lambda} \! \|\mathbf{z_1}\!(0, \! 0) \! \|_{\mathcal{A}_\vartheta} \! \! \left( \! e^{\!-\!\lambda T_{\!i\!-\!1}\!(t,j)} \!\!-\! e^{\!-\lambda T(t,j)} \right), 
\end{equation*}
for all $(t,j) \in \mathcal{I}^u_i$. Then, by resorting to the latter inequality  and the update jump condition, it is possible to determine $T_i \in \mathbb{R}_{>0}$, i.e, the minimum total accumulated time $T(t,j)$ until the $i^{th}$ update of $\mathbf{\hat{q}_1}$, for $i = 2,..., i^*$:
\begin{equation}
    T_i = -\frac{1}{\lambda}\ln\left(1 - \frac{2(i-1)\lambda\alpha}{\kappa \|\mathbf{z_1}\!(0, 0)\|_{\mathcal{A}_\vartheta}} \right) \;\; \textrm{for } \;\; i = 2,..., i^*,
\end{equation}
where $\ln(s):\mathbb{R}_{>0} \mapsto \mathbb{R}$ denotes the natural logarithm function and $i^* = \{i \in \mathbb{Z}_{>0}: 2\lambda \alpha (i-1) > \kappa \|\mathbf{z_1}\!(0, 0)\|_{\mathcal{A}_\vartheta}\}$ ensures that the condition $e^{-\lambda T_{i-1}(t,j)} > \alpha $ holds for $i = 2, ..., i^*$, meaning that, for $(t,j) \in \mathcal{I}^u_i$, the function $\left.\mathrm{dist}(\mathbf{\hat{q}_1}, \!\mathcal{Q}(\mathbf{R_1})\!)\right|_{(t,j)}$ can potentially trigger the update condition. In this way, the maximum number of jumps due to the update of $\mathbf{\hat{q}_1}$, $M_j$, is given by
\begin{equation}
    M_j \!=\! \left\{ \!\!\!\!\begin{array}{cl}
         \min\{i \in \mathbb{Z}_{>0}\!: i \leq i^*, T_i \geq T' \} \!\!\!\! &, \;\; \textrm{if} \;\; T_{i^*} \!>\! T'\\
        i^* \!\!\!\! &, \;\; \textrm{if} \;\; T_{i^*} \!\leq\! T'
    \end{array}\right. 
\end{equation}
With this result in place, for $\mathcal{I}_q \neq \emptyset$ and each solution $\mathbf{z_1}$ satisfying \eqref{eq:FirstExponentialBoundZ1}, the exponential bound \eqref{eq:FirstExponentialBoundZ1} enables writing:
\begin{equation}
    \label{eq:SecondExponentialBoundZ1}
    \|\mathbf{z_1}\!(t,j) \! \|_{\mathcal{A}_\vartheta} \!\leq\!  \kappa^* \!e^{\!-\lambda (t+j)} \! \|\mathbf{z_1}\!(0, 0)\|_{\mathcal{A}_\vartheta} \; \forall \; (t,j) \!\in\! \textrm{dom } \! \mathbf{x_1}
\end{equation}
with $\kappa^* = \kappa e^{\lambda M_j}$. Note that the exponential bound for the case in which $\mathcal{I}_q = \emptyset$ can be obtained from \eqref{eq:SecondExponentialBoundZ1} by considering $M_j = 0$. 
\par To measure the error between the two configurations in $\mathrm{SO}(3)$, $\mathbf{R_1}$ and $\mathbf{I_3}$, consider the following Frobenius norm:
\begin{equation*}
    \|\mathbf{R_1} - \mathbf{I}_3\|_F = \sqrt{2\Tr(\mathbf{I_3}-\mathbf{R_1})}.
\end{equation*}
\noindent Recalling \eqref{eq:rotationmatrixMRP}, one has 
\begin{equation}
    \label{eq:traceRotError}
    \Tr(\mathbf{I_3} \! - \!\mathbf{R_1}(t,j)) = \frac{16\|\left.\boldsymbol{\varphi}^{\!-\!1\!} \! \left(m_1\boldsymbol{\Phi}\!\left(\mathbf{\hat{q}_1}, \!\mathbf{R_1}\!\right)\right)\right|_{(t,j)}\|^2}{(1 + \|\left.\boldsymbol{\varphi}^{\!-\!1\!} \! \left(m_1\boldsymbol{\Phi}\!\left(\mathbf{\hat{q}_1}, \!\mathbf{R_1}\!\right)\right)\right|_{(t,j)}\!\|^2)^2}.
\end{equation}
Then, 
\begin{equation}
    \label{eq:RotationMatrixFrobeniusNorm}
    \|\mathbf{R_1}(t,j) - \mathbf{I_3}\|_F = \frac{4\sqrt{2}\|\left.\boldsymbol{\varphi}^{\!-\!1\!} \! \left(m_1\boldsymbol{\Phi}\!\left(\mathbf{\hat{q}_1}, \!\mathbf{R_1}\!\right)\right)\right|_{(t,j)}\|}{1 + \|\left.\boldsymbol{\varphi}^{\!-\!1\!} \! \left(m_1\boldsymbol{\Phi}\!\left(\mathbf{\hat{q}_1}, \!\mathbf{R_1}\!\right)\right)\right|_{(t,j)}\!\|^2}
\end{equation}
which leads to
\begin{equation}
        \label{eq:UpperBoundFrobeniusNorm}
        \|\mathbf{R_1}(t,j) - \mathbf{I_3}\|_F \leq 4\sqrt{2} \|\left.\boldsymbol{\varphi}^{\!-\!1\!} \! \left(m_1\boldsymbol{\Phi}\!\left(\mathbf{\hat{q}_1}, \!\mathbf{R_1}\!\right)\right)\right|_{(t,j)}\| 
\end{equation}
for all $(t,j) \in \textrm{dom } \! \mathbf{x_1}$. Since $\mathbf{x_1} \in \mathcal{A}_\vartheta$ implies $\boldsymbol{\varphi}^{\!-\!1\!}(m_1\boldsymbol{\Phi}(\mathbf{\hat{q}_1}, \!\mathbf{R_1}\!)\!) = \boldsymbol{0}$, which, in turn, in view of \eqref{eq:rotationmatrixMRP}, results in $\mathbf{R_1} = \mathbf{I_3}$, by combining the latter bound with \eqref{eq:SecondExponentialBoundZ1} one concludes that each solution $\mathbf{x_1}$ to $\mathcal{H}$ verifying \eqref{eq:SolutionX1ICExponential} also verifies
\begin{equation}
        \label{eq:ExpBoundz1*}
        \|\mathbf{z_1^*}\!(t,j\!)\!\|_{\mathcal{A}_R} \!\!\leq\! \kappa^*_{\!R} e^{\!-\!\lambda (t+j)} \! \|\mathbf{z_1^*}\!(0, 0)\!\|_{\mathcal{A}_R} \; \forall \; (t,j) \in \textrm{dom } \! \mathbf{x_1}
\end{equation}
where $\mathbf{z_1^*}(t,j) = \left.(\|\mathbf{R_1} - \mathbf{I_3}\|_F, \boldsymbol{\omega}_{\mathbf{1}}, \boldsymbol{\rho}_{\mathbf{1}})\right|_{(t,j)} \in \mathbb{R}_{\geq 0} \times \mathbb{R}^3 \times \boldsymbol{\Lambda}$ and \begin{equation*}
    \kappa^*_{\!R} = 4\sqrt{2}\kappa^*\|\mathbf{z_1(0, 0)}\|_{\mathcal{A}_\vartheta}\|\mathbf{z_1^*}(0, 0)\|^{-1}_{\mathcal{A}_R}
\end{equation*}
\par Now, let $\mathcal{A}_{\hat{q}} = \{\mathbf{x_1} \in \boldsymbol{\chi}_{\mathbf{1}}: \mathrm{dist}(\mathbf{\hat{q}_1}, \mathcal{Q}(\mathbf{R_1}\!) \!) \leq \alpha\}$ and $\mathcal{A}_{m} = \{\mathbf{x_1} \in \boldsymbol{\chi}_{\mathbf{1}}: \|\boldsymbol{\varphi}^{\!-\!1\!}(m_1\boldsymbol{\Phi}(\mathbf{\hat{q}_1}, \!\mathbf{R_1}\!)\!)\| \leq 1 + \delta\}$. Based on \cite[Theorem 7 - 4)]{mayhew2013} and the proof of \hyperref[lemH*]{Lemma~\ref*{lemH*}}, and since the formulation allows for either jump when both the conditions for the update of the discrete states $\hat{\mathbf{q}}$ and $m$ are concurrently met, for any solution $\mathbf{x_1}$ to $\mathcal{H}_\mathbf{1}$
\begin{equation}
\label{eq:ICMemoryStates}
        \{(t,j) \in \textrm{dom } \! \mathbf{x_1}: \mathbf{x_1} \notin \mathcal{A}_{\hat{q}} \cap \mathcal{A}_m\} \subset \{(0, 0)\}.
\end{equation}
Hence, for any solution $\mathbf{x_1}$ to $\mathcal{H}_\mathbf{1}$
\begin{equation*}
\begin{aligned}
        \{(t,j) \in \textrm{dom } \! \mathbf{x_1}: & \|\mathbf{x_1^*}(t,j)\|_{\mathcal{A}_{\hat{q}}} > 0 \; \cup \\ & \|\mathbf{x_1^*}(t,j)\|_{\mathcal{A}_{m}} > 0\}  \subset \{(0, 0)\},
\end{aligned}
\end{equation*}
with $\mathbf{x_1^*}(t,j) = \left.(\|\mathbf{R_1} - \mathbf{I_3}\|_F, \mathbf{\hat{q}_1}, m_1, \boldsymbol{\omega}_{\mathbf{1}}, \boldsymbol{\rho}_{\mathbf{1}})\right|_{(t,j)} \in \mathbb{R}_{\geq 0} \times \mathbb{S}^3 \times \{-1,1\} \times \mathbb{R}^3 \times \boldsymbol{\Lambda}$, and the set $\mathcal{A}_{\hat{q}} \cap \mathcal{A}_m$ is positively invariant. Furthermore, from these considerations, one can easily infer that, for any given initial condition $\mathbf{x_1}(0, 0)$ satisfying $\left.\mathrm{dist}(\mathbf{\hat{q}_1}, \!\mathcal{Q}(\mathbf{R_1})\!)\right|_{(0, 0)} < 1$, the solution $\mathbf{x_1}(t,j)$ to $\mathcal{H}_{\mathbf{1}}$, for all $(t,j) \in \textrm{dom } \! \mathbf{x_1}$, satisfies
\begin{equation}
    \label{eq:ExponentialBoundAqAm}
    \|\mathbf{x_1^*}(t,j)\|_{\mathcal{A}_{\hat{q}} \cap \mathcal{A}_m} \leq \kappa e^{-\lambda(t+j)} \|\mathbf{x_1^*}(0, 0)\|_{\mathcal{A}_{\hat{q}} \cap \mathcal{A}_m} 
\end{equation}
\par By relying again on the properties of the map $\mathcal{R}_{\boldsymbol{\vartheta}}: \mathbb{\bar{R}}^3 \mapsto \mathrm{SO}(3)$ described in \hyperref[section3]{section~\ref*{section3}}, let $\mathcal{U}^{\boldsymbol{\mu}} \supset \mathcal{A}$ denote an open neighborhood of $\mathcal{A}$ expressed in terms of an open neighborhood of $\mathcal{A}_{\boldsymbol{\vartheta}}$:
\begin{equation}
\begin{aligned}
    \mathcal{U}^{\boldsymbol{\mu}} \!=\! \left\{ \mathbf{x_1} \in  \boldsymbol{\chi}_\mathbf{1} \!: \right. & \| \mathbf{z_1}\!(t,j) \|_{\mathcal{A}_{\boldsymbol{\vartheta}}} \!<\! \mu, \mathrm{dist}(\mathbf{\hat{q}_1}, \mathcal{Q}(\mathbf{R_1}\!) \!) \!\leq\! \alpha \!+\! \mu, \\ & \left. \|\boldsymbol{\varphi}^{\!-\!1\!}(m_1\boldsymbol{\Phi}(\mathbf{\hat{q}_1}, \!\mathbf{R_1}\!)\!)\| \leq 1 + \delta + \mu \right\},
\end{aligned}
\end{equation}
Analogously to the proof of \hyperref[thm:EquivalenceAsymptoticStability]{Theorem~\ref*{thm:EquivalenceAsymptoticStability}}, given that the map $\boldsymbol{\Phi}$ is double-valued for $\mathrm{dist}(\mathbf{\hat{q}_1}, \!\mathcal{Q}(\mathbf{R_1})\!) = 1$, the open neighborhood $\mathcal{U}^{\boldsymbol{\mu}}$ is defined for $\mu < 1 - \alpha$. By observing that the set $\mathcal{A}_R$ does not constraint the state variables $\mathbf{\hat{q}_1}$ and $m_1$, the exponential bound \eqref{eq:ExpBoundz1*} enables writing that each solution $\mathbf{x_1}$ to $\mathcal{H}_\mathbf{1}$ satisfying $\mathbf{x_1}(0, 0) \in \mathcal{U}^{\boldsymbol{\mu}}$, for all $(t,j) \in \textrm{dom } \! \mathbf{x_1}$, also satisfies
\begin{equation}
        \label{eq:ExpBoundx1*Ar}
        \|\mathbf{x_1^*}(t,j)\|_{\mathcal{A}_R} \leq  \kappa^*_{\!R} e^{-\lambda (t+j)}\|\mathbf{x_1^*}(0, 0)\|_{\mathcal{A}_R}
\end{equation}
Since $\mathcal{A} = \mathcal{A}_R \cap \mathcal{A}_{\hat{q}} \cap \mathcal{A}_m$, $\|\mathbf{x_1^*}(t,j)\|_{\mathcal{A}} \geq \|\mathbf{x_1^*}(t,j)\|_{\mathcal{A}_R}$ and $\|\mathbf{x_1^*}(t,j)\|_{\mathcal{A}} \geq \|\mathbf{x_1^*}(t,j)\|_{\mathcal{A}_{\hat{q}} \cap \mathcal{A}_m}$ for all $(t,j) \in \textrm{dom } \! \mathbf{x_1}$. In this way, bearing in mind \eqref{eq:ICMemoryStates}, the combination of the exponential bounds \eqref{eq:ExponentialBoundAqAm} and \eqref{eq:ExpBoundx1*Ar} implies that each solution $\mathbf{x_1}$ to $\mathcal{H}_\mathbf{1}$ verifying $\mathbf{x_1}(0, 0) \in \mathcal{U}^{\boldsymbol{\mu}}$, for each $(t,j) \in \textrm{dom } \! \mathbf{x_1}$, yields 
\begin{equation}
        \label{eq:ExpBoundx1*A}
        \|\mathbf{x_1^*}(t,j)\|_{\mathcal{A}} \leq  \kappa^*_{\!R} e^{-\lambda (t+j)}\|\mathbf{x_1^*}(0, 0)\|_{\mathcal{A}}. 
\end{equation}
Observe that, in virtue of \eqref{eq:ICMemoryStates}, the previous exponential bounds also yields for $\mathbf{x_1}$ to $\mathcal{H}_\mathbf{1}$ verifying $\mathbf{x_1}(0, 0) \in \mathcal{B}$. Therefore, under the assumption that the compact set $\mathcal{A}_\vartheta$ is exponentially stable for $\mathcal{H_2}$ for any given solution $\mathbf{x_2}$ satisfying $\mathbf{x_2}(0, 0) \in \mathcal{B}_{\vartheta}$, it follows from \cite{teel2012} that the compact set $\mathcal{A}$ is exponentially stable for $\mathcal{H}_\mathbf{1}$ for any given solution $\mathbf{x_1}$ verifying $\mathbf{x_1}(0, 0) \in \mathcal{B}$.

\par To prove the converse, suppose now that the set $\mathcal{A}$ is exponentially stable for any given solution $\mathbf{x_1}$ to $\mathcal{H}_{\mathbf{1}}$ satisfying $\mathbf{x_1}(0, 0) \in \mathcal{B}$. In this way, each solution $\mathbf{x_1}$ to $\mathcal{H}_\mathbf{1}$ verifying $\mathbf{x_1}(0, 0) \in \mathcal{B}$, for all $(t,j) \in \textrm{dom } \! \mathbf{x_1}$ yields 
\begin{equation}
        \label{eq:ExpBoundx1*AConverse}
        \|\mathbf{x_1^*}(t,j)\|_{\mathcal{A}} \leq  \kappa e^{-\lambda (t+j)}\|\mathbf{x_1^*}(0, 0)\|_{\mathcal{A}}.
\end{equation}
Given that for $\mathbf{x_1}(0, 0) \in \mathcal{B} \supset \mathcal{U}^{\boldsymbol{\mu}}$ one has $\mathrm{dist}(\mathbf{\hat{q}_1}, \!\mathcal{Q}(\mathbf{R_1})\!)|_{(0, 0)} < 1$, in light of \hyperref[lemSolutionsEquivalence]{Lemma~\ref*{lemSolutionsEquivalence}}, for every solution $\mathbf{x_1}$ to $\mathcal{H}_\mathbf{1}$, there exists a solution $\mathbf{x_2}$ to $\mathcal{H}_\mathbf{2}$ that, for every $(t,j) \in \mathrm{dom}\; \mathbf{x_1}$, \eqref{eq:SolutionsH1H2} holds with $(t,j') \in \mathrm{dom}\;\mathbf{x_2}$ and $j' \in \mathbb{N}$ satisfying $j' \leq j$. Thus, $\mathbf{x_1}(0, 0) \in \mathcal{B}$ implies $\mathbf{x_2}(0, 0) \in \mathcal{B}_{\boldsymbol{\vartheta}}$. Furthermore, as previously demonstrated, each solution $\mathbf{x_1}$ to $\mathcal{H}_\mathbf{1}$ verifying \eqref{eq:ExpBoundx1*AConverse} also satisfies 
\begin{equation}
        \label{eq:ExpBoundz1*Converse}
        \|\mathbf{z_1^*}\!(t,j)\!\|_{\mathcal{A}_R} \!\leq\! \kappa e^{\!-\!\lambda (t+j)}\|\mathbf{z_1^*}(0, 0)\|_{\mathcal{A}_R} \; \forall \; (t,j) \!\in\! \textrm{dom } \! \mathbf{x_1}
\end{equation}
Bearing in mind \eqref{eq:RotationMatrixFrobeniusNorm}, observe that for $\mathbf{x_1} \in \mathcal{A}_m$ the following lower-bound for $ \|\mathbf{R_1}(t,j) - \mathbf{I_3}\|_F$ holds: 
\begin{equation}
     \|\mathbf{R_1}(t,j) - \mathbf{I_3}\|_F \geq \nu \|\left.\boldsymbol{\varphi}^{\!-\!1\!} \! \left(m_1\boldsymbol{\Phi}\!\left(\mathbf{\hat{q}_1}, \!\mathbf{R_1}\!\right)\right)\right|_{(t,j)}\|^2 
\end{equation}
where $\nu \in \mathbb{R}_{\geq 0}$ verifies $\nu = 4\sqrt{2}(1+(1+\delta)^2)^{-1}(1+\delta)$. Consequently, given that $\mathbf{x_1} \in \mathcal{A}_R$ leads to $\mathbf{R_1} = \mathbf{I_3}$, which, in light of \eqref{eq:rotationmatrixMRP}, translates into $\boldsymbol{\varphi}^{\!-\!1\!}(m_1\boldsymbol{\Phi}(\mathbf{\hat{q}_1}, \!\mathbf{R_1}\!)\!) = \boldsymbol{0}$, by recalling \eqref{eq:ICMemoryStates} and \eqref{eq:UpperBoundFrobeniusNorm}, it follows from \eqref{eq:ExpBoundz1*Converse} that any given solution $\mathbf{x_1}$ to $\mathcal{H}_\mathbf{1}$ verifying \eqref{eq:ExpBoundx1*AConverse} also verifies
\begin{equation}
        \label{eq:ExpBoundz1*Converse}
        \|\mathbf{z_1}\!(t,j)\!\|_{\mathcal{A}_\vartheta} \!\!\leq\! \kappa_\vartheta e^{\!-\!\frac{\lambda}{2} (t+j)} \!\|\mathbf{z_1}\!(0, \! 0)\|_{\mathcal{A}_\vartheta} \!\; \forall \; (t,j) \in \textrm{dom } \! \mathbf{x_1}
\end{equation}
 with $\kappa_\vartheta = (4\sqrt{2}\kappa\nu^{-1}\|\mathbf{z_1}\!(0, 0)\|^{-1}_{\mathcal{A}_\vartheta})^{1/2}$. Therefore, it results from \eqref{eq:SolutionsH1H2} that for every solution $\mathbf{x_2}$ to $\mathcal{H}_\mathbf{2}$ satisfying $\mathbf{x_2}(0, 0) \in \mathcal{B}_{\boldsymbol{\vartheta}}$ also satisfies 
 \begin{equation}
             \label{eq:ExpBoundz1*Converse}
        \|\mathbf{x_2}(t,j) \! \|_{\mathcal{A}_\vartheta} \!\!\leq\!\! \kappa_\vartheta e^{\!-\!\frac{\lambda}{2} (t+j)} \!\|\mathbf{x_2}(0, \! 0)\|_{\mathcal{A}_\vartheta} \! \; \forall \; (t,j) \!\in\! \textrm{dom } \! \mathbf{x_2}
 \end{equation}
Hence, based on the definition provided in \cite{teel2012}, supposing that the compact set $\mathcal{A}$ is exponentially stable for every solution $\mathbf{x_1}$ to $\mathcal{H}_{\mathbf{1}}$ verifying $\mathbf{x_1}(0, 0) \in \mathcal{B}$, the compact set $\mathcal{A}_{\boldsymbol{\vartheta}}$ is exponentially stable for every solution $\mathbf{x_2}$ to $\mathcal{H}_{\mathbf{2}}$ verifying $\mathbf{x_2}(0, 0) \in \mathcal{B}_{\boldsymbol{\vartheta}}$. 

\section{Proof of Theorem 3}
\label{appendix:ProofTheorem3}

Let $V_2(\mathbf{\Tilde{x}}_{\boldsymbol{\vartheta}}): \boldsymbol{\chi_\vartheta} \mapsto \mathbb{R}_{\geq 0}$ be given by
\begin{equation}
   V_2(\mathbf{\Tilde{x}}_{\boldsymbol{\vartheta}}) =  2a \ln (1 + \|\boldsymbol{\Tilde{\vartheta}}\|^2) + b\boldsymbol{\Tilde{\vartheta}}^{\!\top}\!\!\boldsymbol{\sigma_{\!\vartheta}}(\mathbf{J}\mathbf{\boldsymbol{\Tilde{\omega}}}) + \frac{a}{2k_{\vartheta}}\boldsymbol{\Tilde{\omega}^{\!\top}}\mathbf{J}\boldsymbol{\Tilde{\omega}}
\end{equation}
\noindent where $\boldsymbol{\sigma_{\!\vartheta}}:\mathbb{R}^3 \mapsto \mathbb{R}^3$ denotes a saturation function with $M_\vartheta \in \mathbb{R}_{>0}$ as saturation level and $a,b \in \mathbb{R}_{>0}$ satisfy 
\begin{equation}
    \label{eq:condition_a1}
    a \!\!>\!\! \frac{b}{2} \!\max \!\! \left\{ \!\!\! \frac{\lambda_{\max}(\mathbf{J}) \alpha_{\!\vartheta}\sqrt{ k_\vartheta}}{\sqrt{\ln(\!1 \!\!+\! \alpha_{\!\vartheta}^2\!)\lambda_{\min}(\!\mathbf{J}\!)}},\! \!\frac{(\!1\!\!+\!(1\!+\!\delta)^2\!) k_\vartheta  \lambda_{\max}(\!\mathbf{J}\!)}{2k_\omega} \!+\! k_\omega \!\!\right\}
\end{equation}
\begin{equation}
    \label{eq:condition_a2}
    a > \frac{\sqrt{3}b M_\vartheta (1+\alpha_\vartheta^2) + \delta \alpha_\vartheta}{4\ln(1+\delta) \alpha_\vartheta}
\end{equation} 
with $\alpha_\vartheta = \max\{\|\boldsymbol{\Tilde{\vartheta}}(0, 0)\|, 1+\delta\}$. The function $V_2$ is continuously differentiable on $\boldsymbol{\chi_{\vartheta}}$ and radially unbounded. Thus, it follows that for any given $\mathbf{\Tilde{x}}_{\boldsymbol{\vartheta}}(0, 0)$, the set $\mathbf{U_2} = \{\mathbf{\Tilde{x}}_{\boldsymbol{\vartheta}} \in \boldsymbol{\chi_\vartheta}: V_2(\mathbf{\Tilde{x}}_{\boldsymbol{\vartheta}}) \leq V_2(\mathbf{\Tilde{x}}_{\boldsymbol{\vartheta}}(0, 0))\}$ is compact. The function $V_2$ is upper-bounded by $V_2 \leq \lambda_{\max}\left(\mathbf{A}_{\boldsymbol{\vartheta_1}}\right) \|\mathbf{\Tilde{x}}_{\boldsymbol{\vartheta}}\|^2$ with 
\begin{equation*}
        \mathbf{A}_{\boldsymbol{\vartheta_1}} = \frac{1}{2k_\vartheta}\left[\begin{array}{cc}
            4ak_\vartheta & bk_\vartheta\lambda_{\max}(\mathbf{J}) \\
            bk_\vartheta\lambda_{\max}(\mathbf{J}) & a\lambda_{\max}(\mathbf{J})
        \end{array}\right]
\end{equation*}
Given the bound $\|\boldsymbol{\Tilde{\vartheta}}\| \leq \alpha_\vartheta$, which is ensured by the hybrid path-lifting algorithm, one has $\ln(1 + \|\boldsymbol{\Tilde{\vartheta}}\|^2) \geq \ln(1+\alpha_\vartheta^2)\alpha_\vartheta^{-2}\|\boldsymbol{\Tilde{\vartheta}}\|^2$. With this bound in mind, $V_2$ is lower-bounded by $V_2 \geq \lambda_{\min}\left(\mathbf{A}_{\boldsymbol{\vartheta_2}}\right) \|\mathbf{\Tilde{x}}_{\boldsymbol{\vartheta}}\|^2$, where
\begin{equation*}
        \mathbf{A}_{\boldsymbol{\vartheta_2}} = \frac{1}{2k_\vartheta}\left[\begin{array}{cc}
            \frac{4ak_\vartheta \ln(1+\alpha_\vartheta^2)}{\alpha_\vartheta^2} & -bk_\vartheta\lambda_{\max}(\mathbf{J}) \\
            -bk_\vartheta\lambda_{\max}(\mathbf{J}) & a\lambda_{\min}(\mathbf{J})
        \end{array}\right]
\end{equation*}
In virtue of \eqref{eq:condition_a1}, the matrix $\mathbf{A}_{\boldsymbol{\vartheta_2}}$ is positive definite and, consequently, the function $V_2$ is positive definite with respect to  $\mathcal{A}_{\vartheta}$. Using the property $4\boldsymbol{\Tilde{\vartheta}}{}^{\!\top}\!\mathbf{T}(\boldsymbol{\Tilde{\vartheta}}) = (1\!+\!\|\boldsymbol{\Tilde{\vartheta}}\|^2)\boldsymbol{\Tilde{\vartheta}}{}^{\!\top}$ (\cite{junkins_2009}) and defining $\boldsymbol{\Lambda} = \diag(\!\boldsymbol{\dot{\sigma}_{\!\vartheta}}(\mathbf{J}\boldsymbol{\Tilde{\omega}})\!)$, $\dot{V}_2$ yields 
\begin{equation*}
    \label{eq:V2dot}
    \dot{V}_2 \!=\! -b k_\omega  \boldsymbol{\Tilde{\vartheta}}^{\!\top}\!\!\!\boldsymbol{\Lambda}\boldsymbol{\Tilde{\omega}} \!-\! b k_\vartheta \boldsymbol{\Tilde{\vartheta}}^{\!\top}\!\!\!\boldsymbol{\Lambda} \boldsymbol{\Tilde{\vartheta}} \!+\! b \boldsymbol{\Tilde{\omega}}^{\!\top} \! \mathbf{T}(\boldsymbol{\Tilde{\vartheta}})^{\!\top}\!\! \boldsymbol{\sigma_{\!\vartheta}}(\mathbf{J}\boldsymbol{\Tilde{\omega}}) \!-\! \frac{a k_\omega} {k_\vartheta}\!\|\boldsymbol{\Tilde{\omega}}\|^2 
\end{equation*}
\noindent By combining the properties detailed in \hyperref[def:SaturationFunction]{Definition~\ref*{def:SaturationFunction}} with the bound on the MRP error norm verified for $\mathbf{x^*_2} \in \mathbf{C_2^*}$ and the equality \cite[p.~123]{junkins_2009}
\begin{equation}
    \label{eq:MRPProperty1}
    \mathbf{T}(\boldsymbol{\Tilde{\vartheta}})^\top\mathbf{T}(\boldsymbol{\Tilde{\vartheta}}) = 4^{-2}(1 + \|\boldsymbol{\Tilde{\vartheta}}\|^2)^2\mathbf{I_3},
\end{equation}
which leads to $\|\mathbf{T}(\boldsymbol{\Tilde{\vartheta}})\| \leq 4^{-1}(1 + (1+\delta)^2)$, the time derivative of $V_2$ is upper-bounded by $\dot{V}_2 \leq - \alpha_\omega\|\mathbf{\Tilde{x}}_{\boldsymbol{\vartheta}}\|^2 \leq 0 \;\; \forall \;\; \mathbf{x^*_2} \in \mathbf{C_2^*}$ with 
\begin{equation*}
    \alpha_\omega = \inf_{t \geq 0} \lambda_{\min}(\mathbf{B}_{\boldsymbol{\vartheta}}(t))
\end{equation*}
\begin{equation*}
    \mathbf{B}_{\boldsymbol{\vartheta}} = \frac{1}{2}\left[\begin{array}{cc}
            b k_\vartheta \boldsymbol{\Lambda} & -b k_\omega \boldsymbol{\Lambda} \\
            -b k_\omega \boldsymbol{\Lambda} & \left(\frac{2ak_\omega}{k_\theta} - \frac{b(1+\alpha_\vartheta)\lambda_{\max}(\mathbf{J})}{2}\right)\mathbf{I_3} 
        \end{array}\right]
\end{equation*}
Since the block $bk_\vartheta\boldsymbol{\Lambda}$ is positive-definite and, in light of \eqref{eq:condition_a1}, its Schur complement it is also positive definite, $\mathbf{B}_{\boldsymbol{\vartheta}}$ is positive-definite for any $\boldsymbol{\Tilde{\omega}} \in \mathbb{R}^3$ and, consequently, $\alpha_\omega \in \mathbb{R}_{>0}$. The discrete evolution of $V_2$ is described by
\begin{equation*}
    V_2(\!\mathbf{G_2^*}(\mathbf{\Tilde{x}}_{\boldsymbol{\vartheta}})\!) - V_2(\mathbf{\Tilde{x}}_{\boldsymbol{\vartheta}}\!) \!=\! -4a\!\ln\!{(\|\boldsymbol{\Tilde{\vartheta}}\|)} - b\boldsymbol{\Tilde{\vartheta}}^{\!\top}\!\!\!\boldsymbol{\sigma_{\!\vartheta}}(\mathbf{J}\mathbf{\boldsymbol{\Tilde{\omega}}}\!)(1 + \|\boldsymbol{\Tilde{\vartheta}}\|^{-2})
\end{equation*}
\noindent Given the MRP error norm bound that stems from the hybrid path-lifting algorithm and \eqref{eq:condition_a2}, one has
\begin{equation}
    \label{eq:V2JumpBehavior}
    V_2(\!\mathbf{G_2^*}(\mathbf{\Tilde{x}}_{\boldsymbol{\vartheta}})\!) \leq e^{-\alpha_\delta} V_2(\!\mathbf{\Tilde{x}}_{\boldsymbol{\vartheta}}\!) \;\; \forall \;\; \mathbf{x^*_2} \in \mathbf{D_2^*}
\end{equation}
\noindent where $\alpha_\delta = -\ln\!\left(1 - \delta/\max\{V_2(0, 0), 2\delta\} \!\right)$. Therefore, $V_2$ strictly decreases during both flows and jumps, which implies that any solution $\mathbf{x_2^*}\!\left(t,\;j\right)$ to $\mathcal{H}^\mathbf{*}_\mathbf{2}$ remains in $\mathbf{U_2}$ for all $(t,j) \in \textrm{dom}\; \mathbf{\Tilde{x}}_{\boldsymbol{\vartheta}}$. Then, since $\mathbf{G_2^*}\left(\mathbf{D_2^*}\right) \subset \mathbf{C_2^*}$ prevents the solutions from jumping out of $\mathbf{C} \cup \mathbf{D}$, each maximal solution to $\mathcal{H}^\mathbf{*}_\mathbf{2}$ is bounded and complete \cite[Proposition 6.10]{goebel_2012}. Moreover, the bound on $\dot{V}_2$ and \eqref{eq:V2JumpBehavior} support the claims
\begin{equation*}
    \dot{V_2} \leq -\lambda_{\theta} V_2 \;\; \forall \;\; \mathbf{x^*_2} \in \mathbf{C_2^*}
\end{equation*}
\begin{equation*}
    V_2(\mathbf{G_2^*}(\mathbf{\Tilde{x}}_{\boldsymbol{\vartheta}})) \leq e^{-\lambda_\vartheta} V_2(\mathbf{x}_{\boldsymbol{\vartheta}}) \;\; \forall \;\; \mathbf{x^*_2} \in \mathbf{D_2^*}
\end{equation*}
with $\lambda_{\theta} = \min\{\alpha_\omega/\lambda_{\max}(\mathbf{A}_{\boldsymbol{\vartheta_1}}), \alpha_\delta\}$. Hence, according to \cite[Theorem 1]{teel2012}, the compact set $\mathcal{A}_\vartheta$ is globally exponentially stable for $\mathcal{H}_\mathbf{2}^\mathbf{*}$.

\section{Proof of Theorem 4}
\label{appendix:ProofTheorem4}

The proof starts with the observation that the hybrid system $\mathcal{H}_\mathbf{1}^{\mathbf{*}}$ behaves similarly to the lifting system
$\mathcal{H}_{\boldsymbol{\vartheta}}^*$ in terms of flows and jumps and that,   
from \hyperref[lemH*]{Lemma~\ref*{lemH*}}, the autonomous system $\mathcal{H}_{\boldsymbol{\vartheta}}^*$, which encapsulates the hybrid dynamic path-lifting algorithm, is well-posed. The flow and jumps sets, $\mathbf{C^*_1}$ and $\mathbf{D^*_1}$, respectively, are closed subsets of $\boldsymbol{\Omega} \times \boldsymbol{\chi}_\mathbf{R}$. Regarding the reference trajectory, the map $\boldsymbol{\omega}_{\mathbf{d}} \mapsto K_\omega \mathbb{B}^3$ is convex and bounded, and a differential equation given by a continuous function governs the flow evolution of $\mathbf{\dot{R}_d}$. Likewise, the underlying differential equation of the attitude tracking dynamics consists of a continuous function. Hence, the flow map $\mathbf{F^*_1} (\mathbf{x_1^*})$ is outer semicontinuous and locally bounded relative to $\mathbf{C^*_1} (\mathbf{x_1^*}) \subset \textrm{dom } \mathbf{F^*_1}$, and $\mathbf{F^*_1} (\mathbf{x_1^*})$ is convex for every $\mathbf{x_1^*} \in \mathbf{C^*_1}$. Furthermore, the reference trajectory $\mathbf{r}(t)$ and the angular velocity tracking error $\boldsymbol{\Tilde{\omega
}}$ remain constant during jumps and, consequently, the respective difference equations are continuous. Thus, combining these notions with the demonstration carried out in the proof of \hyperref[lemH*]{Lemma~\ref*{lemH*}}, the jump map $\mathbf{G^*_1} (\mathbf{x_1^*})$ is outer semicontinuous and locally bounded relative to $\mathbf{D^*_1} (\mathbf{x_1^*}) \subset \textrm{dom } \mathbf{G^*_1}$. Therefore, $\mathcal{H}_\mathbf{1}^{\mathbf{*}}$ meets the hybrid basic conditions enunciated in \cite[Assumption 6.5]{goebel_2012} and, in light of \cite[Theorem 6.30]{goebel_2012}, is well-posed.
\par From \hyperref[thm:StabilityAttitudeTrackingSystemMRP]{Theorem~\ref*{thm:StabilityAttitudeTrackingSystemMRP}}, it follows that the compact set 
$\mathcal{A}_\vartheta := \{\mathbf{x^*_2} \in \boldsymbol{\Omega} \times \boldsymbol{\chi_\vartheta}: \boldsymbol{\Tilde{\vartheta}} = \boldsymbol{0},\,\boldsymbol{\Tilde{\omega}} = \boldsymbol{0}\}$  is exponentially stable for the hybrid attitude tracking system $\mathcal{H}^\mathbf{*}_\mathbf{2}$ for all $\mathbf{x_2^*}(0, 0) \in \mathcal{B}_{\boldsymbol{\vartheta}} = \boldsymbol{\Omega} \times \boldsymbol{\chi_\vartheta}$. Then, based on \hyperref[thm:EquivalenceExponentialStability]{Theorem~\ref*{thm:EquivalenceExponentialStability}}, the compact set $\{\mathbf{x_1^*} \in \mathcal{A}_R: \|\boldsymbol{\varphi}^{\!-\!1\!}(m\boldsymbol{\Phi}(\mathbf{\hat{q}}, \mathbf{\Tilde{R}}))\| \leq 1 + \delta, \; 
\mathrm{dist}(\mathbf{\hat{q}}, \mathcal{Q}(\mathbf{\Tilde{R}})) \leq \alpha\}$,
with $\mathcal{A}_R := \{\mathbf{x^*_1} \in \boldsymbol{\Omega} \times \boldsymbol{\chi}_{\mathbf{R}}: \mathbf{\Tilde{R}} = \mathbf{I_3},\,\boldsymbol{\Tilde{\omega}} = \boldsymbol{0}\}$, is exponentially stable for the hybrid system $\mathcal{H}_{\mathbf{1}}^{\mathbf{*}}$ for all $\mathbf{x_1^*}(0, 0) \in \mathcal{B} = \left\{\mathbf{x_1} \in \boldsymbol{\Omega} \times \boldsymbol{\chi}_{\mathbf{R}}: \mathrm{dist}(\mathbf{\hat{q}}, \!\mathcal{Q}(\mathbf{\Tilde{R}})\!)<1
\right\}$. Hence, the compact set $\mathcal{A}_R$ is globally exponentially stable for $\mathcal{H}_{\mathbf{1}}^{\mathbf{*}}$. Furthermore, since $\mathcal{H}_{\mathbf{1}}^{\mathbf{*}}$ is well-posed, according to \cite[Theorem 7.21.]{goebel_2012}, the global stability result of $\mathcal{A}_R$ for $\mathcal{H}_{\mathbf{1}}^{\mathbf{*}}$ is robust. The robustness margin to perturbations can be quantified through $\mathcal{KL}$ bounds, as detailed in \cite[Definition 7.18.]{goebel_2012}. Thus, the compact set $\mathcal{A}_R$ is robustly globally exponentially stable for $\mathcal{H}_{\mathbf{1}}^{\mathbf{*}}$.

\bibliographystyle{IEEEtran}
\bibliography{IEEEabrv,CDC.bib}

\end{document}